# Triple State QuickSort

## A replacement for the C/C++ library qsort

Ammar Muqaddas
Center of Information Systems, Kuwait University
Shuwaikh Campus, Kuwait
ammar.mu@ku.edu.kw

***Abstract*—An industrial grade Quicksort function along with its new algorithm is presented. Compared to 4 other well known implementations of Quicksort, the new algorithm reduces both the number of comparisons and swaps in most cases while staying close to the best of the 4 in worst cases. We trade space for performance, at the price of *n/2* temporary extra spaces in the worst case. Run time tests reveal an overall improvement of at least 15.8% compared to the overall best of the other 4 functions. Furthermore, our function scores a 32.7% run time improvement against Yaroslavskiy's new Dual Pivot Quicksort. Our function is pointer based, which is meant as a replacement for the C/C++ library qsort(). But we also provide an array based function of the same algorithm for easy porting to different programming languages.**

## I. INTRODUCTION

Since its development by Tony Hoar in 1962 [1], Quicksort has been one of the most widely used sorting algorithms. C/C++ libraries used Quicksort as a general purpose algorithm (qsort function) on most platforms including Unix, Linux and Windows. Despite its $O(n^2)$ worst case and the existence of Mergesort and HeapSort which have much better worst cases of $O(n\ln(n))$. Quicksort, when implemented efficiently, has a better average performance [21], one of the main reasons behind its wide adoption. Although this has been debated recently [5]. Reference [2] gives a good historical background of Quicksort and its different variations.

Although Quicksort is easy to write at its barebones form, it's not straight forward to implement efficiently. This explains historical cases of systems hogging CPU for hours while sorting some inputs (e.g. 2n organ pipe array of 123…nn…321) which should have taken minutes [3]. These simple to cook inputs can cause Quicksort to do $O(n^2)$ comparisons. Unfortunately, as will be discussed later, such bad cases can still be cooked easily on a modern C/C++ library qsort!! despite the fact that Bentley and McIlroy showed how to avoid them twenty years ago [3].

Fortunately, an adapted version of Bentley's excellent qsort [3] has been used in Java for years. It was not until 2009 when Vladimir Yaroslavskiy introduced his new Dual Pivot Quicksort [4] which was compared to Bentley's approach and appeared to perform better. It was not long until Oracle decided to replace the Quicksort function in Java 7 with the dual pivot method. We will show later that our algorithm performs even better than dual pivot Quicksort.

In this work, we take an engineering approach to develop a general purpose, industrial, efficient and scalable Quicksort function that reduces both the number of comparisons and swaps using a rather complex Triple State algorithm. Then we deploy several further optimizations to reduce the run time. In particular, our algorithm is specially designed to reduce the total swaps to about 2/3 of classical Quicksort and even better in common input cases like nearly sorted, nearly reversed, many equals or stair steps and the like.

## II. REDUCING NUMBER OF SWAPS

Given an array ar[ ] to be sorted. In most Quicksort variations that has been developed so far, after choosing a pivot p, two pointers (or index variables) run in opposite direction. **l** from left and **r** from right until two elements **ar[l]** and **ar[r]** are found that need to be swapped (i.e the left element is greater than p and the right element is less than p) then a swap is done. **However, a swap operation needs three memory copy operations**. i.e. **temp=ar[l]; ar[l]=ar[r]; ar[r]=temp**. There exists another method that will almost eliminate the need for that third temporary variable copy. By copying only the first **ar[r]** that is less than p to the temp variable, we create an empty space in the array. Then we proceed scanning from left to find the first **ar[l]** that is greater than p. Then copy **ar[r]=ar[l]**. Now the empty space is at **ar[l]**. We scan from right again then copy **ar[l]=ar[r]** and continue as such. As long as the temp variable hasn't been copied back to the array, the empty space will remain there juggling left and right. The following code snippet explains.

```
// Pre-scan from the right
while (ar[r]>p)
   r--;
temp = ar[r];

// Main loop
while (l<r)
{
   while (l<r && ar[l]<p)
      l++;
   if (l<r) ar[r--] = ar[l];
   while (l<r && ar[r]>p)
      r--;
   if (l<r) ar[l++] = ar[r];
```

```
}
// After loop finishes, copy temp to left side
ar[r] = temp; l++;
if (temp==p) r--;
```

This approach is not new, it has been used before by Paul Hsieh [5], his Quicksort performs reasonably well but unfortunately does badly in reverse order input. Yet this simple method is still very valid and deserves exploitation.

## III. TRIPLE STATE ALGORITHM

Two key points in developing an efficient Quicksort are 1) finding a pivot as closest to the real median as possible and 2) handling of elements equal to the pivot **p**. We will describe our pivot selection method later.

Three way partitioning [6] is a well known method of handling equals, were all elements equal to **p** are kept separately in a third partition to exclude them from being compared again to **p** in subsequent recursive stages. This method reduces heavily the number of comparisons, but the difficulty is **where** to keep those equals. Bentley & McILROY [3] swap them to the sides then swap them back to the middle at the end of the recursive stage. We **keep** or **copy** them to the middle immediately. Consider the following variable, function and macro definitions:

- **l** and **r** are two indexes scanning array **ar[]** from left and right respectively.
- **mid** is the index at the center of the array.
- **p** is the pivot**. pi** is the pivot index where it was picked.
- **ml** and **mr** are two indexes defining the left and right boundaries of the "all equals" partition respectively.
- **m** is the index where the next elemnent equal to **p** should be stored it is either equal to **ml** or **mr** at any time in State 1. Index **m** has also a different porpose when in States 2L, 2R, 3L or 3R, it defines the left or right side of the "all equals" partition after.
- **temp** is a temporary variable.
- **n** is the number of elements in the array
- **a** and **b** are the indexes of the first and last elements to be sorted in the current recursive stage. Initially, **a** = 0 and **b** = **n**-1.
- **cmp**(x,y). Callback comparison function, were x,y are pointers to two array elements. Returns a positive value if *x>*y, a negative value if *x<*y. A value of zero if *x=*y. This is given to the algorithm by the user. C/C++ libraries religiously use this method. This allows qsort to be generic and completely independent of the datatype of the elements in the array[1].
- **lc** stands for "Last Comparison result". It stores the result of the last (latest) comparison operation done.
- **MLEFT( )** and **MRIGHT( )** are two macros that decide which state to go to after state 1 terminates. The code for both will be shown later.
- **MLEFT_CHECKM( )** and **MRIGHT_ CHECKM( )** are identical to **MLEFT( )** and **MRIGHT( )** with minor differences.
- **ar[ ]** is as indicated above the array to be sorted. **tar[ ]** is a temporary buffer array that will be used latter in states 3L and 3R.
- Type **Elem** is a generic type that we use here to represent an element in the array. **Elem** could be a struct or a class containing any number of fields, integers, doubles, strings, etc.

Fig. 1, shows the overall work flow of the algorithm. We will first show the initialization which is done after pivot selection. Then proceed to State 1. Note that for clarity, we ignore handling odd and exception cases in the bellow algorithms. Please refer to the complete algorithm in [12] for the fully working code.

[1] A Callback function, although generic, is not necessarily the most efficient method for doing comparison since it involves pointer function calls. An inline, inplace or macro comparison is preferred when implementing open source or project specific sorting functions.

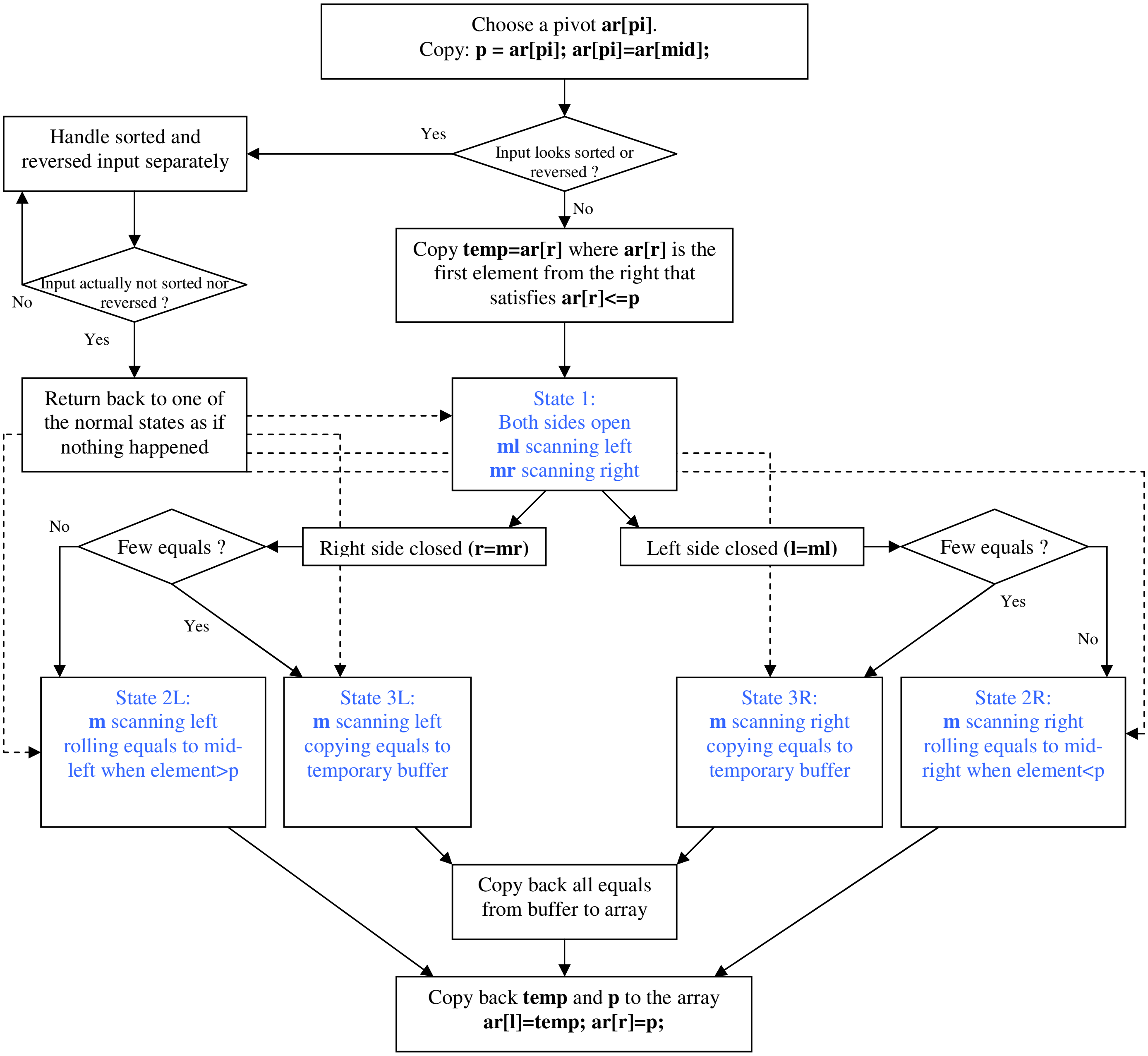

Fig. 1. One recursive stage in Triple State QuickSort. Two elements (**temp** and **p**) are kept outside the array to alow juglling elements without swaps.

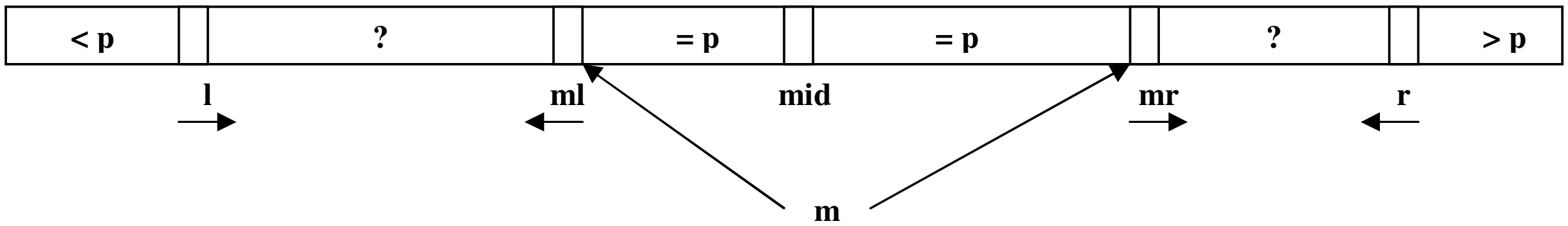

Fig. 2. Shows State 1. Both sides are still open for scanning (comparing). Arrows show movement direction of their indexes.

## A. *Initialization*

```
/* After pivot has been chosen, copy it to p. And make sure
element at index mid is empty. */
p = ar[pi]; ar[pi]=ar[mid];

// Handle special cases separately here ....

// Initialize l,r,m,mr,ml
l=a; r=b;
m=mr=ml=mid;
/* Pre-scan from the right to find first ar[r]<=p and place
it in temp creating an empty space ready for copying to. */
while (cmp(&ar[r],&p)>0)
{   r--;
    if (r==mid)
    {   // Handle special case. See full code...
    }
}

temp=ar[r];
```

## B. *State 1 Algorithm*

At this point the array has two empty spaces, one created by copying **ar[pi]** to **p**, and the other by copying **ar[r]** to **temp**. Those empty spaces are maintained while juggling elements left and right throughout the algorithm until the current recursive stage ends were both elements are copied back from **p** and **temp** to the array.

Fig. 2, shows State 1 processing. Algorithm is as follows:

```
l_scan1:
while ((lc=cmp(&ar[l],&p))<0) // While ar[l]<p
{   l++; if (l==ml) MRIGHT_CHECKM()
}
l_scan1_4:
if (lc==0)
{
l_scan1_3:
    ar[m]=ar[l];
    if (r-mr>ml-l)         // choose largest gap
    {   mr++; goto mr_scan1;
    }
    else
    {   ml--;
        if (ml==l)
        {   MRIGHT();
        }
        else goto ml_scan1;
    }
}
else
{
l_scan1_2:
    ar[r]=ar[l]; r--;
    if (mr==r) MLEFT_CHECKM() else goto r_scan1;
}

mr_scan1:
while ((lc=cmp(&ar[mr],&p))==0)  // While ar[mr]==p
{   mr++; if (mr==r) MLEFT();

}
mr_scan1_2:
if (lc<0)
{   ar[l]=ar[mr]; l++;
    if (ml==l)
    {   ar[mr]=ar[ml]; MRIGHT();
    }
    else
    {   m=mr; goto l_scan1;
    }
}
else
{   ar[r]=ar[mr]; r--;
    if (mr==r) MLEFT()
      else
    {   m=mr;
        goto r_scan1;
    }
}

ml_scan1:
while ((lc=cmp(&ar[ml],&p))==0)  // While ar[ml]==p
{   ml--; if (ml==l) MRIGHT();
}
ml_scan1_2:
if (lc<0)
{   ar[l]=ar[ml]; l++;
    if (l==ml) MRIGHT()
      else
    {   m=ml; goto l_scan1;
    }
}
else
{   ar[r]=ar[ml]; r--;
    if (mr==r)
    {   ar[ml]=ar[mr]; MLEFT();
    }
    else
    {   m=ml; goto r_scan1;
    }
}

r_scan1:
while ((lc=cmp(&ar[r],&p))>0)
{   r--; if (mr==r) MLEFT_CHECKM();
}
r_scan1_2:
if (lc==0)
{   ar[m]=ar[r];
    if (r-mr>ml-l)         // choose largest gap
    {   mr++; goto mr_scan1;
    }
    else
    {   ml--; if (ml==l) MRIGHT() else goto ml_scan1;
    }
}
else
{
r_scan1_3:
    ar[l]=ar[r]; l++;
    if (ml==l) MRIGHT_CHECKM() else goto l_scan1;
}
```

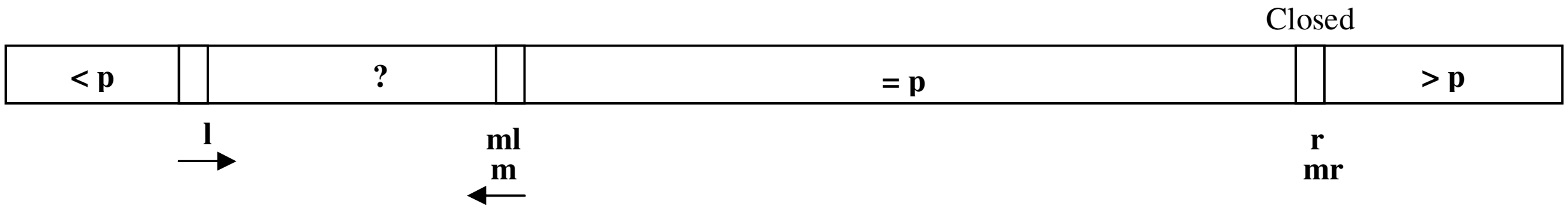

Fig. 3. State 2L. **mr, ml** and **r** are not scanning any more. **r** only moves to left when elements > **p** are found. Only **l** and **m** are scanning in the directions indicated. State 2R is exactly the same but on the oposite direction.

First, we have to say that the above algorithm can be written in a much simpler and readable form without all the gotos and repetitive code, by means of a main loop and conditional statements. However, we implement a case specific approach were the CPU would need to do minimum tests and jumps. It is complex spaghetti code, but generally runs faster. Readability is of minor consideration to us here since this is meant to be a replacement for a library function which very rarely needs to be modified or even looked at.

State 1 algorithm consists of 4 while loop sections, one for each of the 4 index variables **l**,**r**,**ml**,**mr** as they scan the array while comparing the elements with pivot **p** in the directions indicated in Fig. 2. Variable **l** starts scanning from extreme left going right until it finds an element greater than or equal to **p**. If the element is greater than **p**, it is copied to the right partition at **ar[r]** (which we guaranteed to be empty), then scanning starts from **r** going left. If the element is equal to **p** then it is copied to the equals partition at **ar[m]**, were **m** always points to the edge (on the equals partition) that has the largest distance between that edge and the directly facing partition. In other words, if **r-mr** > **ml-l**, then **m** will be equal to **mr**, otherwise **m** is equal to **ml**. Now scanning will proceed from the middle (equals partition) again from the side with the largest distance to its facing partition. This scan will continue, either increasing **mr** or decreasing **ml** (depending on the side chosen) until an element greater than **p** or less than **p** is found. If it's greater than **p** it will be copied to **ar[r]** and scanning starts from **r** going left. If it's less than **p** it will be copied to **ar[l]** and scanning restarts again from **l** going right.

The reader can guess easily now that scanning from **r** going left is identical to the process described above but in reverse direction.

The purpose of the **r-mr** > **ml-l** decision above is to keep State 1 running as long as possible because it's the most efficient state were all equals to **p** are immediately put in the middle and no shifting nor copying to a temporary buffer is required.

### *C. Choosing the next state*

State 1 comes to an end when either side of the equals partition (middle partition) closes. i.e when either **r** = **mr** or **l** = **ml**. Obviously the side that closed can not take more elements that are equal to **p**, they must go to the other side of the equals partition. Furthermore, elements that belong to that closing side (i.e. greater or less than **p**) must still be placed there. Hence a different strategy is needed. We have two methods to tackle this, either states 2L and 2R or states 3L and 3R. States 2L and 2R are more efficient when there are relatively **many** elements that are equal to pivot **p**. States 3L and 3R are more efficient when there are relatively **few** elements that are equal to pivot **p**. To decide, we make a simple *guessing* test as follows.

If the right side closed (i.e. **mr** = **r**) we test if:

$$mr - ml \le \frac{ml - l}{4} \qquad (1)$$

If inequality (1) is found true then there are probably **few** equals to **p**, so proceed to state 3L otherwise proceed to state 2L.

If the left side closed (i.e. **ml** = **l**) we test if:

$$mr - ml \le \frac{r - mr}{4} \qquad (2)$$

If inequality (2) is found true then there are probably **few** equals to **p**, proceed to state 3R otherwise proceed to state 2R.

Here is the code for the **MLEFT( )** and **MRIGHT( )** macros we defined earlier. As noted above these decide which state to go to after state 1.

```
#define MLEFT()                              \
{   do                                       \
    {   ml--; if (ml==l) goto exit2;         \
    } while ((lc=cmp(&ar[ml],&p))==0);      \
    m=ml; ml++;                              \
    if ((mr-ml)<=(ml-l)/4)                   \
       goto m_scan3L_2;                      \
    else                                     \
       goto m_scan2L_2;                      \
}
```

```
#define MRIGHT()                             \
{   do                                       \
    {   mr++; if (mr==r) goto exit2;         \
    } while ((lc=cmp(&ar[mr],&p))==0);      \
    m=mr; mr--;                              \
    if ((mr-ml)<=(r-mr)/4)                   \
       goto m_scan3R_2;                      \
    else                                     \
       goto m_scan2R_2;                      \
}
```

The **MLEFT_CHECKM( )** and **MRIGHT_CHECKM( )** are very similar with minor differences. Refer to complete code in [12].

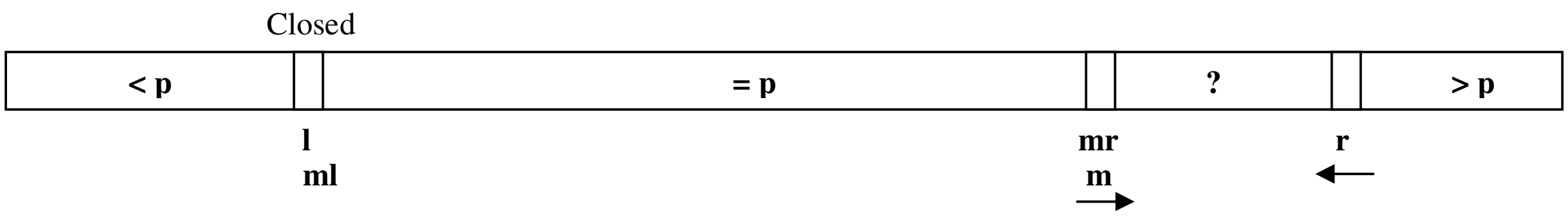

Fig. 4. State 2R. **mr, ml** and **l** are not scanning any more. **l** only moves to right when elements < **p** are found. Only **r** and **m** are scanning in the directions indicated. State 2L is exactly the same but on the oposite direction

### *D. States 2L and 2R Algorithm*

States 2L and 2R are identical to each other except that they work in opposite direction of each other. So it's sufficient to describe one of them in order to understand the other. Fig. 3, shows the workings of state 2L. Notice as we said before we are still maintaining two empty spaces in the array to allow for direct copying without expensive swapping. One of those empty spaces is always guaranteed to be at **r** during state 2L. Algorithm is as follows.

```
// State 2L: m moving left, right side closed
m_scan2L:
while ((lc=cmp(&ar[m],&p))==0)
{   m--; if (m==l) goto exit2;
}
m_scan2L_2:
if (lc<0)
{   ar[l]=ar[m]; l++;
    if (m==l) goto exit2; else goto l_scan2L;
} else
{   ar[r]=ar[m]; r--; ar[m]=ar[r]; m--;
    if (m==l) goto exit2; else goto m_scan2L;
}
l_scan2L:
while ((lc=cmp(&ar[l],&p))<0)
{   l++; if (m==l) goto exit2;
}
if (lc==0)
{   ar[m]=ar[l]; m--;
    if (m==l) goto exit2; else goto m_scan2L;
} else
{   ar[r]=ar[l]; r--; ar[m]=ar[r]; m--;
    if (m==l) goto exit2; else goto m_scan2L;
}
```

As before the algorithm above is written in favor of efficiency over readability. State 2L starts by **m** scanning going left until an element less than or greater than **p** is hit. If less than **p** it will be copied to the left partition at **l** then scanning resumes from **l** going right. If greater than **p** it will copied to the right partition at **r,** then a *roll* is performed so that **r** is decremented by 1, then **ar[r]** (which is part of the equals partition) will be copied to **ar[m]** then scanning continues from **m** going left.

If scanning resumes from **l** going right (when element<**p**) it will continue until an element greater than or equal to **p** is found. If equal to **p,** it will be copied to the equals partition at **m** then scanning resumes again from **m** going left as before. If the element was greater than **p** it will be copied to the right partition at **r**, then the same *roll* is performed exactly as it is described above.

State 2R workings is shown in Fig. 4, the algorithm is as a follows.

```
// State 2R. m moving right, left side closed
m_scan2R:
while ((lc=cmp(&ar[m],&p))==0)
{   m++; if (m==r) goto exit2;
}
m_scan2R_2:
if (lc<0)
{   ar[l]=ar[m]; l++; ar[m]=ar[l]; m++;
    if (m==r) goto exit2; else goto m_scan2R;
} else
{   ar[r]=ar[m]; r--;
    if (m==r) goto exit2; else goto r_scan2R;
}
r_scan2R:
while ((lc=cmp(&ar[r],&p))>0)
{   r--; if (m==r) goto exit2;
}
if (lc==0)
{   ar[m]=ar[r]; m++;
    if (m==r) goto exit2; else goto m_scan2R;
} else
{   ar[l]=ar[r]; l++; ar[m]=ar[l]; m++;
    if (m==r) goto exit2; else goto m_scan2R;
}
```

### *E. States 3L and 3R Algorithm*

The rolling method in states 2L and 2R is not efficient when there are only few equals to **p** because they have to be rolled a relatively long distance (i.e. copied too many times). The best way is to copy those equals to an external temporary buffer and only copy them back to the middle of the array when the recursive stage has completed. Fig. 5, shows the workings of state 3L. Its algorithm is as follows.

```
// State 3L: m moving left, right side closed
m_scan3L:
while ((lc=cmp(&ar[m],&p))==0)
{
    m_scan3L_3:
    tar[i++]=ar[m]; m--; if (m==l) goto exit3L;
}
m_scan3L_2:
if (lc<0)
{
    m_scan3L_4:
    ar[l]=ar[m]; l++;
    if (m==l) goto exit3L; else goto l_scan3L;
} else
{   // Check for a run of elements >p
    k=m;
    do
    {   m--; if (m==l) break;
    } while ((lc=cmp(&ar[m],&p))>0);
```

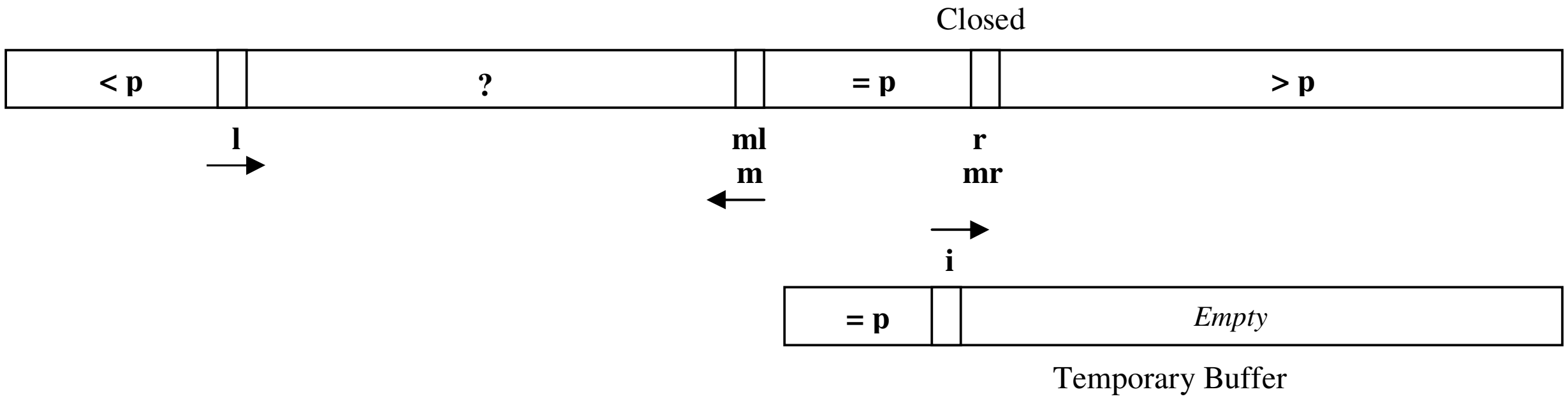

Fig. 5. State 3L. Elements equal to **p** are copied to temporary buffer **tar[ ]**, except for ones that are already found in the array (already in the equals partition), these are copied to **tar**[ ] one by one each time an element > **p** is found. **mr, ml** and **r** are not scanning any more. **r** only moves to left when elements > **p** are found. Only **l** and **m** are scanning in the directions indicated. State 3R is exactly the same but on the oposite direction

```
    k2=m+1;
    // Copy the run to the gap between r and m
    if (k-m<r-k)
        do
        {   ar[r]=ar[k2]; r--;
            if (r>=ml) { tar[i++]=ar[r]; }
        } while ((++k2)<=k);
    else
        for (;;)
        {   ar[r]=ar[k2]; r--;
            if (r>=ml) { tar[i++]=ar[r]; }
            else if (r<=k) { r=k2; break; }
            k2++;
        }
    if (m==l) goto exit3L; else if (lc==0) goto m_scan3L_3;
    else goto m_scan3L_4;

}
l_scan3L:
while ((lc=cmp(&ar[l],&p))<0)
{   l++; if (m==l) goto exit3L;
}
if (lc==0)
{
    l_scan3L_3:
    tar[i++]=ar[l]; m--;
    if (m==l) goto exit3L; else goto m_scan3L;
} else
{
    l_scan3L_2:
    ar[r]=ar[l]; r--;
    // If ar[r] is not empty, make it
    if (r>=ml) { tar[i++]=ar[r]; }
    m--;
    if (m==l) goto exit3L; else goto m_scan3L;
}
```

State 3L starts by **m** scanning going left while copying elements equal to **p** to the temporary buffer **tar[ ]** until an element less than or greater than **p** is found. If less than **p** it will be copied to the left partition at **l** then scanning resumes from **l** going right. If greater than **p**, a scan will be performed by **m** going left to find a run of elements greater than **p**. Then, starting from left at **m**, the run will be copied to the gap between **r** and **m** until the gap is filled or the whole run is exhausted. While doing this, if **r > ml** (i.e if **r** is still crossing the equals partition) then **ar[r]** will be copied to temporary buffer **tar[ ]** at **i** until copying the run has finished. Then scanning continues from **m** going left.

If scanning resumes from **l** going right (when element<**p**) it will continue until an element greater than or equal to **p** is found. If equal to **p,** it will be copied to the temporary buffer at **i** then scanning resumes again from **m** going left as before. If the element was greater than **p** it will be copied to the right partition at **r**, then if **r > ml** (i.e if **r** is still crossing the equals partition) then **ar[r]** will be copied to temporary buffer **tar[ ]** at **i**. Then scanning resumes from **m** going left.

Fig. 6, shows state 3R. It is identical to state 3L except that it works in reverse direction. State 3R algorithm is as follows.

```
// State 3R: m moving right, left side closed
m_scan3R:
while ((lc=cmp(&ar[m],&p))==0)
{
    m_scan3R_3:
    tar[i++]=ar[m]; m++; if (m==r) goto exit3R;
}
m_scan3R_2:
if (lc>0)
{
    m_scan3R_4:
    ar[r]=ar[m]; r--;
    if (m==r) goto exit3R; else goto r_scan3R;
} else
{   k=m;
    // Check for a run of elements <p
    do
    {   m++; if (m==r) break;
    } while ((lc=cmp(&ar[m],&p))<0);

    k2=m-1;
    // Copy the run to the gap between l and m
    if (m-k<k-l)
        do
        {   ar[l]=ar[k2]; l++;
            if (l<=mr) { tar[i++]=ar[l]; }
        } while ((--k2)>=k);
    else
        for (;;)
        {   ar[l]=ar[k2]; l++;
            if (l<=mr) { tar[i++]=ar[l]; }
            else if (l>=k) { l=k2; break; }
            k2--;
        }
    if (m==r) goto exit3R; else if (lc==0) goto m_scan3R_3;
    else goto m_scan3R_4;
}

r_scan3R:
while ((lc=cmp(&ar[r],&p))>0)
{    r--; if (m==r) goto exit3R;
}
if (lc==0)
{   tar[i++]=ar[r]; m++;
    if (m==r) goto exit3R; else goto m_scan3R;
} else
{   ar[l]=ar[r]; l++;
    if (l<=mr) { tar[i++]=ar[l]; }
    m++;
    if (m==r) goto exit3R; else goto m_scan3R;
}
```

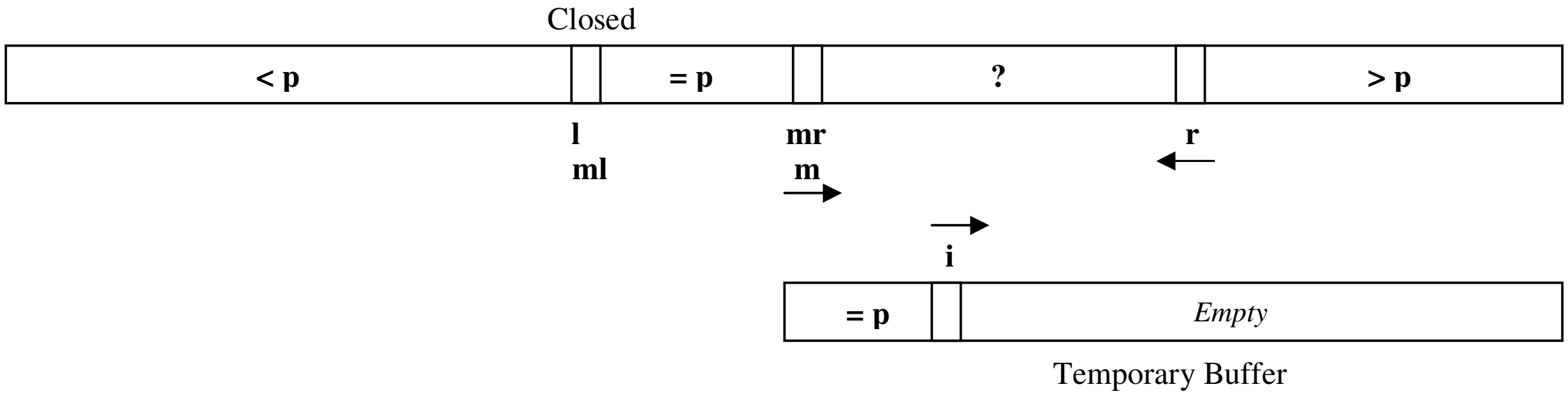

Fig. 6. State 3R. Elements equal to **p** are copied to temporary buffer **tar**[ ], except for ones that are already found in the array (already in the equals partition), these are copied to **tar**[ ] one by one each time an element < **p** is found. **mr, ml** and **l** are not scanning any more. **l** only moves to left when elements < **p** are found. Only **r** and **m** are scanning in the directions indicated. State 3L is exactly the same but on the oposite direction.

### F. *Copying back from temporary storage*

At this point the previous state has finished its loop. If we came from state 3L or state 3R then we need to copy back the contents of the temporary buffer **tar[ ]** to the sorting array **ar[ ]**. All of the elements in **tar[ ]** are equal to **p** so they all should be put in the middle partition of **ar[ ]** which now has an empty gap. This is done by a simple copying while loop. For state 3L:

```
exit3L:
while(i>0)
{   i--; m++;
    ar[m]=tar[i];
}
```

And for state 3R:

```
exit3R:
while (i>0)
{   i--; m--;
    ar[m]=tar[i];
}
```

Also in all cases (i.e after any state), we need to copy back both the **temp** and **p** variables into the right place in **ar[ ]** as follows.

```
exit2:
if ((lc=cmp(temp,&p))>=0) // if temp>=p
{   ar[r]=temp; ar[l]=p;
    l--;
    if (lc==0) r++;
} else
{   ar[l]=temp; ar[r]=p;
    r++;
}
```

Variables **l** and **r** are also adjusted as above so the next recursive stage doesn't include an element that is equal to the current pivot **p** since these elements are already in the correct place.

### G. *A note on temporary storage*

Triple State dynamically and automatically allocates temporary storage requiring *n/2* elements space[2]. However, our implementation favors speed over memory consumption. The temporary storage is only allocated at the start of the sorting and is NOT deallocated after sorting has finished, this is to allow for speedy sorting in the next few sorting calls which can use the same allocated space if **n** is not higher. If the user/programmer knows that there will be no sorting done within the near future, and memory is needed. The programmer can MANUALLY call the FreeTripleStateQuickSortStorage( ) function we provide to free temporary storage.

### H. *Pivot Selection*

Probably the most crucial part of Quicksort is finding a partitioning element. Ideally, the median is the best choice [22]. But finding the real median is expensive; at it basics, it requires the array to be sorted in the first place. There exists better algorithms like the Floyd-Rivest algorithm [8] which has[3] a running time of $O(n)$, specifically it requires about $1.5n$ comparisons. Unfortunately even those fast methods are not applicable for sorting since Quicksort needs to determine a median at every recursive stage in its call tree which would result in more than doubling the number of comparisons at each stage. As in most Quicksort implementations, the practical approach is to find an element that is merely a good *guess* of the median with very little overhead. A Bare bones Qucksort that is found in educational text books usually uses the first or last element, which can lead to the worst case of $O(n^2)$ easily when the array is sorted or reversed. A bit more robust way is to pick a random element as the pivot, this makes the worst case extremely unlikely to occur provided the random generator was seeded using the system timer[4]. Still, this method lacks a clue as to what the real median is, simply choosing one element randomly out of thousands of elements is not necessarily a good guess, since the probability of choosing any element is equal across all elements. We need a method that is a bit more biased towards choosing elements closer to the median. Singleton suggests the median of three method, i.e. the median of the first, middle and last elements [9]. This method turns out to be superior to random pivots [3] and has been used by Sedjewick [7] and the GNU C Library qsort( ) which is what Linux uses. It has several side effect advantages:

- Always picks the exact median in the case of sorted and reversed arrays because the middle (median) element is among the three.
- The three elements can be sorted into place while finding their median which puts the first and last elements into the correct place, so the recursive stage can exclude the first and last element from comparisons.

---

[2] It actually requires *n/2* only in the worst case, but it auto allocates the whole *n/2* to avoid reallocation, favoring speed over memory.

[3] It can actually find the ith smallest element, the median is just a special case were $i=n/2$.

[4] Seeding the random generator with a known constant will always generate a fixed sequence of random numbers, someone might cook an array to consistently cause the worst case of $O(n^2)$ to occur. Hence the better way is to seed it using a totally independent and unknown value like the system timer or the clock. This way every time the program runs, random generation will have a different sequence.

Bentley uses Tukey's "ninther", the median of the medians of three samples [3]. Basically, it samples and sorts 9 elements, hence gives better estimates of the median at the cost of more comparisons. The elements chosen are equally spaced samples. Bentley uses it only for mid sized and large arrays while small arrays are handled using the aforementioned median of three[5]. Bentley shows this yields better results.

Yaroslavskiy's dual pivot Quicksort uses a **5-element sorting network** [4]. It performs a fixed number of comparison operations (9 in this case) to sort 5 samples then take the $2^{nd}$ and $4^{th}$ elements as its two pivots.

Our empirical results show that the more we sample, the closer we are to the median but the improvement is usually not significant, reference [7] confirms this. However, the benefits become more significant when array size is large or very large because then the relatively small overhead of extra samples is overcome by proximity to the median which would eventually lower the overall number of comparisons. Following this notion and the success of the ninther in Bentley's algorithm, we choose a more elaborate method as follows:

- If **n** is very small (i.e tiny arrays) Insertion Sort is used. As has been done by all modern Quicksorts above.
- If **n** is small, use median of three.
- If **n** is medium sized, use median of **five**. (Will be explained shortly).
- If **n** is large use the ninther
- If **n** is very large use the median of five of the median of 3 samples, i.e. the **fifteenther**. (Will be explained shortly).

A reminder here that **n** could be the size of the array or sub-array at any recursive stage in Quicksort.

The median of five method we use above is less aggressive than the ninther and is suitable for arrays sizes that are not quite large. To write an efficient median of five algorithm, a decision tree has to be made (similar to the median of three tree in [3]), were it detects the exact permutation then finally do the swapping. However, the tree is huge since there are 120 (5!) permutations, which would result in a huge algorithm that is tedious to right. We created a program generator to generate the algorithm but the generated code is too lengthy to list here. It can be found in our complete code [12]. We designed the median of five to do the least number of swaps (element copies in our case) as our first objective, then the least number of comparisons as our second objective. It does 6 element copies in the worst case (equivalent to 2 (6/3) swaps) and 8 comparisons in the worst case[6].

The fifteenther above is similar to the nither but samples 15 elements instead of 9. As in the ninther, a median of three is found for every group of three elements, five of them here. Then the median of five is found for the five medians. This method scales better for very large arrays sorting millions of elements.

### *I. Handling special and common cases*

An input array can be described by two properties. First, elements probabilistic distribution, i.e. what elements are there and how many of each. Second, elements ordering, i.e. the permutation of those elements. Real world datasets are obviously not necessarily random data following a uniform distribution. More often, datasets follow the Normal (Gaussian) distribution. Or we might be sorting distinct keys. The array could also be nearly sorted or reversed. Or have sub parts sorted and sub parts reversed. They might be stair steps with many equal elements, i.e. few repeating. And the list goes on. If a case is encountered too often, it is beneficial to handle it separately. In fact, Timsort which is based on Mergesort is built on the basis that many datasets seen in the real world are nearly or fully sorted; it takes advantage of that but with some amount of hit towards less common datasets.

In Triple State Quicksort we handle only two common cases separately. Namely sorted and reversed arrays. These two cases (with the many equals case which was already handled in the main algorithm above) form as the ingredients for many common cases. For example the organ pipes case is formed by combining sorted and reversed sub parts concatenated together. Since Quicksort is recursive, it doesn't matter if the whole of the array is not fully sorted or reversed. If the sub parts are sorted or reversed, Quicksort would eventually reach a level in the recursive tree were it can detect those parts and handle them separately and hence efficiently. Notice that the swapping (copying in our case) that happened earlier in a higher level may not heavily scramble the internal ordering because if a sorted sub part was copied from left to right, it would become reversed, and if a reversed sub part was copied from left to right it would become sorted, and vise versa when copying from right to left.

The tricky part is how to guess[7] a possibly sorted or reversed array while reducing false positives and enduring the minimum overhead possible on the main algorithm. We incorporate our guessing mechanism **within the pivot**

[5] Tiny arrays are handled using insertion sort

[6] Other algorithms exist that can do 7 comparisons in the worst case but most don't optimize for the number of swaps. A 7 comparison worst case with minimal swaps can still be achieved but it involves a more sophisticated method and the benefit is rather insignificant since there are only 16 worst cases out of 120 permutations in our algorithm.

[7] Guaranteed detection is expensive as it needs n comparisons on each recursive stage.

**selection process**. For example, here is our median of three algorithm[8]:

```
inline short MedianOf3S(Elem ar[], int a0, int a1, int a2,
Elem &piv ,int cmp(const void *elem1,const void *elem2))
{   Elem temp;

    if (cmp(&ar[a1],&ar[a0])>=0)
    {    if (cmp(&ar[a2],&ar[a0])>=0)
        {    if (cmp(&ar[a2],&ar[a1])>=0)
            {   // 0 1 2
                piv=ar[a1]; return 1;
            } else
            {   // 0 2 1
                temp=ar[a1]; ar[a1]=ar[a2]; ar[a2]=temp;
            }
        } else
        {   // 2 0 1
            temp=ar[a0]; ar[a0]=ar[a2]; ar[a2]=ar[a1];
            ar[a1]=temp;
        }
    } else
    {    if (cmp(&ar[a2],&ar[a1])>=0)
        {    if (cmp(&ar[a2],&ar[a0])>=0)
            {   // 1 0 2
                temp=ar[a0]; ar[a0]=ar[a1]; ar[a1]=temp;
            } else
            {   // 1 2 0
                temp=ar[a0]; ar[a0]=ar[a1]; ar[a1]=ar[a2];
                ar[a2]=temp;
            }
        } else
        {   // 2 1 0
            temp=ar[a0]; ar[a0]=ar[a2]; ar[a2]=temp;
            piv=ar[a1]; return -1;
        }
    }
    piv=ar[a1];
    return 0;
}
```

The function above will return 1 if it finds that the three samples were already sorted. It will return -1 if it finds the three samples reversed. And will return 0 otherwise. 1 or -1 means the array could be sorted or reversed respectively so we will handle it separately but fall back to the main algorithm if it was found not so. 0 means that the array might not be sorted or reversed so we will process it normally in the main algorithm. This guess will become more accurate at reducing false positives when applied to the median of five, nither and fifteenther because those involve more samples.

The following is the handler for possibly sorted arrays, i.e. if pivot selection returns 1.

```
// Possibly Sorted handler
while ((lc2=cmp(&ar[l],&p))<0)
{   l++;
   if (l==mid)
   {   while ((lc2=cmp(&ar[r],&p))>0)
      {   r--;
          if (r==mid) { r++; l--; goto bypass; }
      }
      if (lc2==0)
      {   mr=mid+1;
          if (mr==r) { r++; l--; goto bypass; }
          while ((lc=cmp(&ar[mr],&p))==0)
          {   mr++;
             if (mr==r) { r++; l--; goto bypass; }
          }
          temp=ar[r]; ml=mid; MRIGHT_NOSCAN();
      } else
      {   temp=ar[r];
          ml=mr=mid;
          MRIGHT();
         }
      }
   }
}
if (lc2==0)
{   while ((lc=cmp(&ar[r],&p))>0)
   {   r--;
       if (r==mid)
      {   ml=mid-1;
          if (ml==l) { r++; l--; goto bypass; }
          while ((lc=cmp(&ar[ml],&p))==0)
          {   ml--;
             if (ml==l) { r++; l--; goto bypass; }
          }
          temp=ar[l];  mr=mid; MLEFT_NOSCAN();
      }
   }

   if (lc==0)
   {  ml=mid;
      do
      {   ml--;
          if (ml==l)
          {  mr=mid+1;
             if (mr==r) { r++; l--; goto bypass; }
             while ((lc=cmp(&ar[mr],&p))==0)
             {   mr++;
                 if (mr==r) { r++; l--; goto bypass; }
             }
             temp=ar[l]; ar[mid]=ar[r]; MRIGHT_NOSCAN();
          }
      } while ((lc=cmp(&ar[ml],&p))==0);
      mr=m=mid; temp=ar[l]; ar[mid]=ar[r];
      goto ml_scan1_2;
   } else
   {   temp=ar[l];  m=mr=ml=mid; goto r_scan1_3;
   }
}

while ((lc2=cmp(&ar[r],&p))>0)
{   r--;
    if (r==mid)
    {   use_temp=0; m=ml=mid;
        goto checkp11;
     }
}
temp=ar[r];  m=mr=ml=mid; goto l_scan1_2;
```

**MLEFT_NOSCAN( )** and **MRIGHT_NOSCAN( )** above are identical to **MLEFT( )** and **MRIGHT( )** but without the equals scanning, please refer to complete code in [12].

The "possibly sorted" handler above does not strictly check if the array is sorted! It only checks if all elements on the left are less than pivot **p** and that all elements on the right are greater than **p**. deeper recursive stages will check further into the left and right in the same manner. If the handler detects that the array is not sorted, it will immediately fallback to the appropriate place in the main algorithm with minimal overhead. The main algorithm will not start over, it will take the current progress to its advantage. The following is the handler for possibly reversed arrays, i.e. if pivot selection returns -1.

```
// counters. cl for ar[l]<p and cr for ar[r]>p
cl=1; cr=1;
for (;;)
{  if ((lc2=cmp(&ar[l],&p))>0)
   {
      checkp2:
      if ((lc=cmp(&ar[r],&p))<0)
      {  SWAP_AR(ar,l,r,temp2);
         if (l+2==r) { goto bypass; }
         l++; r--;
         if (l==mid) { mr=mid+1; goto checkp7; }
         else if (r==mid)
         {  ml=mid;
            do
            {   ml--;
                if (ml==l)
```

---

[8] A side effect of this function is that it sorts the three elements. This is unrelated to the discussion here.

```
            {  // Special case short cut.
               if ((lc=cmp(&ar[l],&p))>0)
               {  ar[mid]=ar[l]; ar[l]=p; l--;
                  goto bypass;
               }
               else if (lc==0) { l--; r++; }
               else r++;
               goto bypass;
            }
         } while ((lc=cmp(&ar[ml],&p))==0);
         lc2=(cmp(&ar[l],&p)>0);
         temp=ar[l]; mr=mid; MLEFT_NOSCAN();
      }
   }
   else if (lc>0)
   {   r--;
      if (r==mid)
      {  ml=mid;
         do
         {   ml--;
            if (ml==l)
            {  // Special case short cut.
               ar[mid]=ar[l]; ar[l]=p; l--;
               goto bypass;
            }
         } while ((lc=cmp(&ar[ml],&p))==0);
         temp=ar[l]; mr=mid; MLEFT_NOSCAN();
      }
      if (cr>REVERSE_TOLERANCE)
      {  temp=ar[l]; m=mr=ml=mid; goto r_scan1;
      }
      else { cr++; goto checkp2; }
   }
   else   //lc==0
   {  mr=mid;
      do
      {  mr++;
         if (mr==r)
         {  ml=mid;
            do
            {  ml--;
               if (ml==l)
               {  SWAP_AR(ar,l,r,temp2); l--;
                  goto bypass;
               }
            } while ((lc=cmp(&ar[ml],&p))==0);
            ar[mid]=ar[r]; temp=ar[l];
            RESETRFL();
            MLEFT_NOSCAN();
         }
      } while ((lc=cmp(&ar[mr],&p))==0);
      temp=ar[l]; ar[mid]=ar[r]; ml=mid;
      goto mr_scan1_2;
   }
}
else if (lc2<0)
{   l++;
   if (l==mid)
   {  mr=mid;
      do
      {   mr++;
         checkp7:
         if (mr==r)
         {  // Special case short cut.
            if ((lc=cmp(&ar[r],&p))<0)
            {  ar[l]=ar[r]; ar[r]=p; r++; goto bypass;
            }
            else if (lc==0) { l--; r++; }
            else l--;
            goto bypass;
         }
      } while ((lc=cmp(&ar[mr],&p))==0);
      while (MORE_SLC(ar[r],p,lc2))
      {  r--;
         if (mr==r)
         {  if (lc<0)
            {   ar[l]=ar[r]; ar[r]=p; r++; goto bypass;
            }
            else l--;
            goto bypass;
         }
      }
      temp=ar[r]; ml=mid; MRIGHT_NOSCAN();
      }
      if (cl>REVERSE_TOLERANCE) { goto main_algo; }
      else { cl++; continue; }
   }
   else   //lc2==0
   {   ml=mid;
      do
      {  ml--;
         if (ml==l)
         {  mr=mid;
            do
            {  mr++;
               if (mr==r)
               {  if ((lc=cmp(&ar[r],&p))<0)
                  {    SWAP_AR(ar,r,l,temp2); r++; }
                  else { l--; if (lc==0) r++; }
                  goto bypass;
               }
            } while ((lc=cmp(&ar[mr],&p))==0);
            while ((lc2=cmp(&ar[r],&p))>0)
            {   r--;
               if (mr==r)
               {  if (lc<0)
                     { SWAP_AR(ar,mr,ml,temp2); r++; }
                  else l--;
                  goto bypass;
               }
            }
            temp=ar[r]; ar[mid]=ar[ml]; MRIGHT_NOSCAN();
         }
      } while ((lc2=cmp(&ar[ml],&p))==0);
      ar[mid]=ar[l]; temp=ar[ml];
      m=ml; mr=mid; goto r_scan1;
   }
}
```

For reversed arrays, it's best to just swap the elements on the left with elements on the right one after the other until middle element is reached. Otherwise, trying to reduce swaps with the copying method we described in section II will cause a shift in elements that need further recursion copying to be corrected. The "possibly reversed" handler above checks if all elements on the left are greater than **p** and all elements on the right are less than **p** while doing the swapping. Deeper recursion will either find that the array has been sorted in the first stage or that it requires further sorting for the sub parts.

The "possibly reversed" handler has a *tolerance* mechanism. Were if the array is found to be NOT reversed, the handler will tolerate some amount of none-reversed elements until a predefined threshold is exceeded at which point the handler will fallback to the appropriate place in the main algorithm with minimal overhead. This tolerance mechanism is important since some of the pivot selection functions we describe above actually sort the samples which would render a reversed array not completely reversed. The tolerance is easy to implement and is useful for nearly reversed arrays too.

## IV. SORTING TINY ARRAYS AND SUB PARTS

Past Quicksort implementations have religiously used Insertion sort to tackle small to very small sub parts that are encountered deep in Quicksort's recursive tree [3][7]. The GNU library qsort, dual pivot Quicksort and Bentley's Quicksort all use Insertion sort for small or tiny arrays. Insertion sort has a lower overhead than Quicksort when **n** is very small. Some implementations do insertion sort per recursive stage once **n** is less than a predefined threshold. Others (like GNU) simply stop Quicksort at **n** < threshold, then when Quicksort finishes do a single pass of Insertion sort across the whole array. Another method exists were the tiniest parts (**n** < 5) are sorted using decision tree functions [5], i.e. a separate function for n=3, another for n=4 and so on and Insertion sort is used to sort 5 ≤ **n** < threshold.

We have done several tests and found Insertion sort to be indeed advantageous over pure Quicksort for tiny arrays. We designed a special Insertion sort function that reduces the number of swaps again by doing direct element copies instead of full swaps as follows:

```
// Insertion Sort. Sort elements from r to b
for (k=r+1; k<=b; k++)
{   if (cmp(&ar[k],&ar[k-1])<0)
    {   j=k; temp=ar[j];
        ar[j] = ar[j-1];
        if (--j>r)
            while (cmp(&temp,&ar[j-1])<0)
            {   ar[j] = ar[j-1];
                if (--j==r) break;
            }
        ar[j] = temp;
    }
}
```

This concludes the algorithmic part of Triple State Quicksort.

## V. Asymptotic Analysis & Worst Case Mitigation

### A. Average case

On the average, a classical Quicksort uses:

comparisons $\approx 2n\ln(n) - 2.8456n$ [16] (3)

swaps $\approx 0.33n\ln(n) - 0.58n$ [17] [18] (4)

Before proceeding with the analysis for our algorithm, we should note here that most of these asymptotic estimates in the literature make the following assumptions:

1. The input array is a uniformly distributed random permutation [1] [15] [16] . i.e. each permutation (ordering of elements) is equally likely to occur.
2. There are no equal elements. The array has to be a collections of distinct keys [1] [15] [16].
3. If assumption 1 is not met, the pivot has to be selected at random [1].

These assumptions are made to simplify analysis since considering all input types and all aspects of an algorithm is usually intractable. In addition to the assumptions above, there are a few other considerations:

4. An advanced pivot selection method (such as median of three or so) is usually not considered in the analysis (EXCEPT by sedgewick in [17]). Usually, pivot is either assumed to be the first or last element or a randomly selected element.
5. Insertion sort for tiny arrays and other improvements likewise are also usually not considered (EXCEPT by sedgewick [17] and recently by Wild and Nebel in [15] with their rigorous analysis).

While the above might be acceptable analytically, it might not suffice a software engineer working with real world data. Therefore, it is advised to take these estimates as general aid, not as a conclusive decision maker.

The analysis in [16] can be applied reasonably to Triple State, leading to the same average number of comparisons *2n ln(n) – 2.8456n.*

For the number of swaps, Our algorithm does NOT do swaps literally but element copies instead (most of the time), the only way to compare it to other algorithms (that do swaps) is to have an equivalent measure. Since a swap is composed of 3 element copy operations, we can say that *swaps = copies* / 3. So we can convert the number of copy operations that our algorithm does to equivalent **virtual** swaps by dividing over 3. Surely this equivalence is not exact since one might argue that copying from/to a temporary variable (in a swap) is different from copying from/to memory locations. But it still gives a reasonable indication.

It has been shown in [1] that the number of swaps in a single Quicksort recursive stage is:

$$\frac{n}{6} + \frac{5}{6n} \qquad (5)$$

Since each swap involves two elements. The number of elements copied in our algorithm is twice as much swaps. Adding one final copy operation to copy back the **temp** element then converting the number of copies to virtual swaps by dividing by 3. Gives:

$$\frac{1}{3}\left[2\cdot\left(\frac{n}{6} + \frac{5}{6n}\right) + 1\right] = \left(\frac{n}{9} + \frac{5}{9n}\right) + \frac{1}{3}$$

Notice here that we don't need to add 1 copy for copying back pivot **p** since that's already included in (5) part of a full swap. The average number of swaps $S_n$ can be expressed as:

$$S_n = \left(\frac{n}{9} + \frac{5}{9n}\right) + \frac{1}{3} + \frac{1}{n}\sum_{k=0}^{n-1}(S_k + S_{n-1-k})$$

Since $\sum_{k=0}^{n-1} S_k = \sum_{k=0}^{n-1} S_{n-1-k}$ , we can rewrite:

$$S_n = \left(\frac{n}{9} + \frac{5}{9n}\right) + \frac{1}{3} + \frac{2}{n}\sum_{k=0}^{n-1} S_k$$

Multiplying by $n$.

$$nS_n = n\left(\frac{n}{9}+\frac{5}{9n}\right)+\frac{n}{3}+2\sum_{k=0}^{n-1}S_k \qquad (6)$$

substituting $n$ by $n-1$

$$(n-1)S_n = (n-1)\left(\frac{n-1}{9}+\frac{5}{9(n-1)}\right)+\frac{n-1}{3}+2\sum_{k=0}^{n-2}S_k \qquad (7)$$

Subtracting (7) from (6) and simplifying would finally yield

$$S_n = \left(1+\frac{1}{n}\right)\left(\frac{2}{9}+S_{n-1}\right)$$

We know that $S_2 = 0.5$ since a two elements random array will have the same probability of being already sorted or reversed. Solving the recurrence relation above with initial condition $S_2 = 0.5$ results:

$$S_n = \frac{2}{9}(n+1)\sum_{k=1}^{n}\frac{1}{k}-\frac{1}{6}(n+1)$$

We know that $H_n = \sum_{k=1}^{n}\frac{1}{k}$ were $H_n$ is the nth harmonic number. We can rewrite $S_n$ as:

$$S_n = \frac{2}{9}(n+1)H_n - \frac{1}{6}(n+1) \qquad (8)$$

We also know that $H_n$ can be approximated:

$$H_n \approx \ln(n)+\gamma+\frac{1}{2n}$$

Were $\gamma \approx 0.5772$ is the Euler–Mascheroni constant. Substituting back in (8).

$$S_n = \frac{2}{9}(n+1)\left(\ln(n)+\gamma+\frac{1}{2n}\right)-\frac{1}{6}(n+1)$$

$$=\frac{2}{9}n\ln(n)+n\cdot\left(\frac{2\gamma}{9}-\frac{1}{6}\right)+\frac{2}{9}\ln(n)+\frac{1}{9n}+\frac{2\gamma}{9}-\frac{1}{18}$$

$$S_n \approx 0.222n\ln(n)-0.038n \qquad (9)$$

It can be shown by trivial math that $S_n$ in (9) is less than the average swaps in (4) of classical Quicksort for all $n > 139$.

### *B. Worst case*

Quicksorts worst case is of $O(n^2)$. This is true for any Quicksort including our algorithm as long as it does pivot selection and has a divide and conquer algorithm. The worst case happens when the pivot is **consistently** equal or near the maximum or minimum of the elements in the array in every recursive stage. However, for an advanced Quicksort that handles equal elements correctly, the probability of the worst case happening is pretty slim in practice. For example, if the pivot selection was done at random or input array is a random permutation. Then the probability that the pivot is maximum or minimum in a **single** stage is ***2/n***. Since there are about ***n*** recursive stages in a Quicksort binary tree. The probability of all stages selecting minimum or maximum pivot will be:

$$p[worst_case] \approx \left(\frac{2}{n}\right)^n = \frac{2^n}{n^n}$$

$n^n$ is much larger than $2^n$ for even as low as $n=5$. Hence the extremely low probability[9].

### *C. Worst case mitigation*

Despite the low probability in practice, someone might cook a worst case input array that causes the $O(n^2)$ behavior. McILROY showed in [19] a program to construct a killer input array that results in such quadratic time. His method works on any Quicksort function with minimal requirements, even dual pivot Quicksort. A hacker might inject such constructed data into say a database or an accounting software using typical data entry methods then sort the data causing high CPU usage and ultimately a system crash. The same input array will always cause $O(n^2)$ time every time it is sorted.

Introsort is a well known method to totally prevent Quicksort from falling into $O(n^2)$ by switching to Heapsort whenever the depth of the recursive tree is too large. This comes for a price of some overhead and possibility of going into Heapsort too quickly.

If Introsort's overhead is not desirable, then the only way to effectively lower $O(n^2)$ possibility is to use random pivots that are **seeded with the system timer or clock[10]**. However, random pivoting lacks the great advantages of using advanced pivoting like the one described in subsection III.H.

[9] In the worst case, the depth of the recursive tree will become equal to the number of stages **n**.

[10] Reference [19] says that the killer program will work for random pivots. We claim that this is true only when the same seed is used prior to sorting. This is not the case in reality, the moment the killer input is constructed, the random generator sequence would have normally moved to a different start, rendering the killer input ineffective.

We propose a method that partially randomizes our pivoting system with minimal overhead. The idea is to shift the positions of intermediate samples (i.e. NOT the first, middle nor last samples) randomly to the left or right. This is done using a single random number that is generated only at the start of the sort. For example, the 2nd and 4th samples in our median-of-five are shifted by this random amount, while taking care not to cross the boundaries of the remaining 3 samples. Algorithm as follows.

First we start by seeding an independent internal random generator. This is done only once at program start, not on every sorting operation, hence the **static** keyword is used.

```
// Internal Random Generator
// Using Park-Miller random generator
// Done only when program starts.
static unsigned int const r_a=16807,r_m=2147483648;
static unsigned int zgen = clock();
double dran,dran2;
```

The random generator is internal to the algorithm to avoid disturbing the rand() function distribution and to prevent a hacker from an attempt to seed it with a pre-known value. Then every time the sorting function is called, we do:

```
// Generate internal random number
zgen=(r_a*zgen) % r_m;
dran = 0.5+(double) zgen/r_m;
dran2 = 1 + dran;
```

Then at each recursive stage, a random amount is added to the sampling positions before the appropriate pivot finding function is called. For example, here is the median-of-five case (were **n** is medium sized).

```
gap = (n/4)*dran; // Without mitigation gap=(n/4)

switch (MedianOf5S(ar,a,a + gap,mid,b - 
gap,b,p,cmp))
{
   ....
}
```

Complete code for all cases of pivoting is in [12]. We tried the Quicksort killer against this method and it effectively prevented the killer from causing quadratic time. This method also preserves all the advantages of median-of-five, ninther, and fifteenther since the first, middle and last samples are not touched.

The only case where this method wont work is if **n** is small (i.e. **n**<70, when using median-of-three) since there are no internal samples to randomize. However, we provide the compiler directive MEDOF3_SMALL that if disabled will disable the median-of-three case and use median-of-five instead which prevents this worst case from happening if necessary.

Our worst case mitigation method can be enabled/disabled simply by defining/undefining the compiler directive RANDOM_SAMPLES.

## VI. COMPARING COMPARISONS AND SWAPS

Before going through further implementation related optimization. We conduct several tests only to measure number of comparisons and swaps.

Our test is similar to what's in [3] and [4] generating the 5 adverse inputs: Sawtooth, Random, Stagger, Plateau and Shuffle exactly as defined in [3] except for the Random input case were we use a better certified random generator [11] than the C/C++ library's rand() function. We also add two more input types to harden our tests, namely: Hill and Organ Pipes defined as follows:

```
// Park-Miller random generator

unsigned int const a=16807,c=0,m=2147483648;

int GenRandom2(int smallest, int largest)
{   return GenRand()*(largest-smallest+1)+smallest;
}

double GenRand()
{  // Park-Miller random generator
   zgen=(a*zgen+c) % m;
   return ((double) zgen/m);
}

void Hill(Elem ar[], int size, int arange)
{   int i;
    for (i=0; i<size; i++)
       ar[i] = min(i<(size>>1)? i:size-i,arange);
}

void OrganPipes(Elem ar[], int size, int arange, int 
op_max_distance, int op_add)
{   int i, j,dist;
    double step,v1,v2;
    v1 = GenRandom2(1,arange);

    for (i=0; i<size;)
    {   v2 = GenRandom2(1,arange)+op_add;
        dist = GenRandom2(1,op_max_distance);
        step = ((double) v2-v1)/dist;
        while (dist!=0 && i<size)
        {   ar[i++] = v1;
            v1+=step; dist--;
        }
        v1=v2;
    }
}
```

Were **arange** is the range of the elements generated. **op_max_distance** and **op_add** are parameters specific to organ pipe generation. Running time tests are shown later in this work after non-algorithmic optimization is done. After each input type is tested, we test for 6 reorderings, 5 of them are as mentioned in [3] namely: Sorted, Reversed, Front half reversed, Back half reversed and Dither[11]. We also add an additional reordering: "Fort", that is designed to deeply and recursively reverse a sorted array to cause a high number of swaps as follows:

```
void Fort(Elem ar[],int a,int b,const int minl)
{
    int i,h;
```

[11] Dither is not strictly a reordering, it add an increasing integer value to the original array. It's more of a modification than an reordering

```
    Reverse(ar,a,b);

    if (b-a+1>minl)
    {   h = (a+b)>>1;
        Fort(ar,a,h,minl);
        Fort(ar,h+1,b,minl);
    }
}
```

Were function **reverse( )** is as follows:

```
void Reverse(Elem ar[],int a,int b)
{   int i, j=b ,h = (a+b)>>1;
    Elem temp;
    for (i=a; i<=h; i++, j--)
    {   temp = ar[i];
        ar[i] = ar[j];
        ar[j] = temp;
    }
}
```

We test Triple State Quicksort against 4 other famous or very good Quicksort algorithms: Bentley & McILROY, GNU library, Microsoft's Visual C++ std:sort() and Yaroslavskiy's dual pivot Quicksort.

We should note here that VC++ has two very different Quicksort functions, the C library qsort() and the C++ library std:sort() function which is the one we are testing. Although the former qsort() function often performed better in our preliminary tests, it's easily vulnerable to exhibiting $O(n^2)$ time using an input array of reverse Plateau or Hill. The std::sort() is not susceptible to such effect[12].

We should also note that we are using the latest version of dual pivot Quicksort which appeared in Yaroslavskiy's paper on September 22, 2009 [4][13]. We ported the code from Java to C/C++ with minimal changes merely to use a common callback comparison function for test fairness. We also added counters for swaps and comparisons.

Fig. 7, shows the Random array input test for medium sized arrays. Element values range from 1 to 500 (**arange**=500)[14], both Bentley's and Triple State perform closely and outperform other algorithms while Triple State begins to excel at higher array sizes due to our aggressive pivot selection.

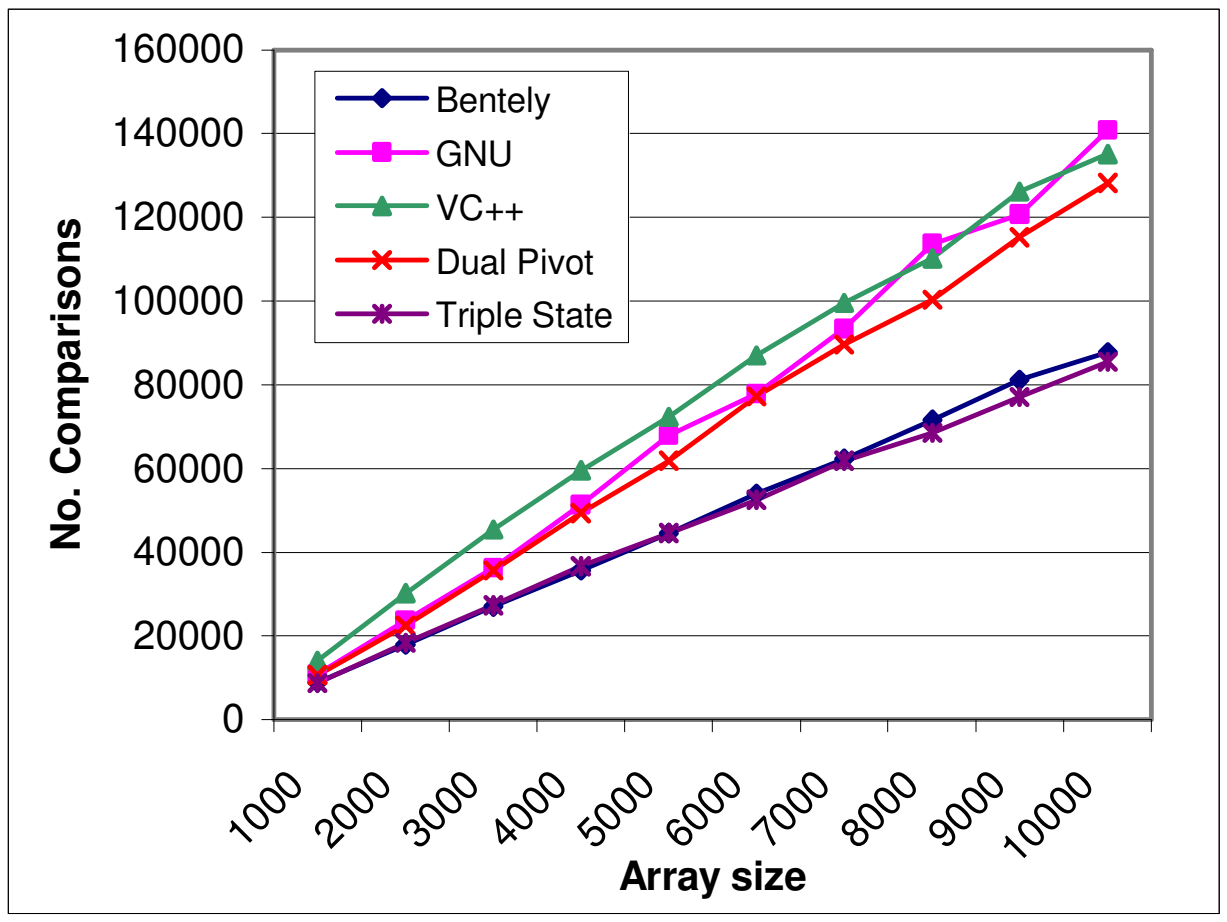

Fig. 7. Number of comparisons. Random input.

Fig. 8, Shows the number of swaps performed for the same input array. Triple State obviously does less swaps in all array sizes.

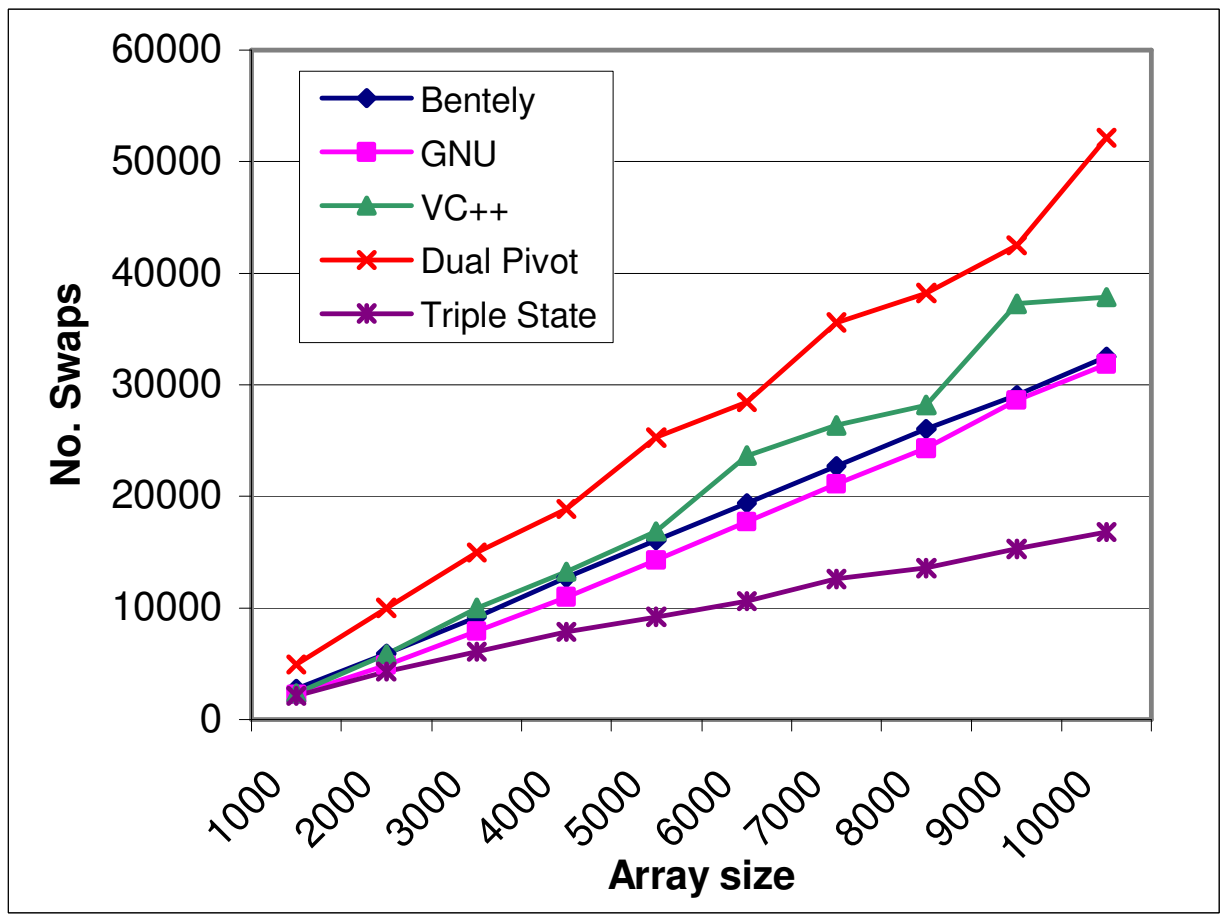

Fig. 8. Number of swaps. Random Input.

Further results with different orderings of the same arrays can be seen in Appendix B.

[12] It seams that std::sort() uses Introsort which is basically a Quicksort that switches to Heapsort whenever a bad case input is found to go extensively deep in the Quciksort recursive tree. This effectively avoids $O(n^2)$ cases.

[13] There is another earlier version that he posted in Gmain mailing list [13].

[14] Jon Bentley refers to this as vairable **m**. We call it **arange**

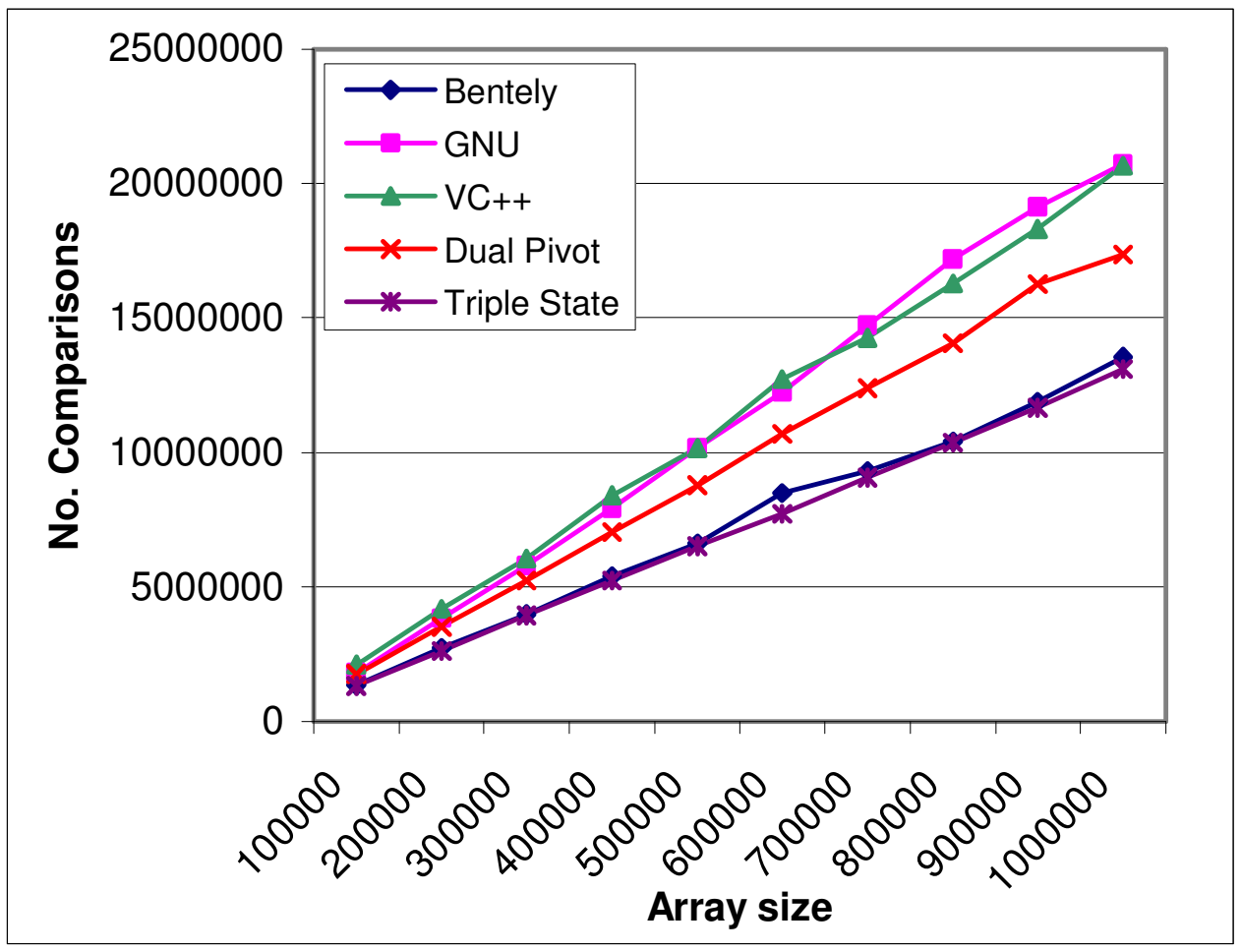

Fig. 9. Number of comparisons. Random input. Large array sizes.

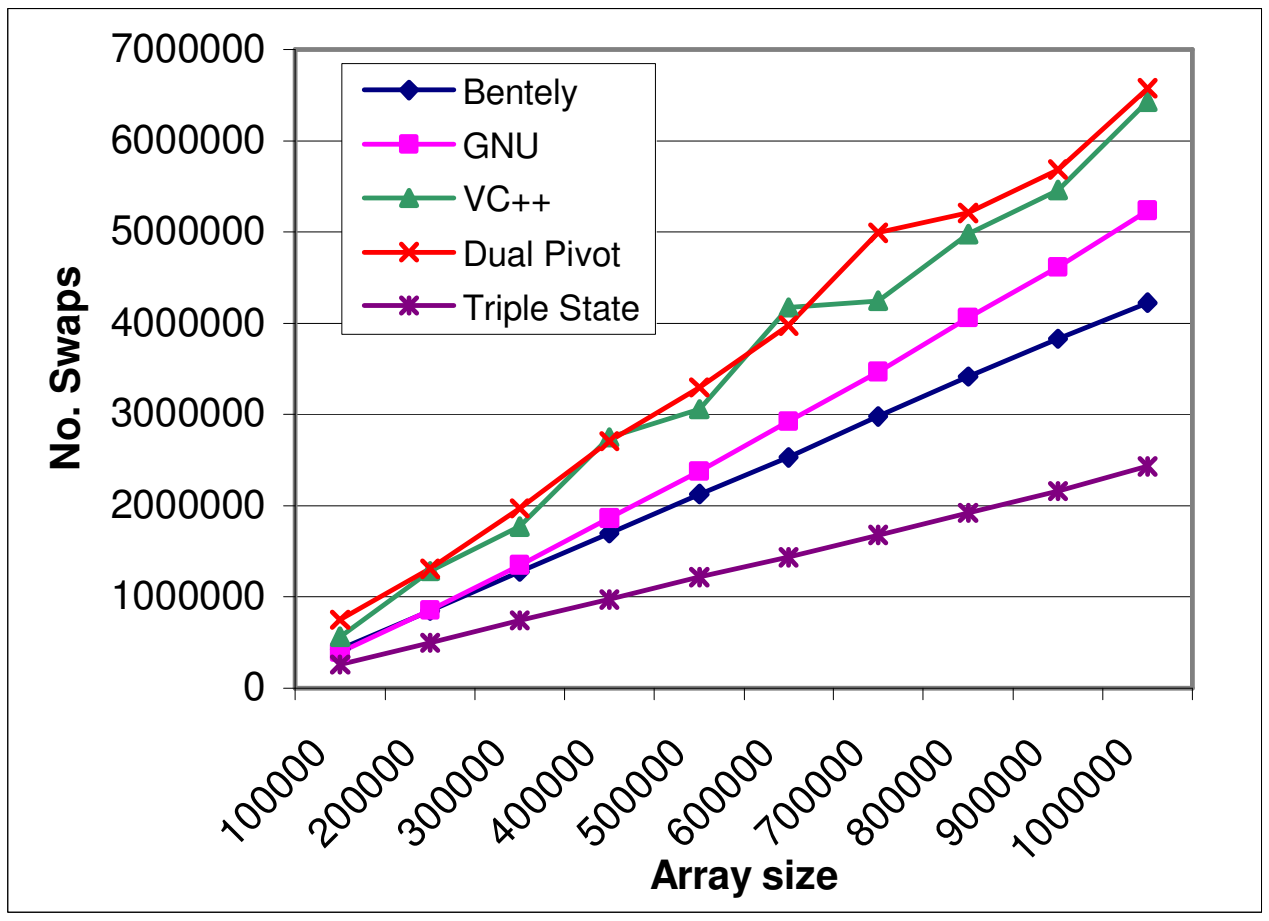

Fig. 10. Number of swaps. Random input. Large array sizes.

Test results for large array sizes while elements range is from 1 to 9000 (**arange**=9000) are shown in Fig. 9, and Fig. 10. Triple State is best both in comparisons and swaps in all cases. Further results shown in Appendix B. Our complete detailed results can be found in reference [12]. Fig. 11, shows comparisons results for the 6 adverse inputs mentioned above for arrays of 2 million elements. Triple State is best in all cases except in Shuffle and Hill were it comes in second place. Swaps results in Fig. 12, also show that Triple State is best in all cases except shuffle were it's slightly more than GNU and in plateau where it slightly more than VC++.

Taking a big-picture look after observing all the results so far. One would realize that all the algorithms have a weak case somewhere except for Triple State Quicksort. In almost all cases, Triple State is either the best or very close to the best. This is a feature essential to a general purpose sorting algorithm were there is little variance in performance towards different types of inputs.

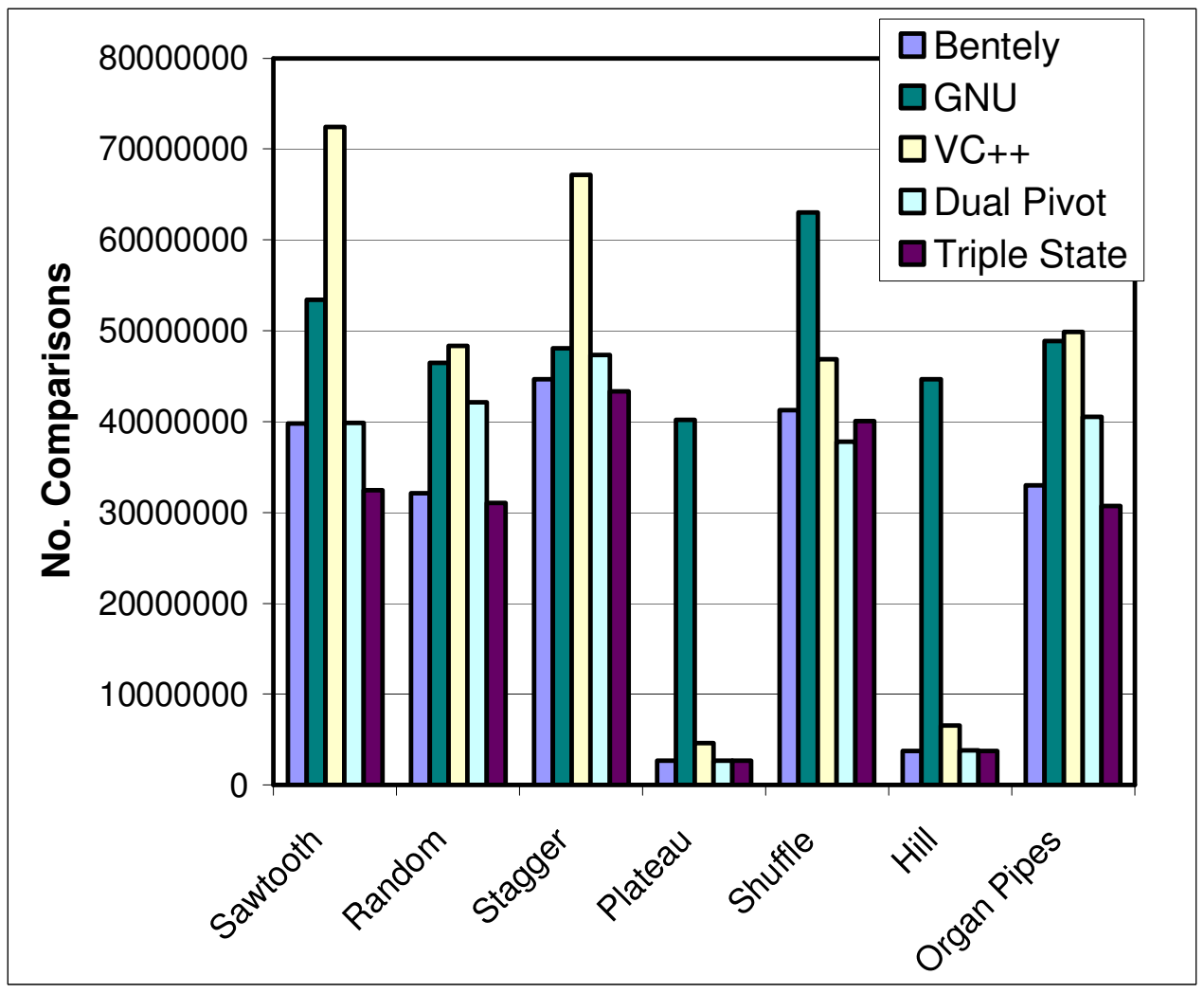

Fig. 11. Number of comparisons. Multi adverse type input test. 2 million elements array.

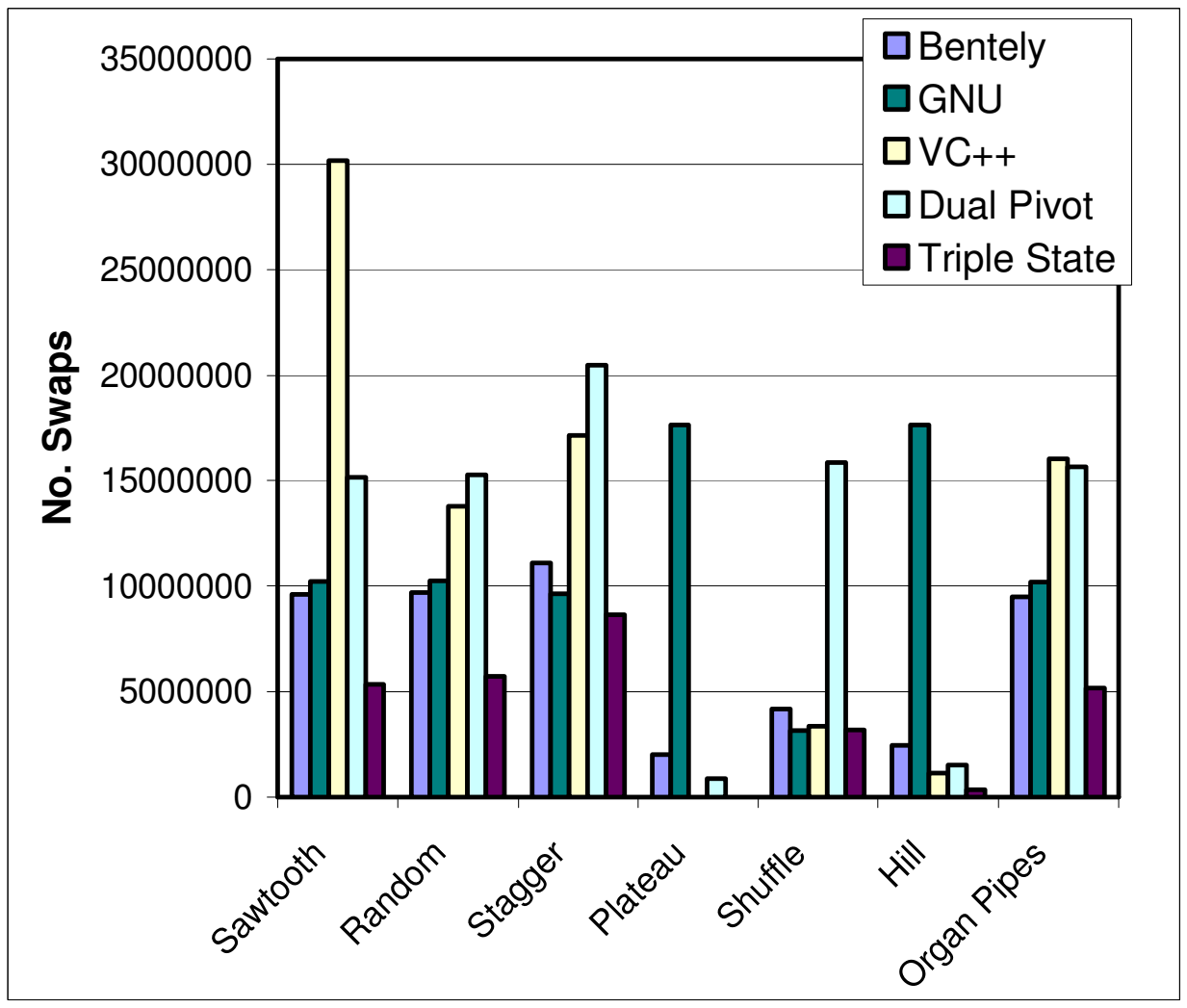

Fig. 12. Number of swaps. Multi adverse type input test. 2 million elements array.

## VII. INTRODUCTION TO THE UNKNOWN

There are two unknown factors when designing a generic sorting algorithm such as a library function as qsort:

- Comparison function complexity
- Memory size of each element

In most C/C++ libraries, the comparison function is a pointer to a callback function given by the user/programmer. The memory size in bytes of each element is usually given as a parameter to qsort. So both of these factors are unknown at compile time of qsort (not at compile time of user code). This obviously places the library qsort function at a disadvantage.

On the other hand, a programmer could take an efficient implementation of Quicksort and adapt it to a none generic form, removing the overhead of the callback comparison function and tuning a macro based swapping/copying function that is designed to run fast for his/her element data type size. This method is recommended yet it is not expected to be used heavily by most programmers.

Returning back to our generic qsort, there is not much that can be done with the comparison function since that comes solely from the user/programmer. However, the swapping or copying functions can be optimized but it is very sensitive to variations in hardware and compilers. Furthermore, after long experiments we found that even minor changes to a copying function can result in significant changes in running time. We came to a conclusion that the copying function should be fine tuned per platform, compiler and if relevant, per hardware.

We implemented two versions of Triple State Quicksort. The first is an **array based** version that assumes that both comparison function and elements size is known at compile time. This version is suitable for open source libraries or for porting the code into other languages like Java. The second version is **pointer based** which assumes both comparison and element size to be unknown. It is designed mainly for C/C++ libraries. The two versions use exactly the same algorithm consuming the same number of comparisons and swaps. The only differences are in the implementation.

## VIII. Copying and Swaping Functions Optimization

This section and all its subsections apply only to the pointer based implementation of Triple State Quicksort since copying operations are intrinsic in the array based implementation, i.e. provided by the compiler.

For copying, one could use the built in function memcpy(). Unfortunately memcpy() is inefficient in copying **large** amounts of **small elements**. memcpy() is mainly designed to copy large blocks of memory a few thousand times, not small blocks of memory millions of times. Sorting needs the later, strictly not the former.

### A. *Copying in chunks*

We'll start with a simple copying function then improve on it gradually.

```
inline void ecopyChar(char *d, char *s , size_t
sesv)
{
    for (;sesv>0; sesv--,d++,s++)
        *d= *s;
}
```

Were **s** and **d** are the source and destination addresses respectably. **sesv** is the element size. Function `ecopyChar` above copies source to destination byte by byte (i.e. char by char) which is rather slow. It has been shown that copying memory word by word is faster than copying byte by byte [3]. And it's not hard to see why. Modern CPU registers are at least 32bits long (word) if not 64bits (double words). Copying between memory locations is done using those registers, so copying say an integer (32bit) word-wise would require two machine instructions:

```
Mov eax, [s]
Mov [d], eax
```

While copying an integer byte-wise would require 8 instructions:

```
Mov al, [s]
Mov [d], al
Mov ah, [s+1]
Mov [d+1], ah
Mov bl, [s+2]
Mov [d+2], bl
Mov bh, [s+3]
Mov [d+3], bh
```

Were **s** and **d** are the source and destination addresses respectably. Hence, copying Word-wise can potentially take 1/4 the CPU cycles of byte-wise copying[15]. The problem is the size of an element may not be a multiple of words. The remaining fraction of a word has to be copied either byte-wise (8bits) or as a combination of 16bits and 8bits. The following code explains, but first we make a few definitions:

```
// Types
#define SWORD short    // 16 bit
#define LWORD long     // 32 bit
#define LLWORD long long  // 64 bit

#define WC sizeof(char)
#define WS sizeof(SWORD)
#define WL sizeof(LWORD)
#define WLL sizeof(LLWORD)
```

Copying function:

```
inline void ecopyA32(char *d, char *s , size_t sesv)
{
    // Copy words.
    for (;sesv>=WL; sesv-=WL,d+=WL,s+=WL)
        *(LWORD *) d= *(LWORD *)s;

    // Copy remaining chars. This seams to be not
    // neccessary in VC++ when type containes an int
    // field. It seams to be forced
    // by compiler to be multiples of WL
    for (;sesv>0; sesv-=WC,d+=WC,s+=WC)
        *d= *s;
}
```

Function `ecopyA32` copies words then the remaining fraction in characters (bytes). We could also copy 16bits before copying bytes but the overhead of the additional for loop (or if statement) seams to cancel the benefit. Another way is to copy double words wise (64bits) then copy the remaining fraction of words and chars. This is more efficient when the target processor and operating system is 64bit but does not degrade the performance when used in 32bit platforms. As follows:

```
inline void ecopyA64(char *d, char *s , size_t sesv)
{
    // Copy double words
```

[15] This is not a strict rule since modern processors can have multiple pipelines. It's rather a general and very good assumption to make.

```
    for (;sesv>=WLL; sesv-=WLL,d+=WLL,s+=WLL)
        *(LLWORD *) d= *(LLWORD *)s;

    // Copy remaining word.
    if (sesv>=WL)
    {   *(LWORD *) d= *(LWORD *)s;
        sesv-=WL; d+=WL; s+=WL;
    }

    // Copy remaining chars.
    for (;sesv>0; sesv-=WC,d+=WC,s+=WC)
        *d= *s;
}
```

## B. *Loop unfolding*

Again, it's important to note that **any small improvements** here can have a **significant impact** on running time yet those improvements are **system and compiler dependant**. Loop unfolding has been used before to lessen the number of comparisons and looping operations. i.e. lessen the branches and jump instructions executed by the CPU. The following shows unfolding for the first loop in `ecopyA64`:

```
    // Copy double words in chunks of 8
    for (;sesv>=8*WLL; sesv-=8*WLL)
    {   *(LLWORD *) d= *(LLWORD *)s;
        d+=WLL; s+=WLL;
        *(LLWORD *) d= *(LLWORD *)s;
        d+=WLL; s+=WLL;
        *(LLWORD *) d= *(LLWORD *)s;
        d+=WLL; s+=WLL;
        *(LLWORD *) d= *(LLWORD *)s;
        d+=WLL; s+=WLL;
        *(LLWORD *) d= *(LLWORD *)s;
        d+=WLL; s+=WLL;
        *(LLWORD *) d= *(LLWORD *)s;
        d+=WLL; s+=WLL;
        *(LLWORD *) d= *(LLWORD *)s;
        d+=WLL; s+=WLL;
        *(LLWORD *) d= *(LLWORD *)s;
        d+=WLL; s+=WLL;
    }

    // Copy ramaining double words
    for (;sesv>=WLL; sesv-=WLL,d+=WLL,s+=WLL)
        *(LLWORD *) d= *(LLWORD *)s;
```

The first "for loop" copies in chunks of 8 which would minimize loop overhead. Essntially, this reduces the number of times **sesv** is decremented and tested, and also reduces the number of jumps to beginning of the loop. We can improve on this further. Modern proccssors have **offset addressing** ability which allows to do arithmetic addition within another insruction. Example:

```
Mov eax, [s+4]

Mov [d+4], eax
```

Modern compilers can optimize to that instead of producing separate addition instructions. We can exploit this as follows:

```
   // Copy double words in chunks of 8
    for (;sesv>=8*WLL; sesv-=8*WLL)
    {   *(LLWORD *) d= *(LLWORD *)s;
        *(LLWORD *)(d+WLL)= *(LLWORD *)(s+WLL);
        *(LLWORD *)(d+2*WLL)= *(LLWORD *)(s+2*WLL);
        *(LLWORD *)(d+3*WLL)= *(LLWORD *)(s+3*WLL);
        *(LLWORD *)(d+4*WLL)= *(LLWORD *)(s+4*WLL);
        *(LLWORD *)(d+5*WLL)= *(LLWORD *)(s+5*WLL);
        *(LLWORD *)(d+6*WLL)= *(LLWORD *)(s+6*WLL);
        *(LLWORD *)(d+7*WLL)= *(LLWORD *)(s+7*WLL);
        d+=8*WLL; s+=8*WLL;
    }

    // Copy ramaining of double words
    for (;sesv>=WLL; sesv-=WLL,d+=WLL,s+=WLL)
        *(LLWORD *) d= *(LLWORD *)s;
```

Since **WLL** is known at compile time, any modern compiler will do all multiplication above at compile time. Then addition will be done using offset addressing except for the line `d+=8*WLL; s+=8*WLL;` where addition can only be done separtly. The code above is only effcient when element size sesv ≥ 8*WLL. Smaller elements are still copied entirly through the second loop above. We can unroll the second loop using a switch. So the function becomes:

```
inline void ecopyA64_2(char *d, char *s , size_t
sesv)
{   // Copy double words in chunks of 8
    for (;sesv>=8*WLL; sesv-=8*WLL)
    {   *(LLWORD *) d= *(LLWORD *)s;
        *(LLWORD *)(d+WLL)= *(LLWORD *)(s+WLL);
        *(LLWORD *)(d+2*WLL)= *(LLWORD *)(s+2*WLL);
        *(LLWORD *)(d+3*WLL)= *(LLWORD *)(s+3*WLL);
        *(LLWORD *)(d+4*WLL)= *(LLWORD *)(s+4*WLL);
        *(LLWORD *)(d+5*WLL)= *(LLWORD *)(s+5*WLL);
        *(LLWORD *)(d+6*WLL)= *(LLWORD *)(s+6*WLL);
        *(LLWORD *)(d+7*WLL)= *(LLWORD *)(s+7*WLL);
        d+=8*WLL; s+=8*WLL;
    }

    // Copy ramaining of double words
    switch (ses/WLL%8)
    {   case 7:
        *(LLWORD *)(d+6*WLL)=*(LLWORD *) (s+6*WLL);
        case 6:
        *(LLWORD *)(d+5*WLL)=*(LLWORD *) (s+5*WLL);
        case 5:
        *(LLWORD *)(d+4*WLL)=*(LLWORD *) (s+4*WLL);
        case 4:
        *(LLWORD *)(d+3*WLL)=*(LLWORD *) (s+3*WLL);
        case 3:
        *(LLWORD *)(d+2*WLL)=*(LLWORD *) (s+2*WLL);
        case 2:
        *(LLWORD *)(d+WLL)=*(LLWORD *) (s+WLL);
        case 1:
        *(LLWORD *)d = *(LLWORD *)s;
        case 0:
    }

    // Copy remaining word.
    if (sesv>=WL)
    {   *(LWORD *) d= *(LWORD *)s;
        sesv-=WL; d+=WL; s+=WL;
    }


    // Copy remaining chars.
    for (;sesv>0; sesv-=WC,d+=WC,s+=WC)
        *d= *s;
}
```

Although the switch above removes loop overhead it adds a fixed overhead of the switch statement. Unfortunatly this resolves to about 5 intructions in Microsoft VC++ compiler, though other compilers might differ[16]. The improvment is NOT significant for small remainders but improves for larger remainders.We can increase the chunk size to 16, this will only improve more and add almost no overhead.

[16] The switch here does not compare every value in each case. It jumps immediately to the correct case using an indexed

### C. Redundant increments

There is a redundant increment to **s** and **d** done at the end every time the copying function is called (i.e. either by `d+=8*WLL; s+=8*WLL;` or by `d+=WC,s+=WC`). This can be eliminated by a more sofisticated method but it's too lengthy for this paper. Please refer to the complete algorithm in [12].

### D. Alternative similar methods

Aleternativly one could choose a less aggressive method of unfolding as follows:

```
inline void ecopyA64_3(char *d, char *s, size_t
sesv)
{
    // Copy double words
    for (;sesv>=WLL; sesv-=WLL)
    {   *(LLWORD *) d = *(LLWORD *) s;
        if ((sesv-=WLL)<WLL) break;
        *(LLWORD *)(d+WLL)=*(LLWORD *)(s+WLL);
        if ((sesv-=WLL)<WLL) break;
        *(LLWORD *)(d1+2*WLL)=*(LLWORD *)(s1+2*WLL);
        if ((sesv-=WLL)<WLL) break;
        *(LLWORD *)(d1+3*WLL)=*(LLWORD *)(s1+3*WLL);
        if ((sesv-=WLL)<WLL) break;
        *(LLWORD *)(d1+4*WLL)=*(LLWORD *)(s1+4*WLL);
        if ((sesv-=WLL)<WLL) break;
        *(LLWORD *)(d1+5*WLL)=*(LLWORD *)(s1+5*WLL);
        if ((sesv-=WLL)<WLL) break;
        *(LLWORD *)(d1+6*WLL)=*(LLWORD *)(s1+6*WLL);
        if ((sesv-=WLL)<WLL) break;
        *(LLWORD *)(d1+7*WLL)=*(LLWORD *)(s1+7*WLL);
        if ((sesv-=WLL)<WLL) break;
        d1+=BATCH_SIZE*WLL; s1+=BATCH_SIZE*WLL;
    }

    // Copy remaining words.
    if (sesv>=WL)
    {   *(LWORD *) d= *(LWORD *)s;
        sesv-=WL; d+=WL; s+=WL;
    }

    // Copy remaining chars.
    for (;sesv>0; sesv-=WC,d+=WC,s+=WC)
        *d= *s;
}
```

Function `ecopyA64_3` reduces the number of loops and allows the compiler to use offset addressing while still bieng simple and without adding additional overhead.

We provide two other methods in the full algorithm totaling 4 methods [12]. Our compiler directive REPCOPYSWAP _METHOD can be used to choose which of the 4 is used. The directive REPCOPYSWAP_PREVENT_LAST_INC can be used to prevent the last redundant increment from hapenning. An engineer should try all 4 methods to determine whats best for his/her compiler, target CPU and hardware.

table of addresses. Modern compilers are capable of such optimization

### E. Copying the remainder

In both `ecopyA64_2` and `ecopyA64_3` the remainder of the element (after copying double words) is copied using an "if" and "loop" statements. There is a way to avoid this overhead. The remainder size can range from 0 to 7 bytes. We can write 8 dedicated copying functions one for each remainder, each of those functions will act exactly the same except on the remainder copying part. One of those functions will be chosen at the beginning of the algorithm. We implement a similar but more aggressive method in Triple State Quicksort using 24 dedicated copying functions. Please refer to the complete code in [12].

### F. Using SSE2 instruction set

**Streaming SIMD Extensions 2** is an intel based supplementary assembly language instruction set introduced with Pentium 4 in 2001. SSE2 uses the 128bit sized XMM registers that can do memory operations in parallel since there are 8 registers[17]. SSE2 is primarily used in speeding up copying large memory blocks in real-time graphics demanding applications. An example is a memcpy() version by William Chan and Google [10] which is claimed to be 30-70% faster than memcpy() in Microsoft Visual Studio 2005.

Unfortunately, In order to use SSE2, copied memory blocks have to be 16byte aligned. This is easy to guarantee for large memory blocks using the library function `_aligned_malloc()`. This is not the case in sorting were small blocks (elements) are copied, it requires that the element size be a multiple of 16bytes which is a total memory wastage for small element arrays[18].

However, it is possible that element size as well as the starting address of the array just happen to be a multiple of 16 bytes, in which case SSE2 operations are valid. This is not a far shot since some compilers do quantize memory sizes to the nearest 4 bytes when data types contain 4 byte primitive types like integers. The following is our SSE2 based copy function.

```
inline void AlignedMemcpySSE2(const void *d, const
void *s, const size_t sesv)
{
    __asm
    {   mov esi, s;     //src pointer
        mov edi, d;     //dest pointer
        mov ebx, sesv;  //counter

        // Copy 128bit chunks

        loop_sse2_copy:
        movdqa xmm0, [esi];
        movdqa [edi], xmm0;
        sub ebx,16;
        jz sse2_end;

        movdqa xmm1, 16[esi];
        movdqa 16[edi], xmm1;
        sub ebx,16;
```

[17] XMM is different than MMX. The latter is an older technology

[18] Infact there are SSE2 instructions that can handle non-aligned addresses but they are not even remotely as efficient as the aligned instructions

```
        jz sse2_end;

        movdqa xmm2, 32[esi];
        movdqa 32[edi], xmm2;
        sub ebx,16;
        jz sse2_end;

        movdqa xmm3, 48[esi];
        movdqa 48[edi], xmm3;

        add esi, 64;
        add edi, 64;
        sub ebx,16;
        jnz loop_sse2_copy

        sse2_end:
    }
}
```

Notice the above uses inline assembly which uses 4 XMM registers to exploit parallelism. Triple State Quicksort uses this method whenever it detects that element size and the array starting address are 16byte aligned. This decision is made only once at the beginning of the algorithm.

### G. *Why swapping pointers is not healthy ?!!*

Element size can sometimes be large or very large, such as in a database server sorting a table with 10 or 20 fields in each row (record). A well known trick is to create an array of pointers (or indexes) that point to the elements in the original array, then sort this array of pointers instead of sorting the original array, and use the pointer array in the remaining of the program to access the original array in sorted order. This method will reduce swapping time significantly since pointers (32bit) are much smaller than the large elements. The small cost endured is the pointer indirection of access to the original array, i.e. the comparison function and program access to the original array will be a little bit slower.

However, there is a rather unobvious and quite terrible side effect of using this trick. To illustrate, consider the following program:

```
#include <iostream>
#include <time.h>
using namespace std;

void main()
{
    clock_t tim,tim2;
    int size,i,j,rn,sum,*temp,**pi,*vl,repeat;
    bool shuffle;

    cout << "Size? "; cin >> size;
    cout << "Repeat? "; cin >> repeat;
    pi = (int**) malloc(size*sizeof(int*));
    vl = (int*) malloc(size*sizeof(int));

    cout << "Shuffle? (1=Yes, 0=No)";
    cin >> shuffle;

    // fill from 1 to size-1
    cout << "Preparing arrays...\n";
    for (i=0; i<size; i++)
    {   vl[i]=i; pi[i]=&(vl[i]);
    }

    // shuffle randomly
    if (shuffle)
        for (i=0; i<size; i++)
        {   // Choose random index from i to size-1
            rn =
((double)rand()/(RAND_MAX+1))*(size-i)+i;
            temp = pi[i]; pi[i]=pi[rn]; pi[rn]=temp;
        }

    // Start timer
    cout << "Timer started...\n";
    tim = clock();

    for (j=0; j<repeat; j++)
    {   sum=0;
        for (i=0; i<size; i++)
            sum+=*(pi[i]);
    }

    // Stop timer
    tim2 = clock() - tim;

    cout << "time = " << tim2 << "\nsum=" << sum <<
"\n";
   }
```

The purpose of the above program is to test how long it takes to iterate an array of integers **sequentially** from first to last integer as opposed to how long it takes to iterate the same array **randomly**. An array vi[ ] is filled with values from 0 to `size`. Then a second pointer array pi[ ] is set to point to each element in vi[ ] sequentially. We run the program twice, once with `shuffle=false` which will do sequential interation. And a second time with `shuffle=true` which will shuffle the array of pointers pi[ ] randomly before traversal starts. Iteration of vi[ ] is done indirectly through the pointers in pi[ ] in both runs. Time is calculated only while iterating, so shuffling and filling does not have an effect in our test[19].

The test bed is an Intel Core i7 2.93Ghz with 4Gb of RAM. Results are eyebrow lifting, as shown in TABLE I. Once the array **size** starts to grow larger than 50000 intergers, random iteration time becomes twice as much as sequential iteration and continues to increase rapidly as the **size** is increased reaching upto 12 (11.9) times slower than sequential iteration as **size** aproaches 200,000,000. The only sound explanation for this behaiviur is **cache misses**. Proccessor caches are obviusly much faster than RAM and are desgined to automatically fetch nearby memory locations which are expected to be accessed shortly. In Sequential access, once an array element is accessed, nearby locations will be readily available in the cache hence hastening access to the next few array elements. Random iteration abuses the cache since it accesses elements that are very far from each other which significatly increases the number of cache misses and forces direct access and fetching from RAM to cache most of the time.

The i7 CPU has an L1 cache of 32 Kbytes per core, which is enough to hold 8192 integers. It can be seen in TABLE I. that random performance does NOT degrade as long as size < 8000 which is just enough for the cache to hold the entire array.

[19] We use clock() for clarity and ease of use, the error is marginal since we count higher magnitudes. But high performance counters like QueryPerformanceCounter() can also be used leading to similar result.

TABLE I. SEQUENTIAL VS. RANDOM ACCESS

| Size | Iteration Time (ms) (repeat=15000) | |
|---|---|---|
| | ***Sequential*** | ***Random*** |
| 4000 | 47 | 47 |
| 8000 | 93 | 94 |
| 10000 | 110 | 140 |
| 50000 | 577 | 1202 |
| 100000 | 1170 | 2839 |
| 500000 | 5991 | 16146 |
| 1000000 | 12012 | 65239 |
| | **(repeat = 500)** | |
| 2000000 | 1154 | 6864 |
| 5000000 | 2533 | 22963 |
| 10000000 | 5897 | 55021 |
| 20000000 | 11793 | 141305 |

Back to the main subject, sorting pointers whitout sorting the actual elements can cause the same type of slowness as in random access. The sort may be fast, but the whole program performance will suffer afterwards. Note here that this conclusion relies on the fact that the array is usually accessed more frequently than being sorted. Otherwise, for the odd case were the array is being sorted so frequently and accessed only oncassionally, then pointer swaping is still an effective method.

### *H. Late swaping*

For a large element size, we propose a method to take advantage of the swift pointer swaping without incurring the slowness side effects described in sub section G. Were the pointers array is only temporary, once the pointers are sorted, the original array elements are swaped into sorted order using the information in the pointers array, this proccess can be done in *O(n)* swaps at a maximum of $n/2$ swaps in original array[20]. The algorithm is as follows.

```
#define MALLOC_ALIGN    32
int i,j;
char **platepc=NULL;
char *pa,*pa2;
size_t size;
static char *temp=(char *)
_aligned_malloc(ses,MALLOC_ALIGN);

// Allocate temporary pointer array
size = (n+2)*sizeof(*platepc);
platepc = (char **) _aligned_malloc(size,MALLOC_ALIGN);

// Set every pointer to point to its corresponding real
element.
pa = a;
for (i=0; i<n; i++)
{   platepc[i]=pa; pa+=ses;
}

// Sort array of pointers
_TripleStateQuickSortPCLate((char *)
platepc,n,sizeof(*platepc),cmp);

// Start Late Swaping (copying)
pa = a;
for (i=0; i<n; i++)
{   // If the "pointed to" element hasn't been put into
place yet AND if it's not already inplace
    if (platepc[i]!=NULL && platepc[i]!=pa)
    {   ECOPY(temp,pa,ses);
        pa2=pa; j=i;
        do
        {   ECOPY(pa2,platepc[j],ses);
            pa2 = platepc[j]; platepc[j]=NULL;
            j=(pa2-a)/ses;
        } while (platepc[j]!=pa);
        ECOPY(pa2,temp,ses); platepc[j]=NULL;
    }
    pa+=ses;
}

_aligned_free(platepc);
```

ECOPY(a,b,ses) above copies element **b** into element **a** were **a** and **b** are pointers to elements and **ses** is element size. This algorithm will only be triggered if element size is larger than a predefined threshold; were the benefit of pointer swapping will start to overweigh the overhead taken above. The threshold is found empirically.

## IX. RUN TIME RESULTS

Our test bed is an Intel Core i5 2.93Ghz with 6Gb of RAM. The test done here is exactly the same test in section VI but showing run time. We do the same test first with small elements then with larger sized elements. Then we do several other difereing tests.

### *A. Sorting small elements*

First, we put our algorithm at a disadvantage by sorting small sized elements (integers). Swaps are less important in this case since they contribute the least to running time thus any algorithm optimizing swaps would be at a disadvantage. We repeat the test 5000 times to show significant running times (in millimseconds) instead of counting microseconds which could have included system noise and increased measurement error. Results in Fig. 13, still show that Triple State is the best while Bentley's is very close.

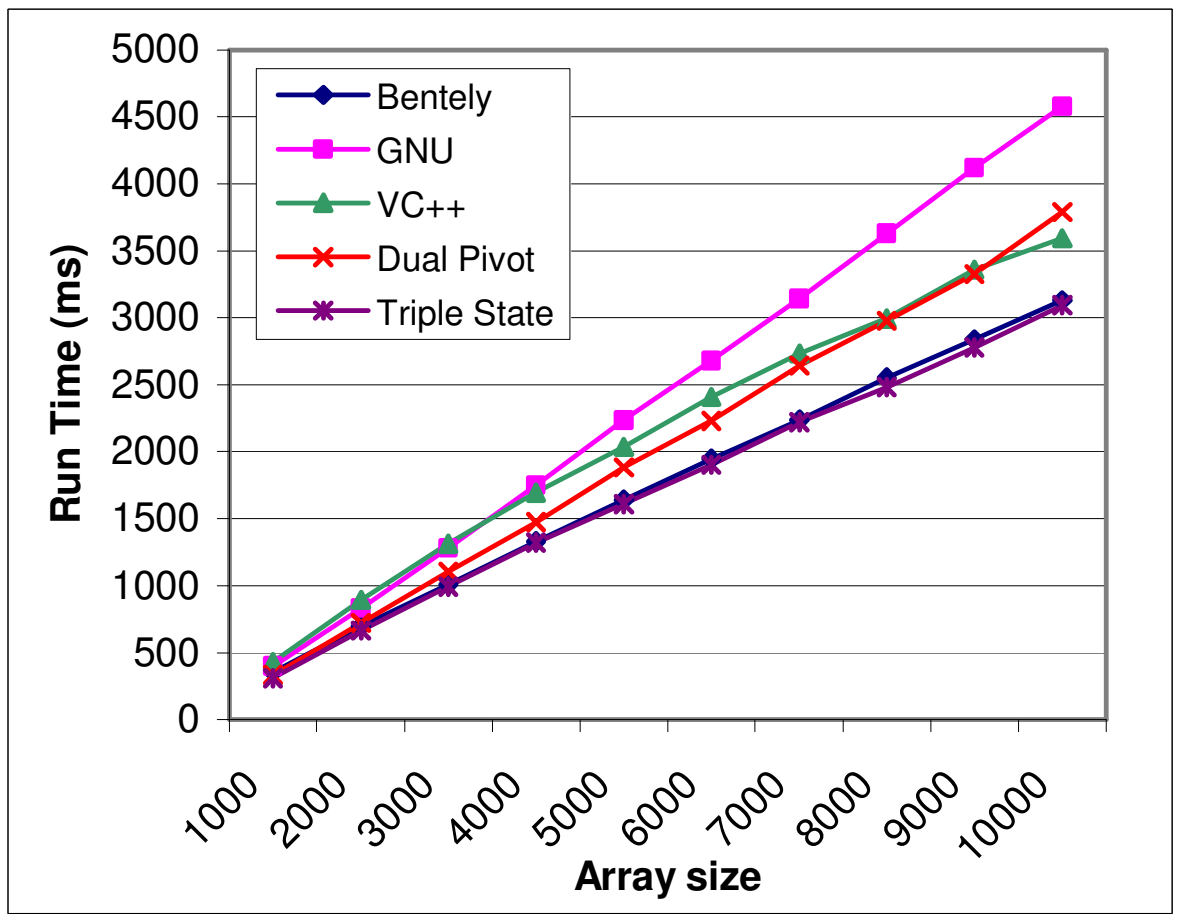

Fig. 13. Run time. Random Elements. Integers.

[20] This is done after *0.222nln(n)* of swapping pointers in sorting as indicated earlier.

Different orderings of the same random input arrays confirm lower times for Triple State in almost all cases shown in Fig. 14, Fig. 15, Fig. 16, and Fig. 17. The rest of the results are in Appendix C.

Tests for large arrays are repeated 40 times. Results in Fig. 18, were Triple State comes either best or second to best. Dithered ordering is shown in Fig. 19 with much lesser times for Triple State. The difficult Fort ordering in Fig. 20, still shows Triple State being best while Bentley's is very close. The rest of the results are in Appendix C and all confirm Triple State performed in lowest time.

Fig. 21, shows times for the 6 adverse inputs as done in section VI. Triple State has lowest run time in all cases.

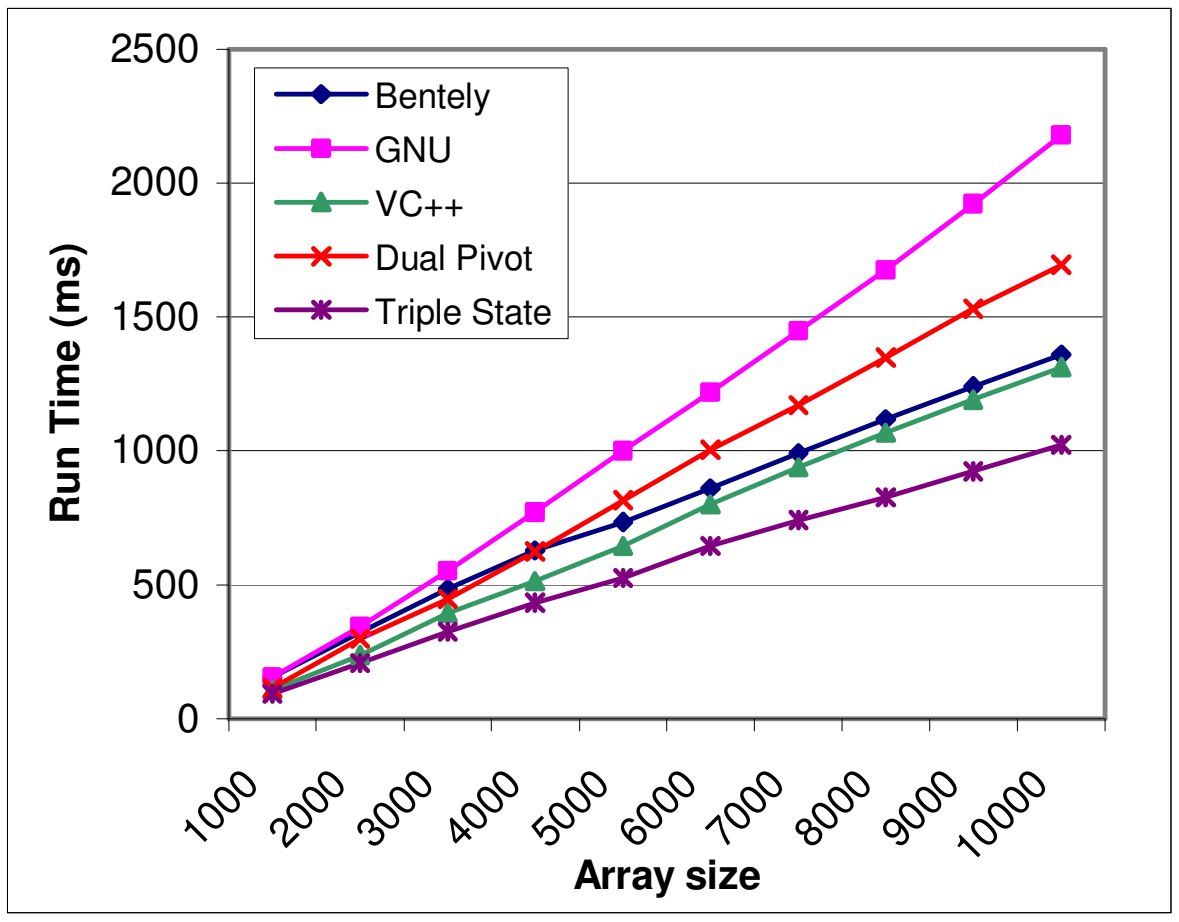

Fig. 14. Run Time. Sorted Input. Random Elements. Integers.

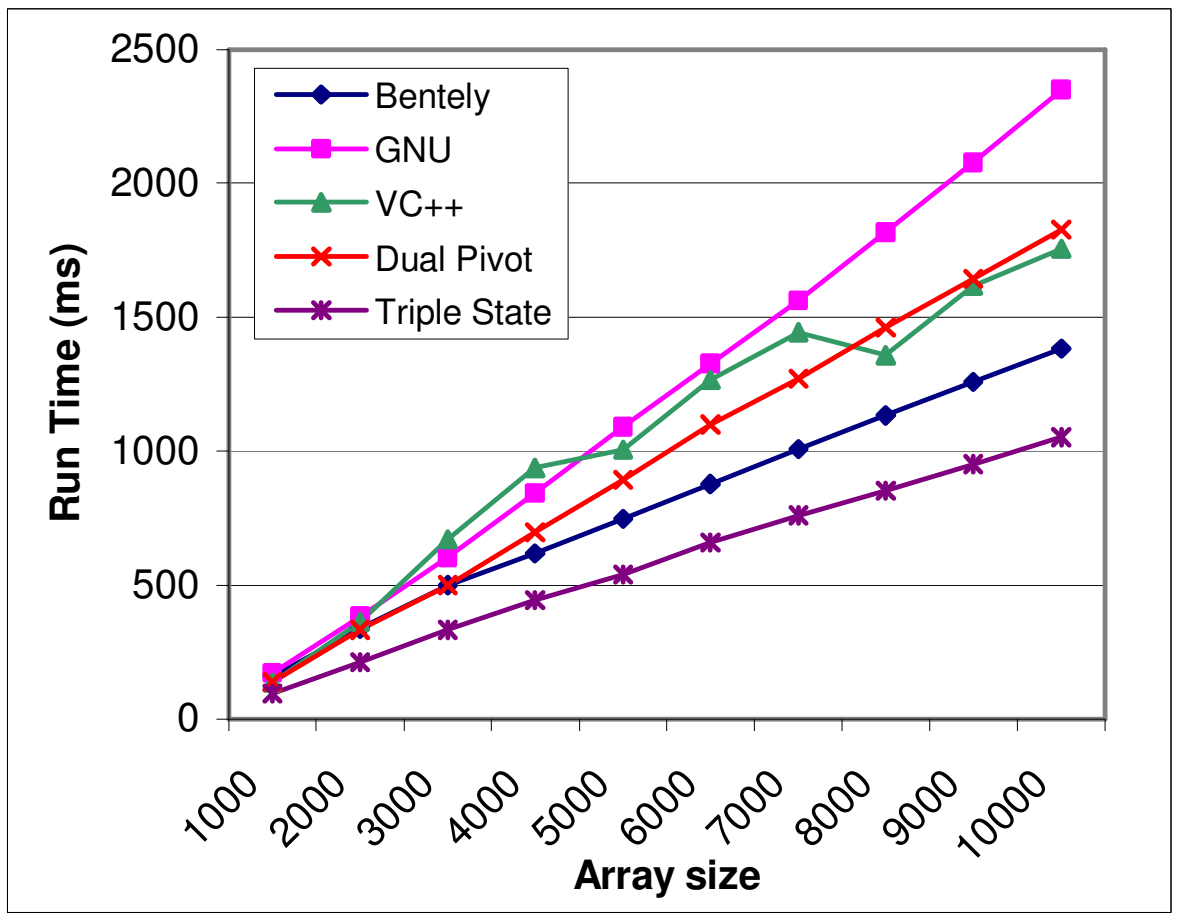

Fig. 15. Run Time. Reversed Input. Random elements. Integers.

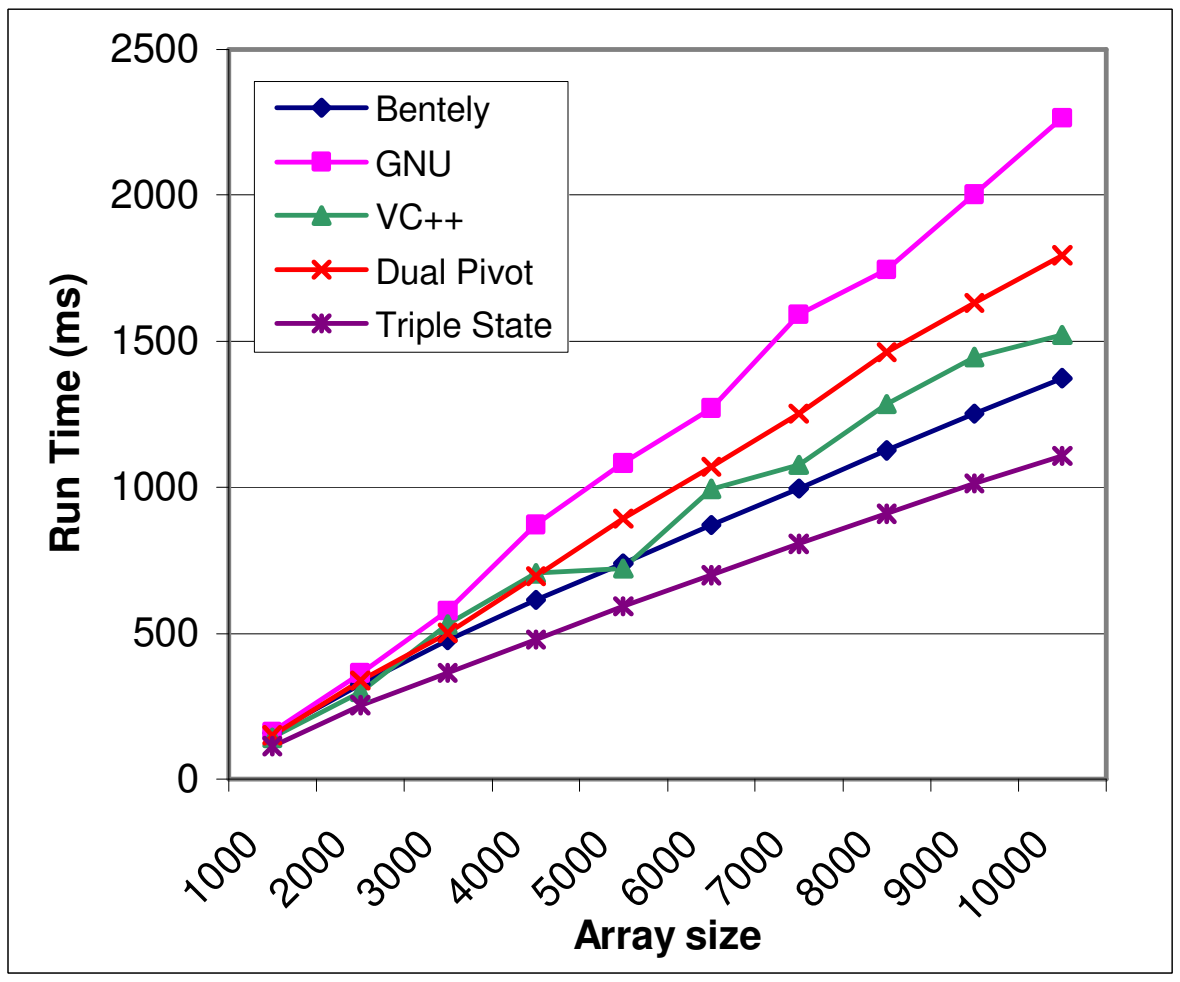

Fig. 16. Run Time. Front half reversed. Random elements. Integers.

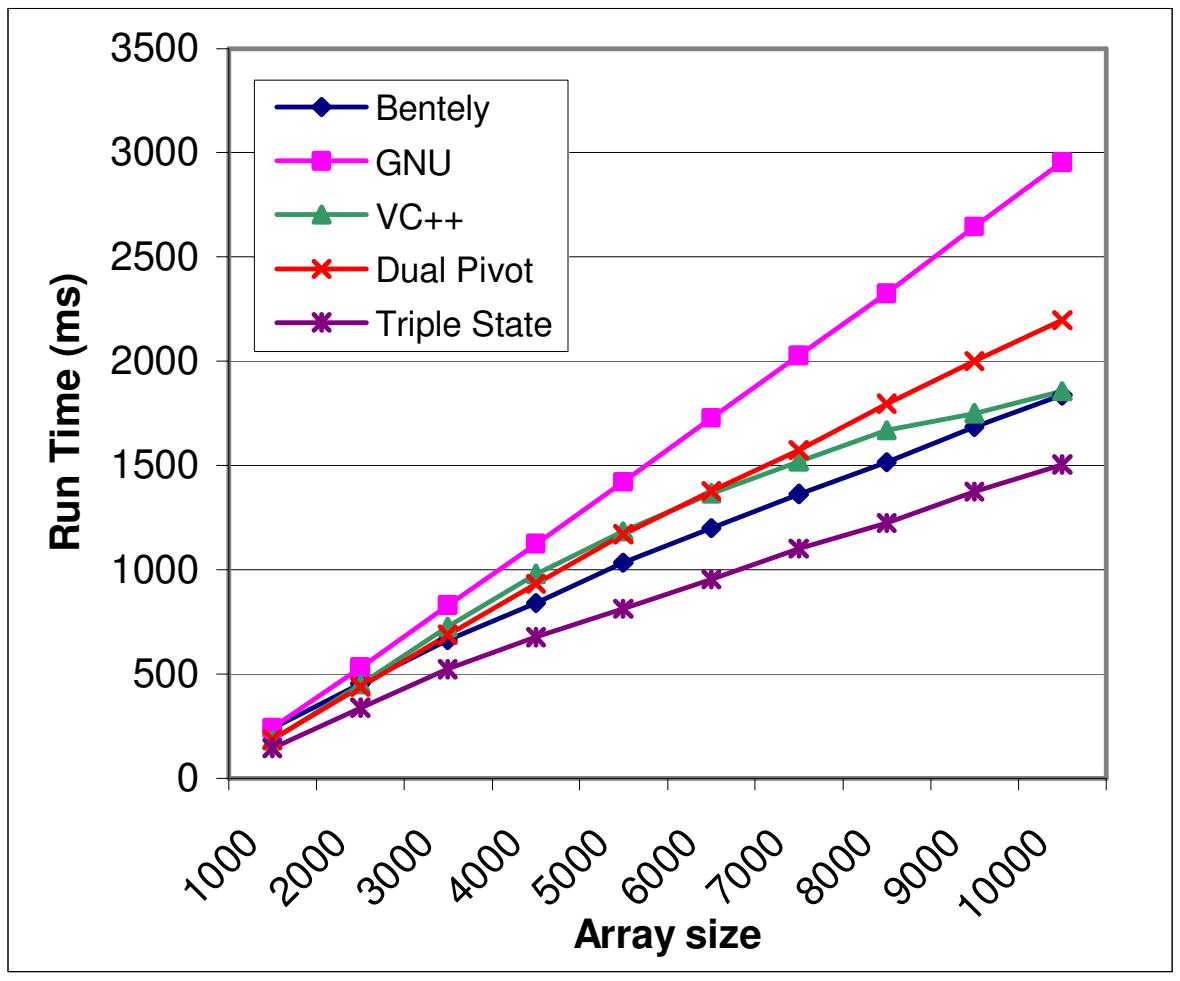

Fig. 17. Run time. Dithered input. Random elements. Integers.

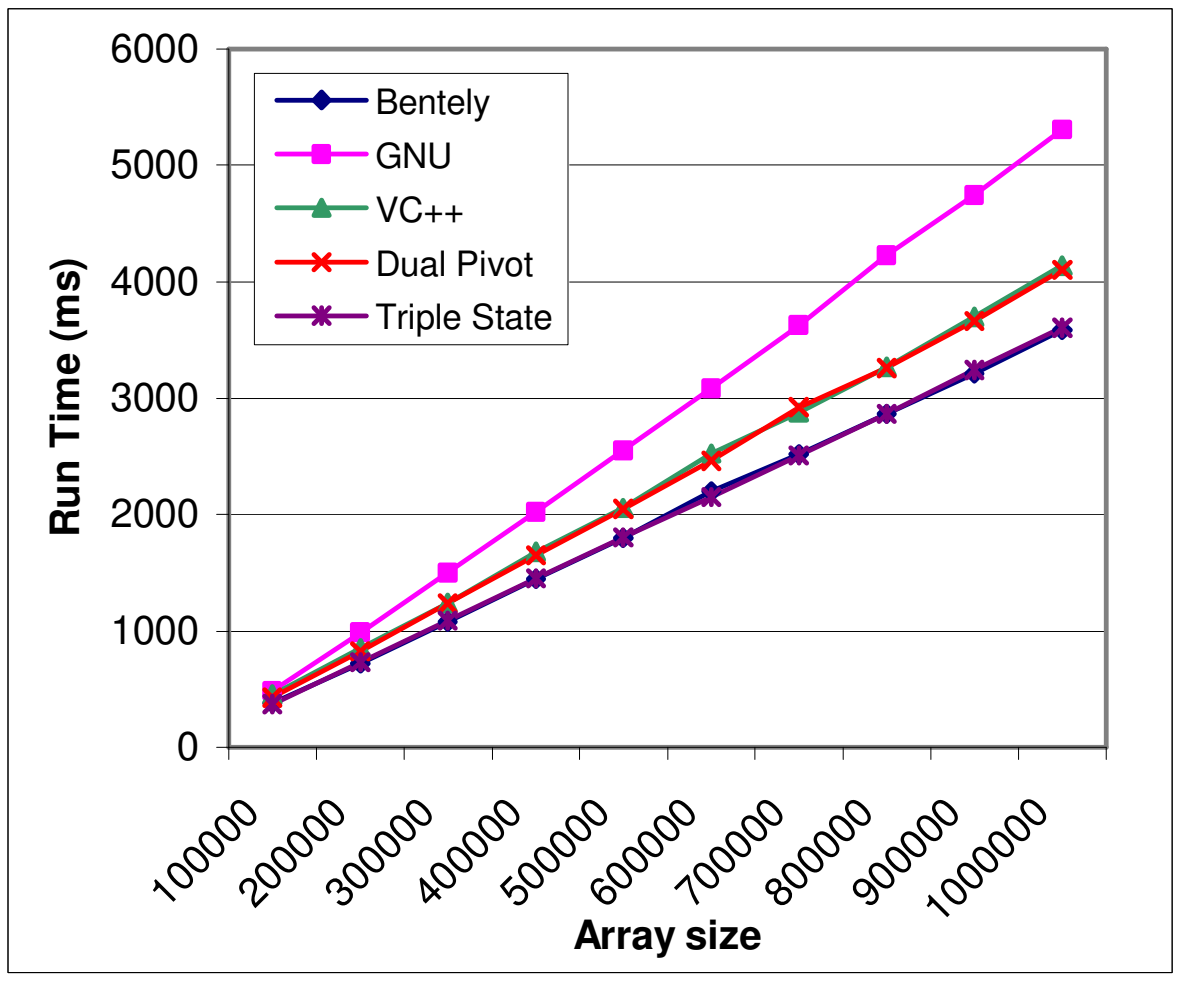

Fig. 18. Run time. Random input. Large arrays. Integers.

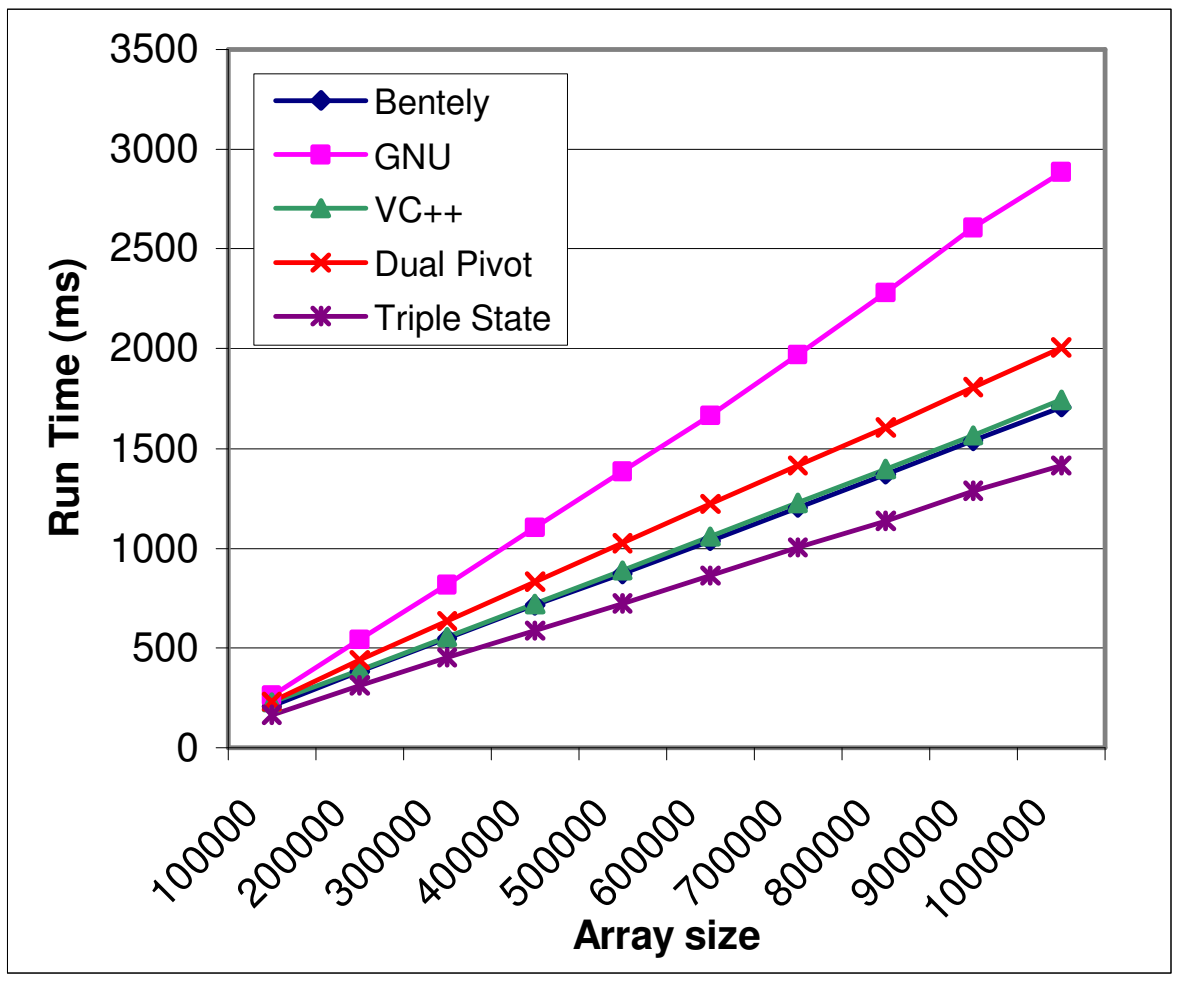

Fig. 19. Run time. Dithered input. Random Elements. Large arrays. Integers.

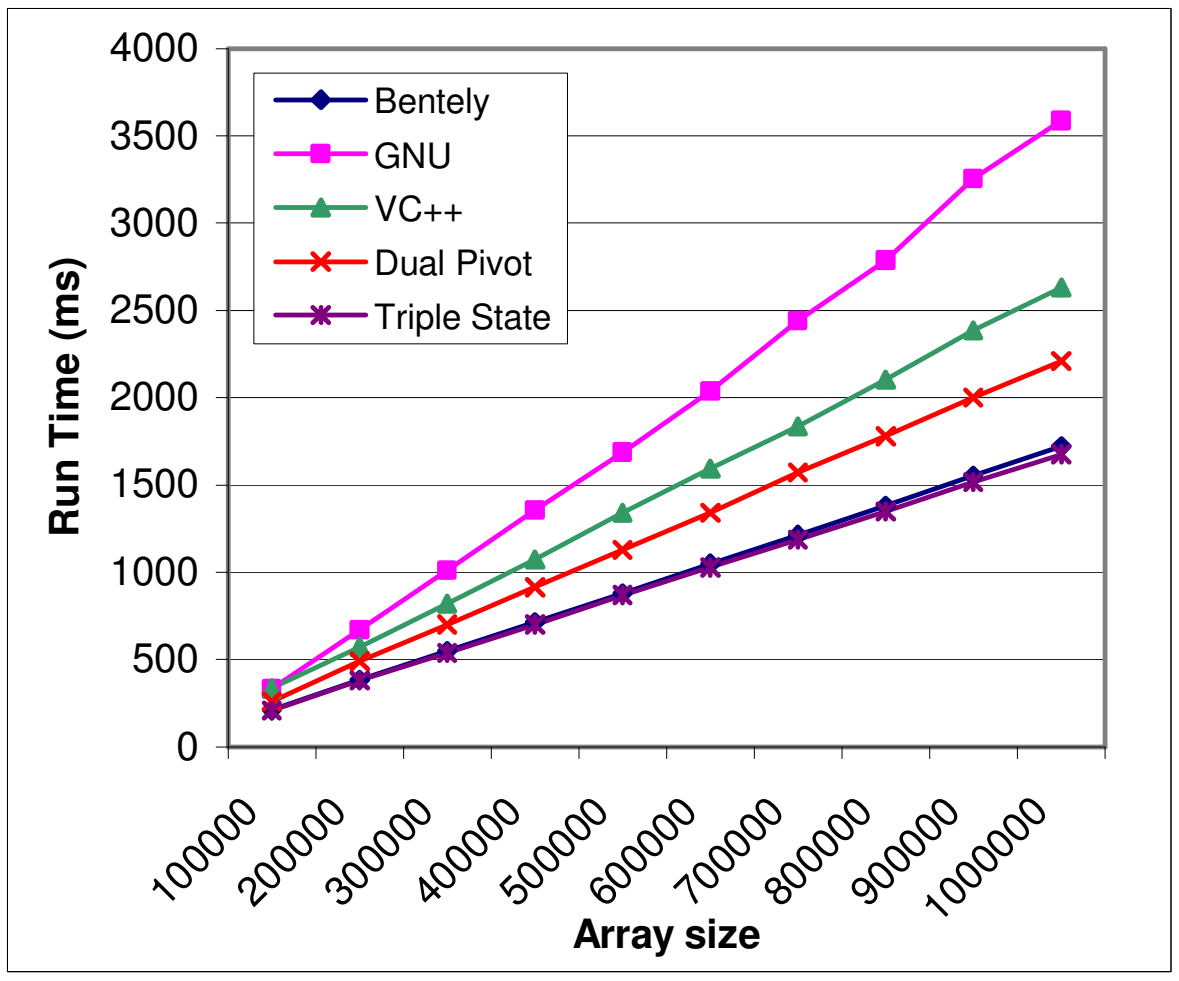

Fig. 20. Run time. Fort input. Random elements. Large arrays. Integers.

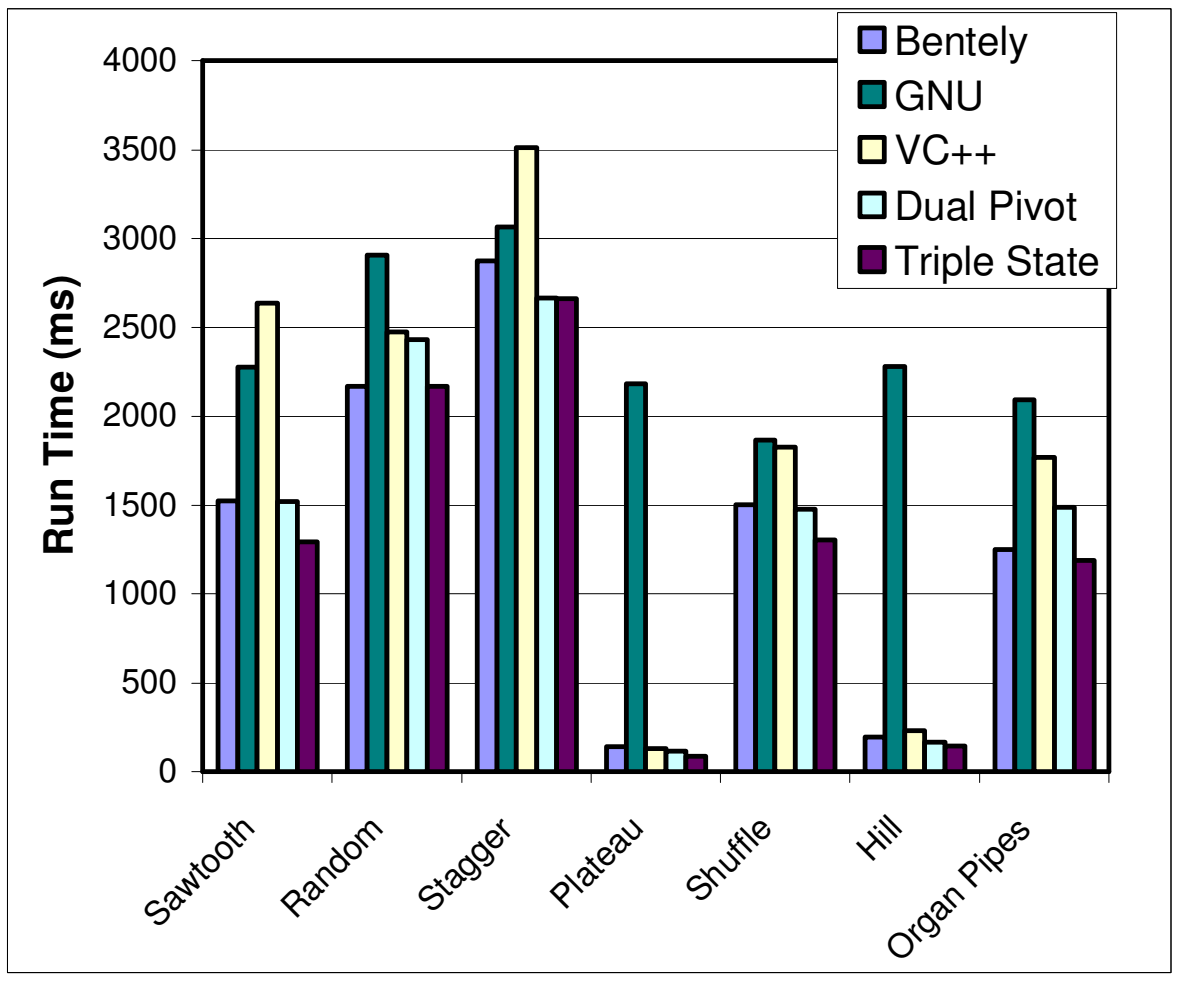

Fig. 21. Run time. Multi adverse type input test. 2 million elements array. Integers.

## *B. Sorting Large elements*

As the elements become larger, the advantage of lowering the swaps slowly becomes more prominent. Fig. 22, shows run times for random input, large arrays with 56 byte sized elements. Each element is a record of 5 integers, 3 doubles, 1 float and 1 short. As shown, Triple State is still best in all cases. Notice here that 56 bytes is not a multiple of 16 bytes which means that Triple State is not even using SSE2 instructions as described in subsection VIII.F. Otherwise it would have performed even better. 56 bytes is also less than our late swapping threshold (i.e 320 bytes) so it's not using late swapping as described in subsection VIII.G. Comparing with Fig. 13, we can see how other algorithms (except Bentley's) degrade in sorting large elements. Other orderings inputs generally result in much lower times for Triple State. Fig. 23, also confirms these finding for other adverse inputs.

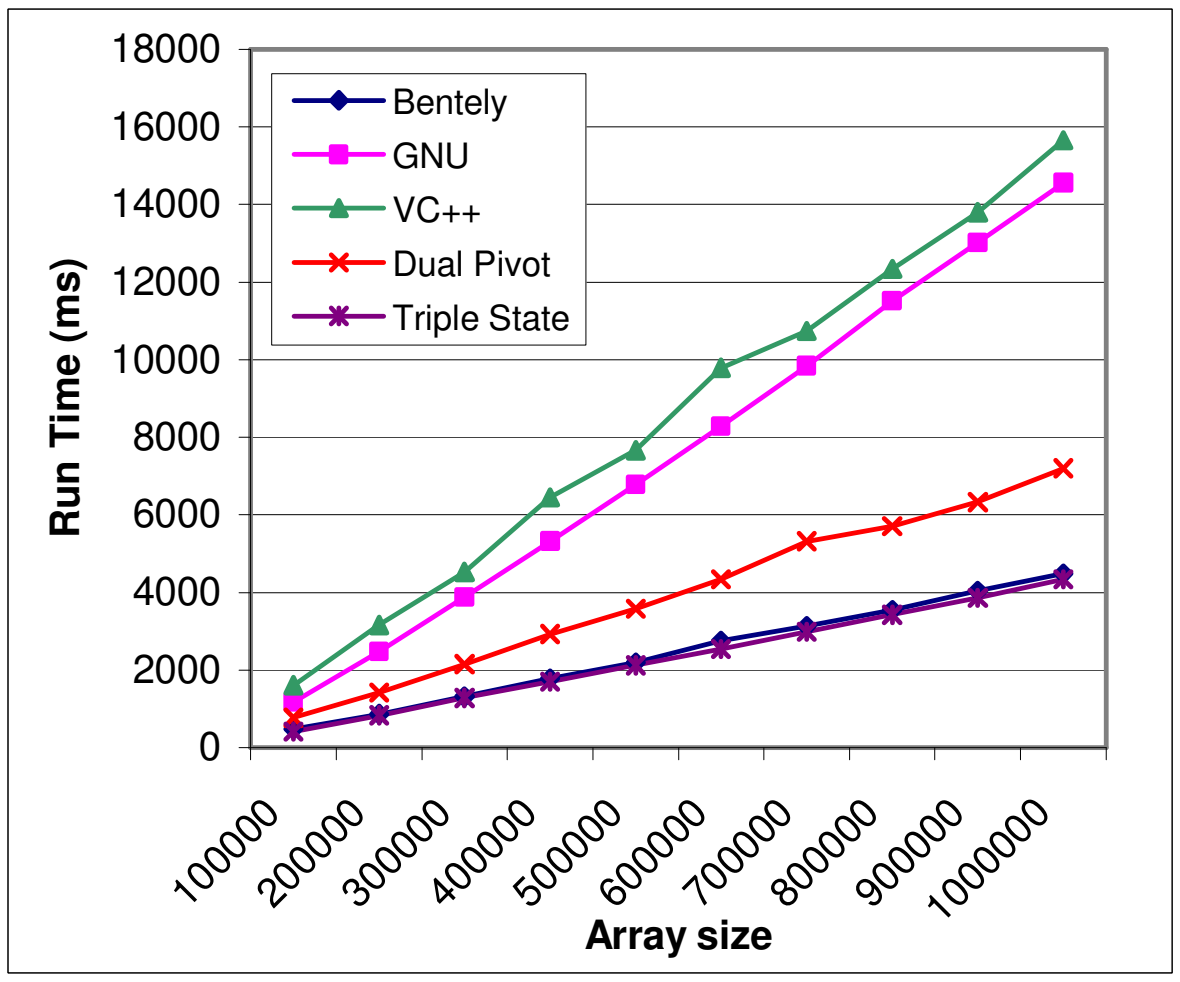

Fig. 22. Run time. Random input. Large arrays. 56 byte elements.

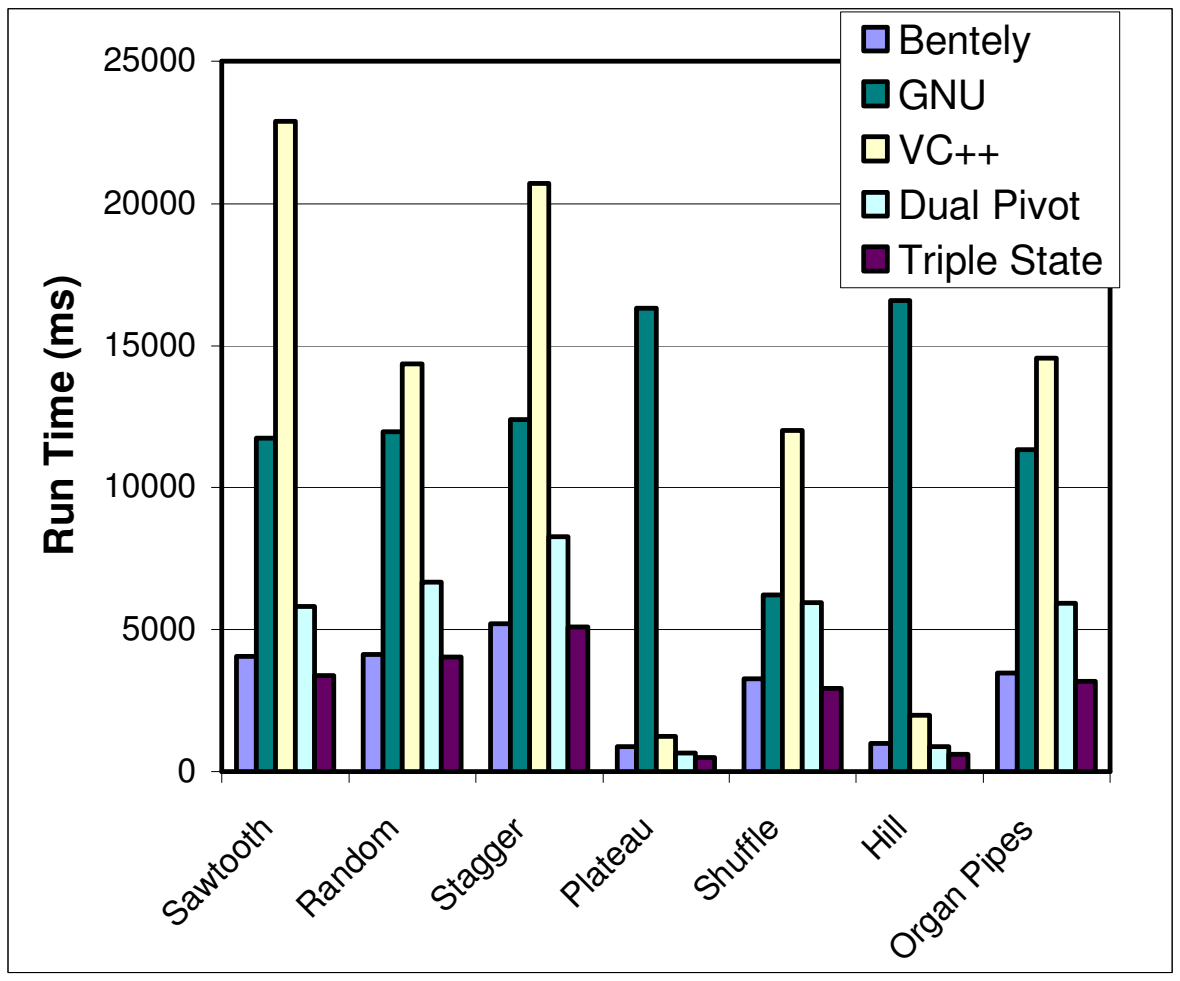

Fig. 23. Run time. Multi adverse type input test. 2 million elements array. 56 byte elements.

*C. Varying the range*

So far we have seen how the 5 algorithms behave when **n** is variable while the range of elements (i.e **arange**) is fixed at relatively low value. This allows for a fair amount of equal values to occur within the input arrays. We do this because equal elements is known to make Quicksort misbehave, so it's worth testing for. Nevertheless, real world input can be completely unique values, like for example sorting civil IDs or customer numbers.

This test sorts 100000 16 byte size elements (containing 2 integers and 1 double) while varying the range from a low 100 for many equals to a very high 2 billion for an almost no-equals array[21]. The test is repeated 200 times to get significant run times. Random input shown in Fig. 24, Triple Sate is best in all cases. Notice how VC++ qsort significantly worsens its performance the more unique the elements become, which signifies the importance of such test in fool proofing an engineers decision.

The rest of the results are in [12] and all show Triple State being either best or very close to the best. For example Fig. 25, shows the front-half-reversed ordering with significant improvement in Triple State allover the range.

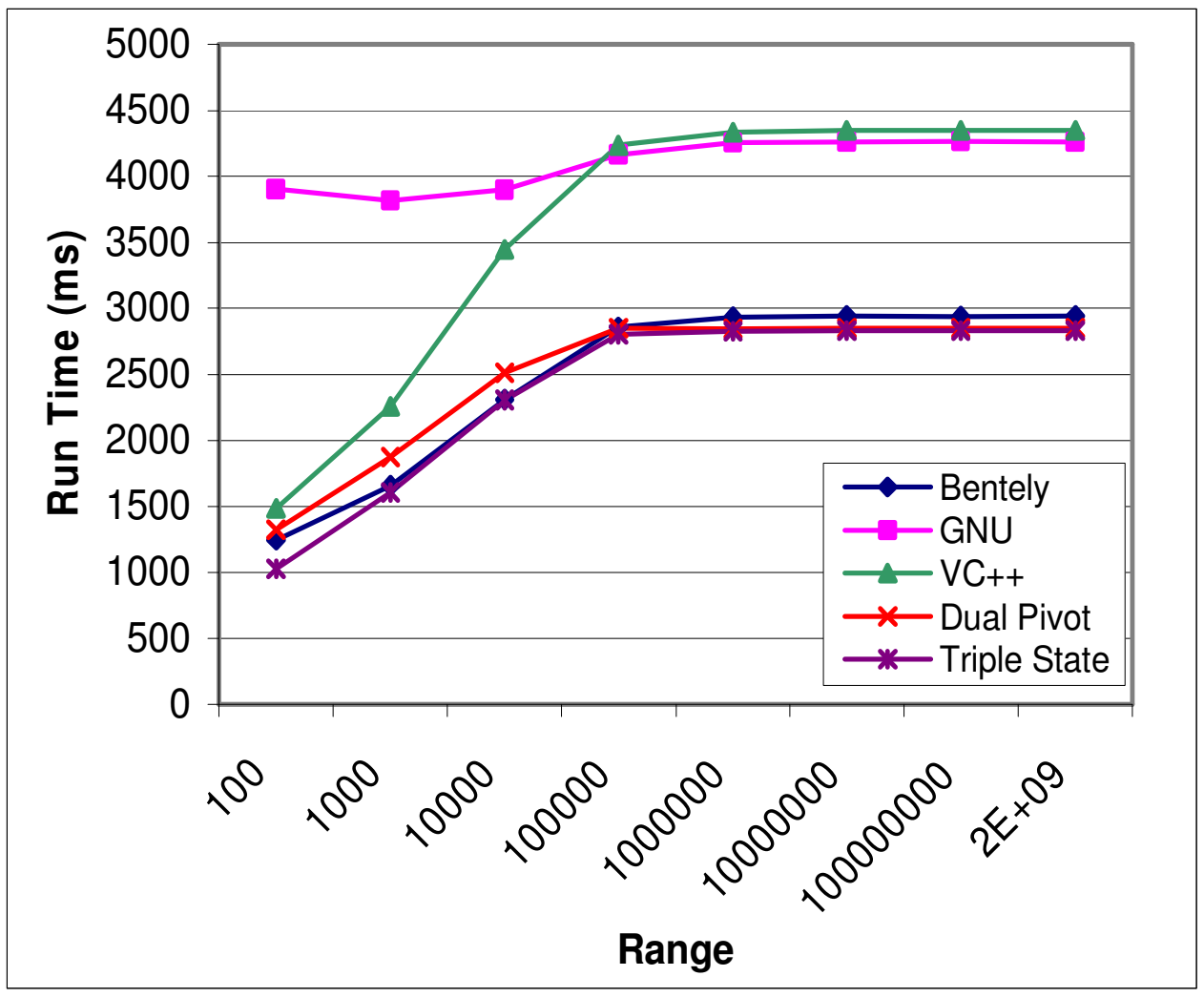

Fig. 24. Run time. Random input. 100000 elements. Variable range. Element size is 16 byte.

[21] Using a range of 2 billion to achieve almost unique elements is much easier than producing random guaranteed unique elements. And the result is almost the same. Thus we do it the easier way.

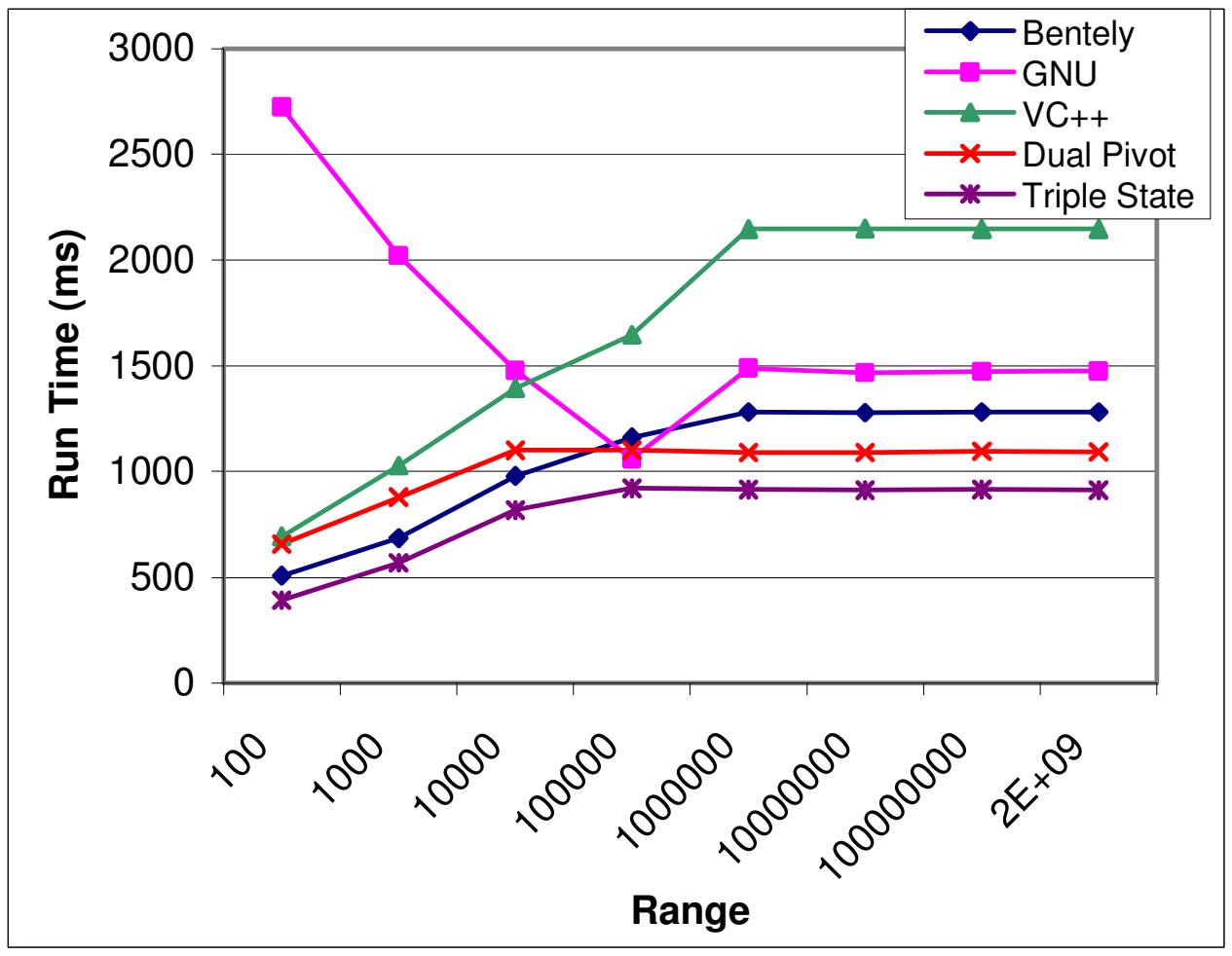

Fig. 25. Run time. Random input, front half reversed. 100000 elements. Variable range. Element size is 16 byte.

*D. Sorting Strings*

The dominant time consuming factor in sorting strings is the number of comparisons since the comparison operation can take extensive CPU time, while comparing character by character in alphabetic order which can take much more than what integer and floating point comparison would take.

Regarding string swap time. There are two types of strings with regard to storage. *Fixed length strings*, which can be stored in the array itself. And *Dynamic length strings,* which have varying sizes and are usually terminated with a null character to indicate their lengths. Dynamic length strings, are not stored in the array itself. Instead, the array being sorted would contain pointers that point to the individual strings in other locations in memory. Both types of strings are used in today's common database systems. For example in SQL, CHARACTER(n) and VARCHAR(n).

The effective result of the above is: sorting fixed length strings is essentially similar to sorting large elements as discussed in sub-section IX.B with the difference of the comparisons taking more CPU. While sorting dynamic length strings is essentially sorting 32bit (or 64bit) pointers again with the difference of the comparisons being the dominant time consumer.

We do a test for dynamic length strings. For a proper test, we sort a randomly generated list of real English/American full names. We use two freely available databases of English/American first and last names [14] to generate our array[22]. Every element contains a pointer to a string composed of a randomly chosen first name from the first-name database and a randomly chosen last name from the last-name database. The two names are concatenated together to form a full name string that is pointed to by the pointer in an element in the array that is to be sorted. For a more realistic sort, the element also includes a redundant integer which is obviously not used in the comparison. It should also be noted here that the "dithered" ordering is not possible in this test since it's supposed to add an integer amount to the string which obviously has no meaning. Fig. 26, shows run time results where Dual pivot, Bentley and Triple State have lowest times closely.

[22] The first-name database had some duplicates!! so we removed those.

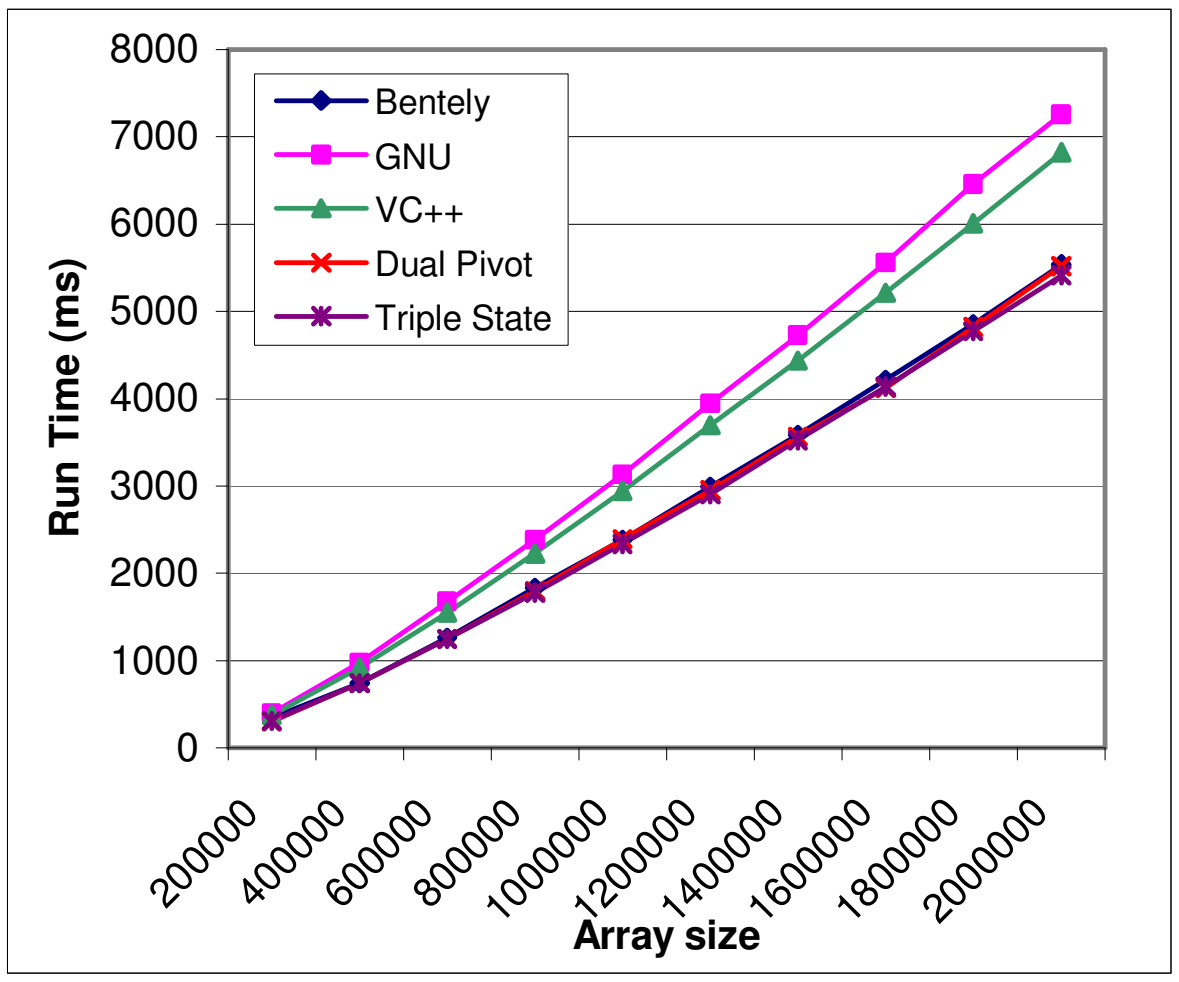

Fig. 26. Run time. Random string full names. Large arrays up to 2 milion elements.

Dual pivot Quicksort exhibits lowest run times in Sorted ordering input, since it has significantly low number of comparisons in this cases. Fig. 27, shows Sorted case. Triple State is best in Fort and Reversed. Fig. 28, shows Fort case. In Front half reversed and Back half reversed, Triple State and Dual pivot are both best and very close in almost all cases. Again, our complete results are in [12].

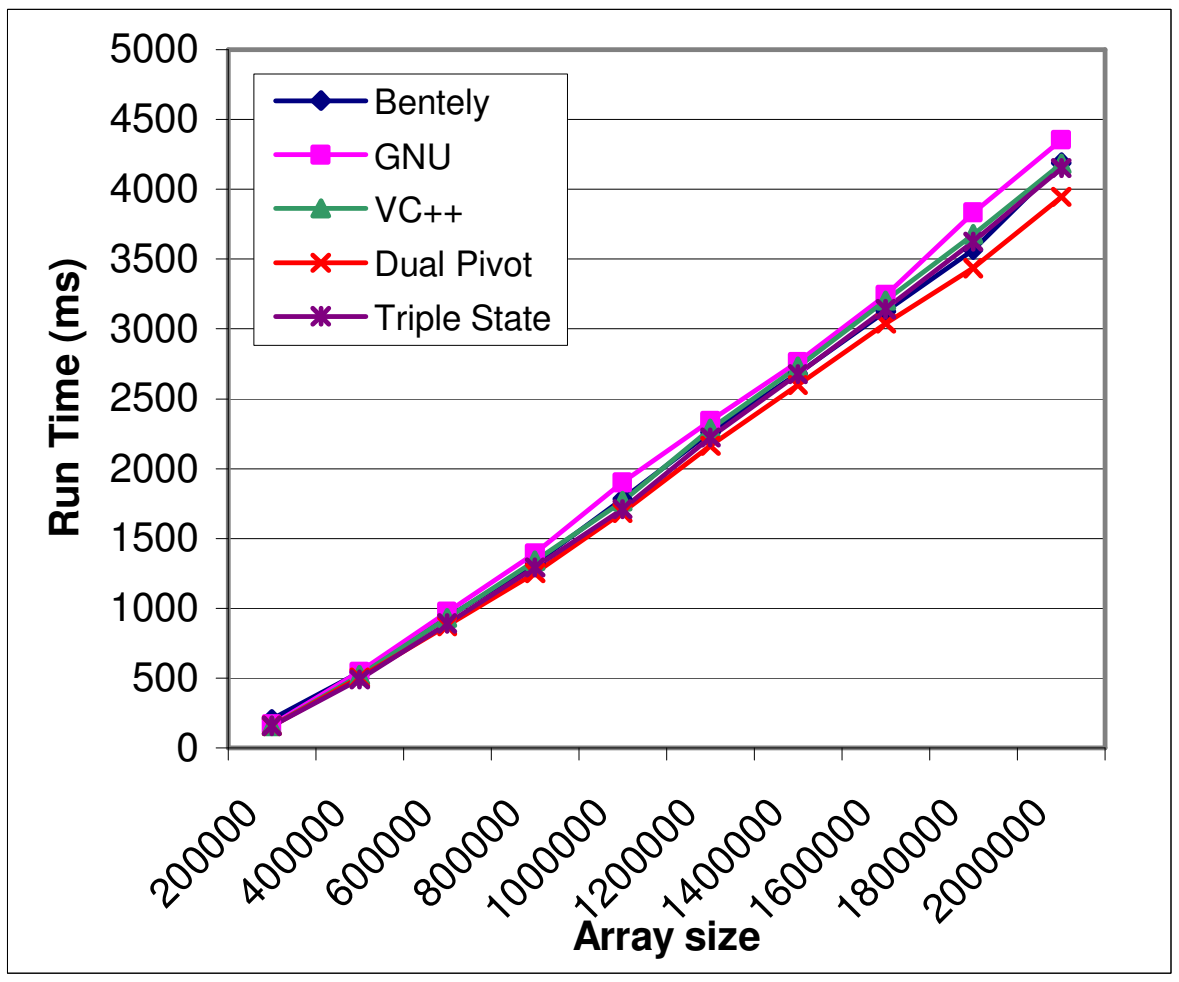

Fig. 27. Run time. Random string full names. Sorted input. Large arrays up to 2 milion elements.

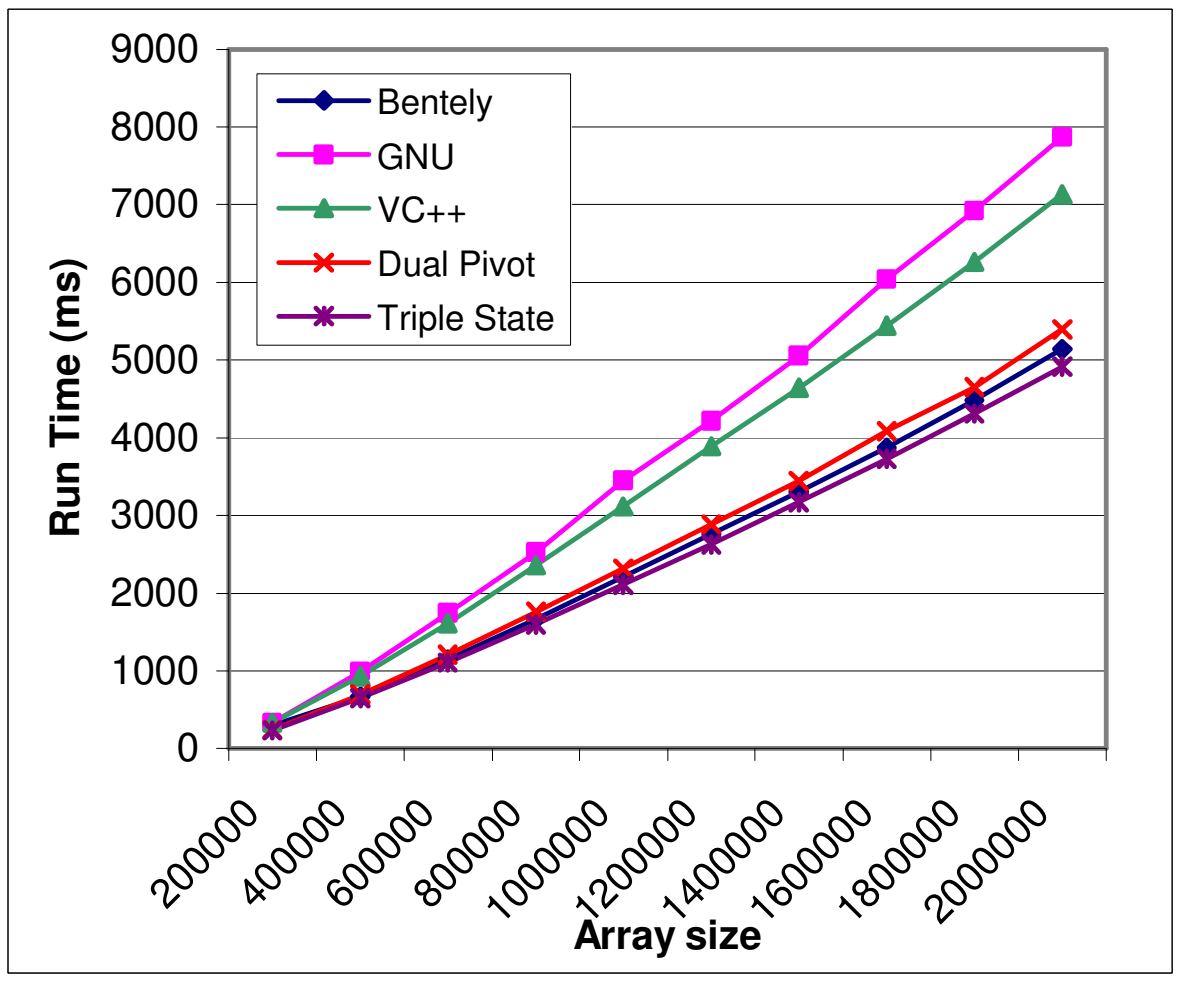

Fig. 28. Run time. Random string full names. Fort input. Large arrays up to 2 milion elements.

### *E. Overhead test*

When a sorting function is run millions of times on small or tiny arrays, the coding overhead like variable initialization, function call time and temporary memory setup become much more pronounceable and affects the total running time. Obviously, code overhead is not immediately measurable from neither the comparisons nor swaps results. A special type of test needs to be done, where comparison and swap time is minimized relative to code overhead. We run a test with small 50 elements (integers) arrays run 400000 times. Fig. 29, shows the results, notice the significant improvement that Triple State exhibits.

One might wonder how a complex algorithm like Triple State can do well in this test. The main reason relies behind the fact that speed of code overhead depends upon which part of the code is executed how often, not necessarily the size of the code.

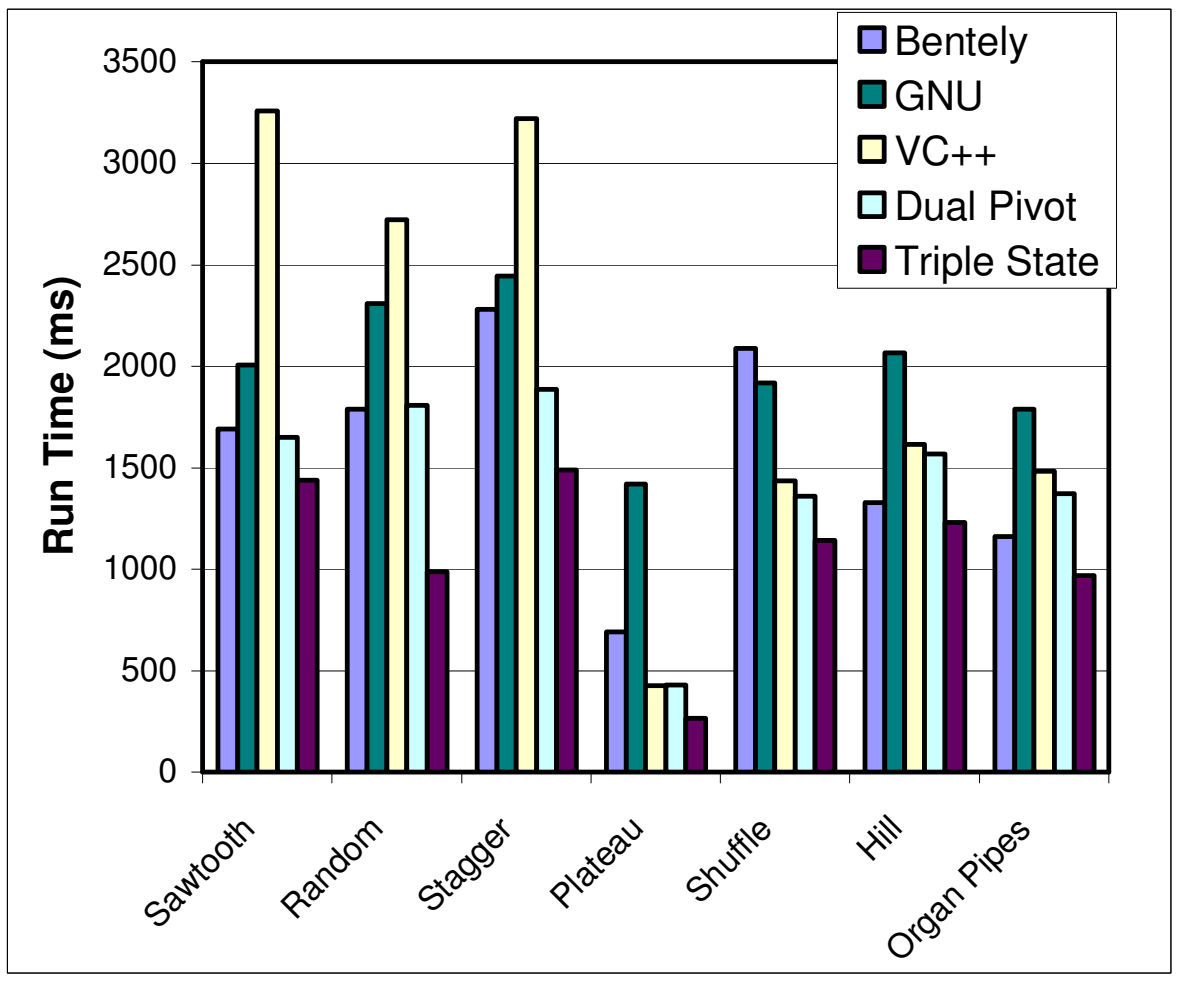

Fig. 29. Overhead test. Multi adverse type input test. 50 elements array, arange=15. repeated 1200000 times. Integer elements.

### F. Overall and Average Percentage Improvement

Percentage improvement is trivially calculated as follows:

$$improvement = \frac{T_{triple} - T_o}{T_o} \times 100\%$$

Were $T_{triple}$ is run time of Triple State Quicksort. $T_o$ is the run time of any other algorithm. We sum all run times of each algorithm for all tests we did above and calculate the **overall** percentage improvement shown in TABLE II.

TABLE II. OVERLL PERCENTAGE IMPROVEMENT OF TRIPLE STATE COMPARED TO OTHER ALGORITHMS

| | Bentley | GNU | VC++ | Dual Pivot | Triple State |
|---|---|---|---|---|---|
| **Total Run Time (ms)** | 702857 | 1474837 | 1423174 | 879461 | 591163 |
| **Improvement** | 15.8% | 59.9% | 58.4% | 32.7% | --- |

Although the above method of calculation is straight forward, it is rather biased towards longer tests, for example when n is large. Essentially, short tests have a smaller impact on the improvement figure which could be considered unfair by an engineer if short sorts were frequently repeated in the real world.

A less biased approach of calculation is the **average** percentage improvement. Were we calculate an improvement for each test we did individually, then take the average of all those improvement figures, shown in TABLE III.

TABLE III. AVERAGE PERCENTAGE IMPROVEMENT OF TRIPLE STATE COMPARED TO OTHER ALGORITHMS

| | Bentley | GNU | VC++ | Dual Pivot | Triple State |
|---|---|---|---|---|---|
| **Improvement** | 19.5% | 48.3% | 43.4% | 29.2% | --- |

## X. CONCLUSION

The goal of this work is to design and implement an efficient general purpose sorting function based on Quicksort. The function should cope with the increasing large database demands on modern personal computers or servers while facilitating the new advances of modern processors and memory architectures. Triple State Quicksort improves significantly on other current Quicksort implementations as seen in subsection IX.F at the price of a temporary *n/2* elements extra memory.

The rationale behind the acceptance of the *n/2* memory overhead can be summarized as follows:

1. Memory is usually cheaper and more easily expandable in today's computers more than processor speed.
2. In a single thread, the same temporary *n/2* memory is used no matter how many times the sorting is repeated or how many *n* elements arrays are sorted.

Therefore, we recommend Triple State Quicksort as a replacement for any C/C++ library qsort function. We also recommend the array based version of Triple State to be ported and tested under other programming languages to be considered as a replacement candidate.

## APPENDIX A

### A FEW NOTES ON DUAL PIVOT QUICKSORT

It has been debated recently that Yaroslavskiy's Dual Pivot Quicksort is superior to Bentely's implementation of Quicksort [4][13]. This finding has obviously been verified by Oracle as they replaced their implementation of Quicksort with Yaroslavskiy's new Dual Pivot Quicksort in the new Java 7. However, recent <u>analytical</u> studies suggest that this may not be the case [15][20] mainly due to the high number of swaps that Yaroslavskiy's method exhibits. From our study we can confirm the following:

1. Dual Pivot Quicksort algorithm definitely results in significantly high number of swaps. See Fig. 8, Fig. 10.
2. Dual Pivot Quicksort has the advantage of lower comparisons when compared to a moderately good Quicksort like GNU or VC++ in many cases. However, when compared to a superior Quicksort like Bentely's, this advantage doesn't hold. See Fig. 9.
3. The Java 6 former implementation of Quicksort used an <u>adapted</u> version of Bentely's Quicksort. It has to be said that Bentely's Quicksort was initially designed for C. It is pointer based. It uses C macros. And has a swapping function that will work only in C.

For the reasons above and looking deeply at our test results, it would be a far stretch to say that Dual Pivot Quicksort is best. However, we have to be fair and say that the Dual Pivot function we tested was also <u>adapted</u> from Yaroslavskiy's Java implementation with minimal changes to work in C/C++. The algorithm definitely has its advantages, especially in sorting dynamic strings. Further research would be of interest to try to reduce the high number of swaps in Dual Pivot Quicksort.

# APPENDIX B

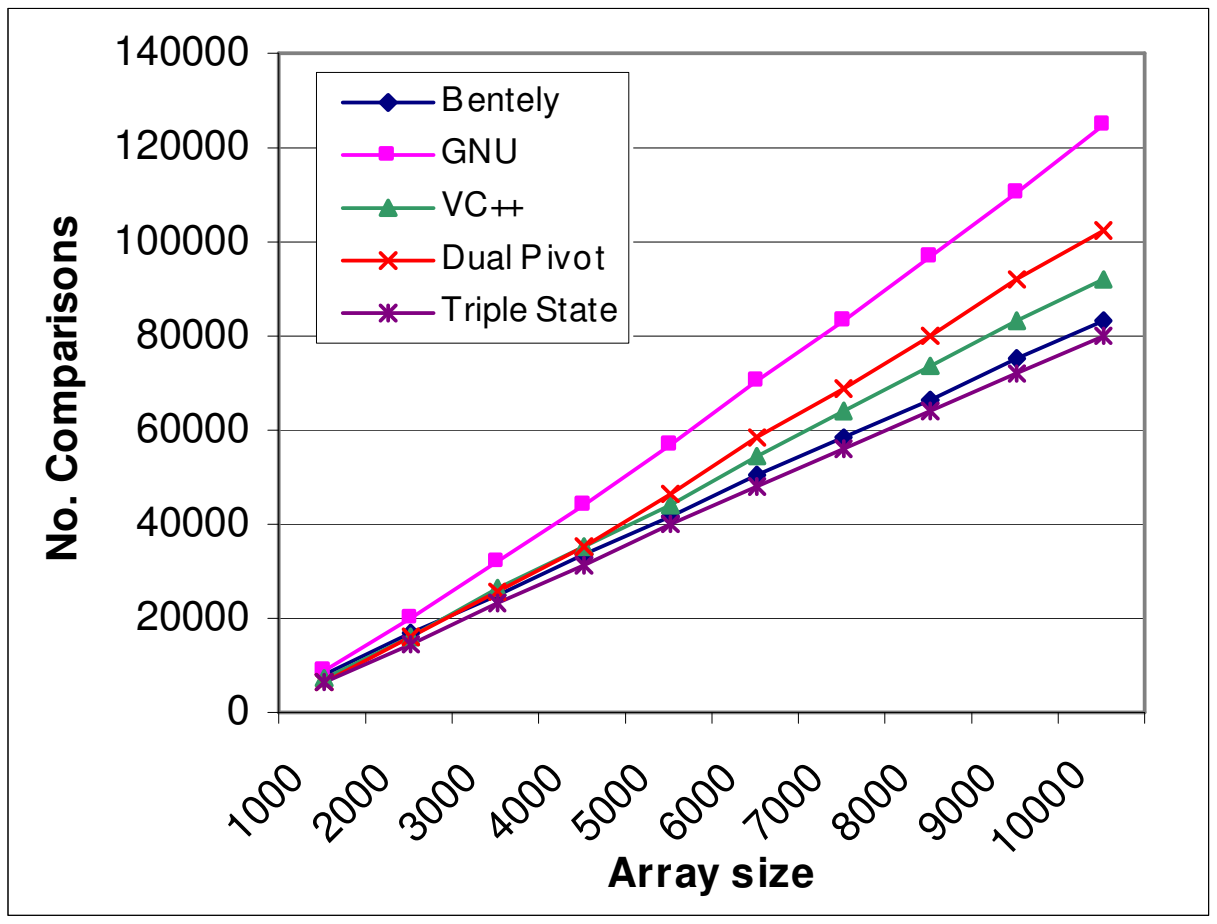

Fig. 30. Number of Comparisons, Sorted input, random elements

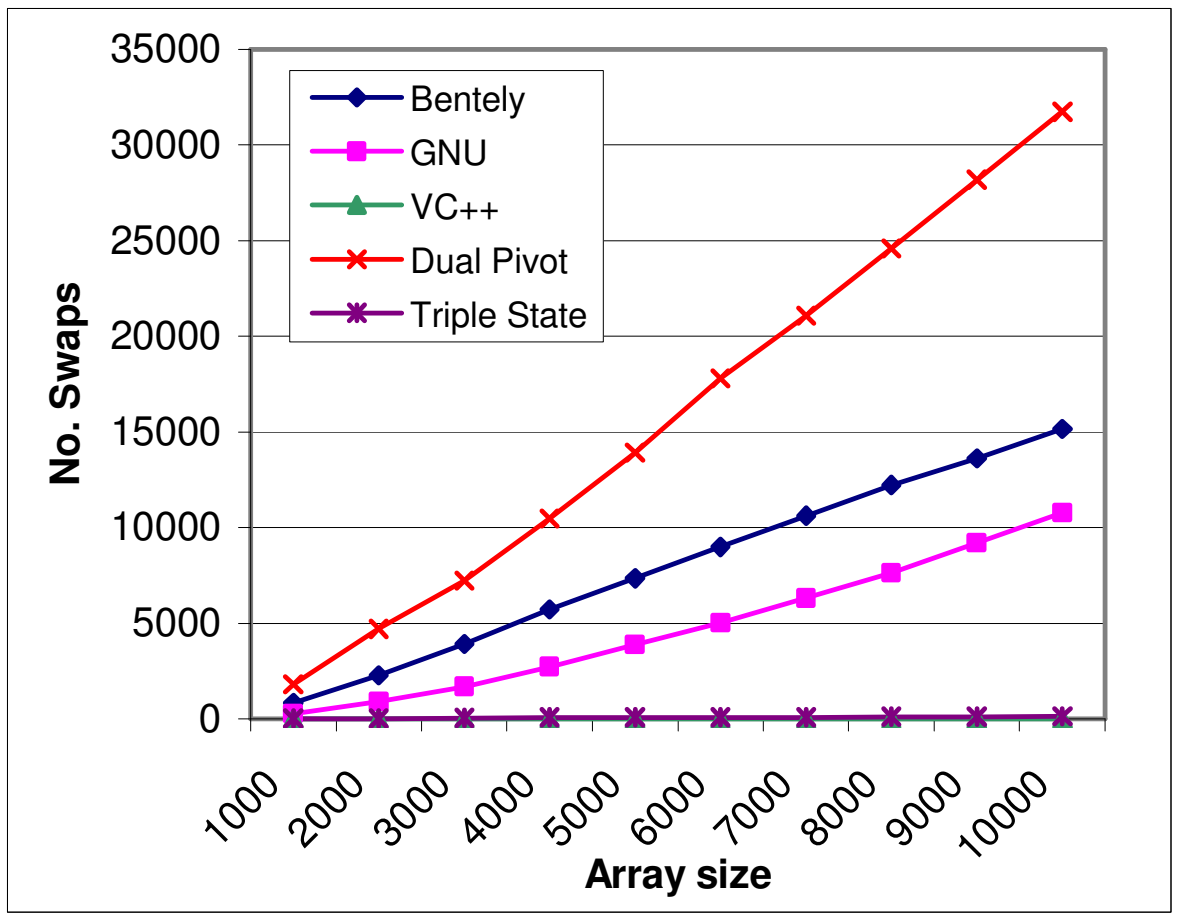

Fig. 31. Number of Swaps. Sorted input, random elements. (Triple State is very low, VC++ is zero. Note here that Triple State actually does zero swaps too but we are counting copy operations to the **p** variable)

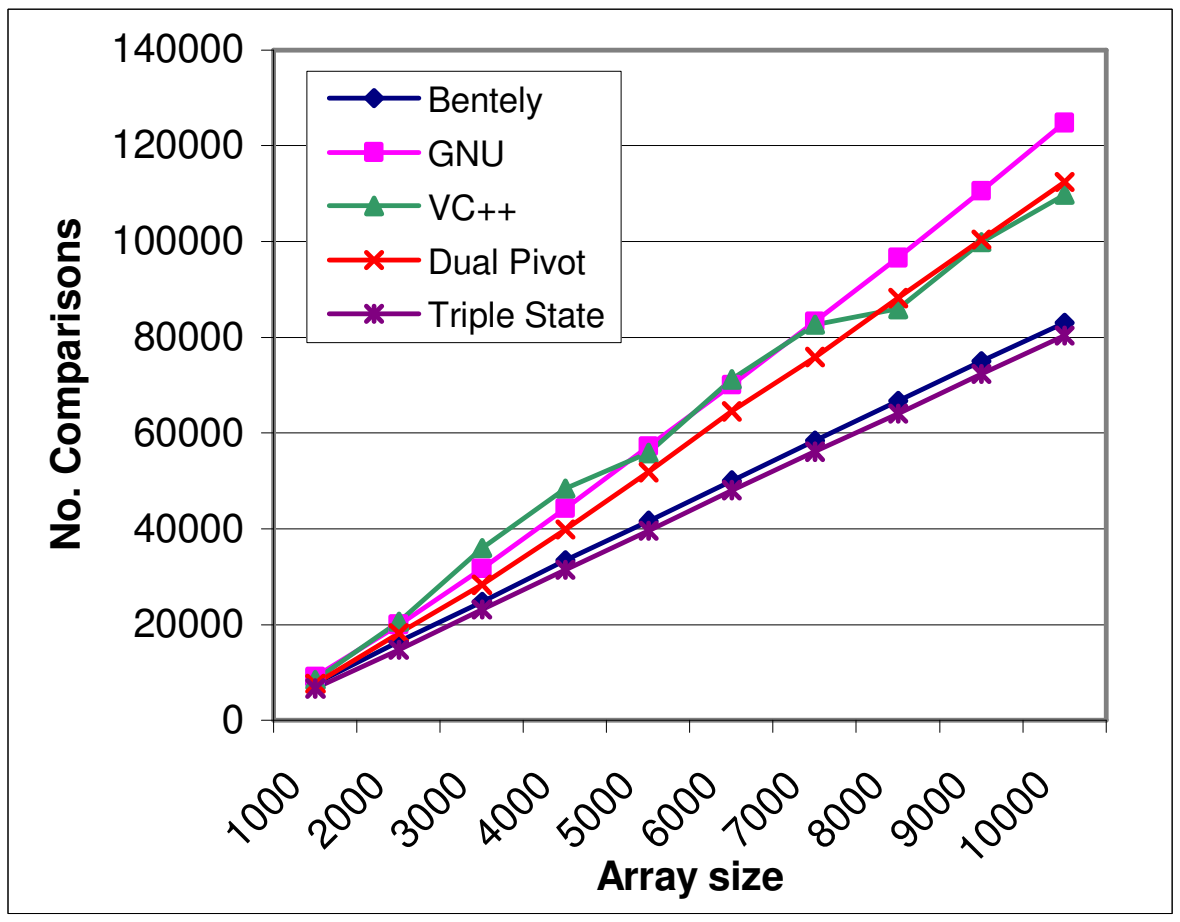

Fig. 32. Number of Comparisons, Reversed input. Random elements

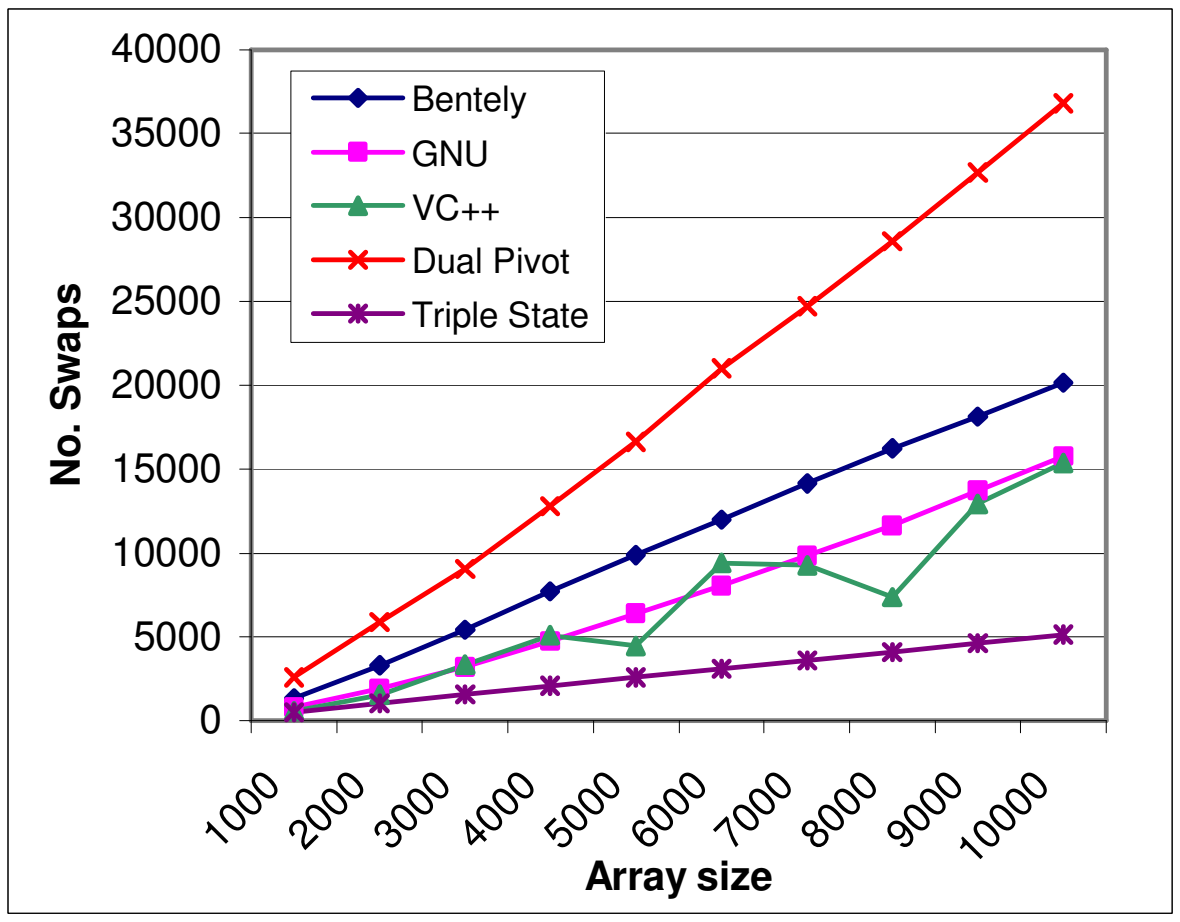

Fig. 33. Number of Swaps. Reversed input, random elements.

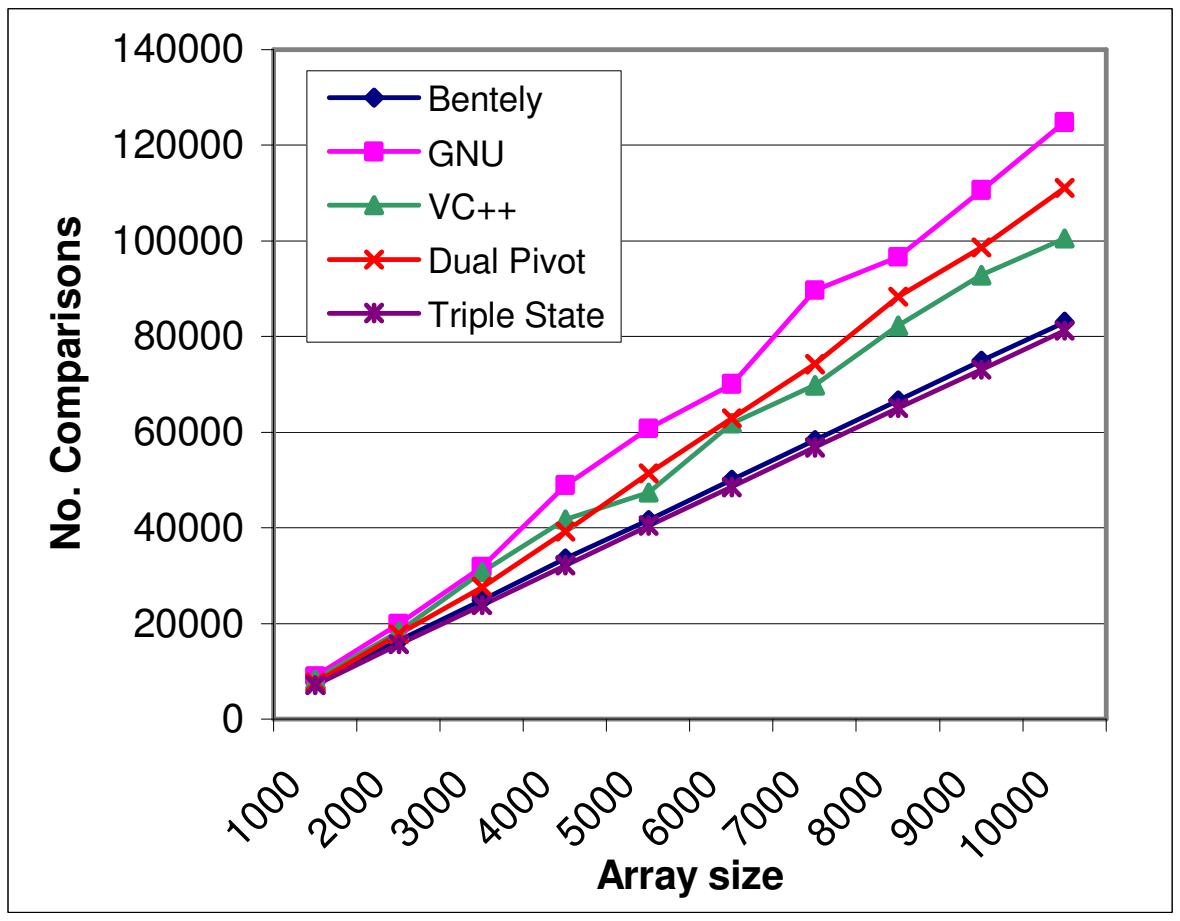

Fig. 34. Number of Comparisons, Front half reversed input, random elements

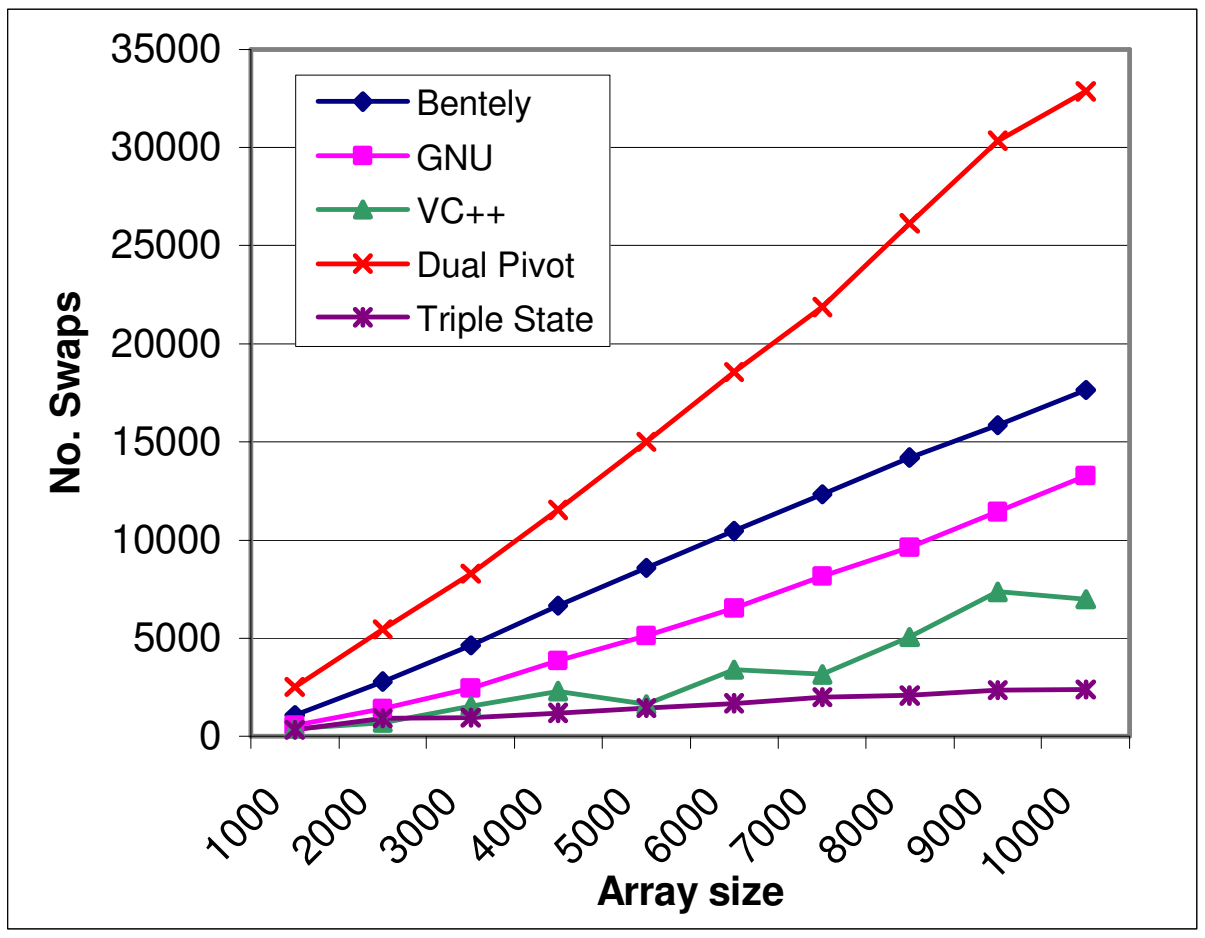

Fig. 35. Number of swaps. Front half reversed input, random elements.

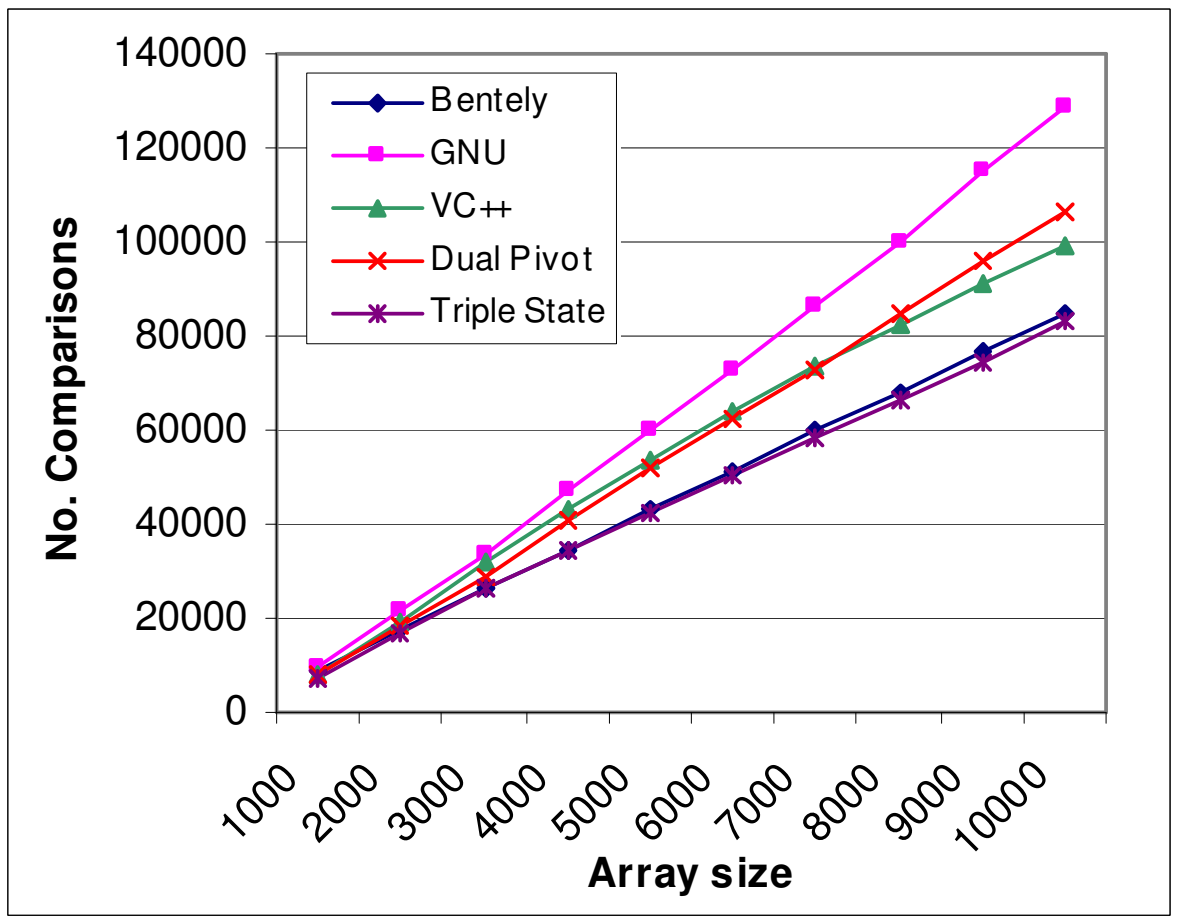

Fig. 36. Number of comparisons. Dithered input, random elements.

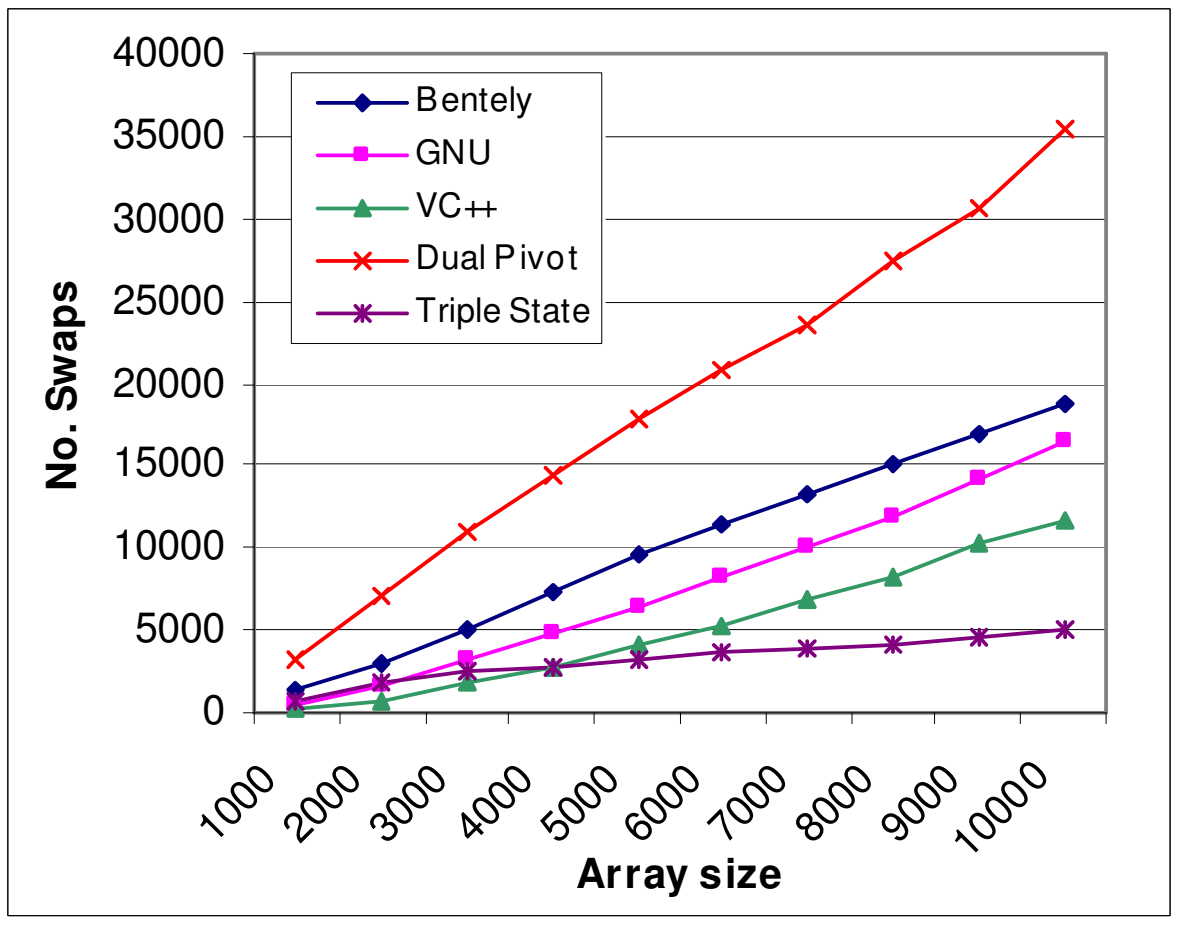

Fig. 37. Number of swaps. Dithered input, random elements.

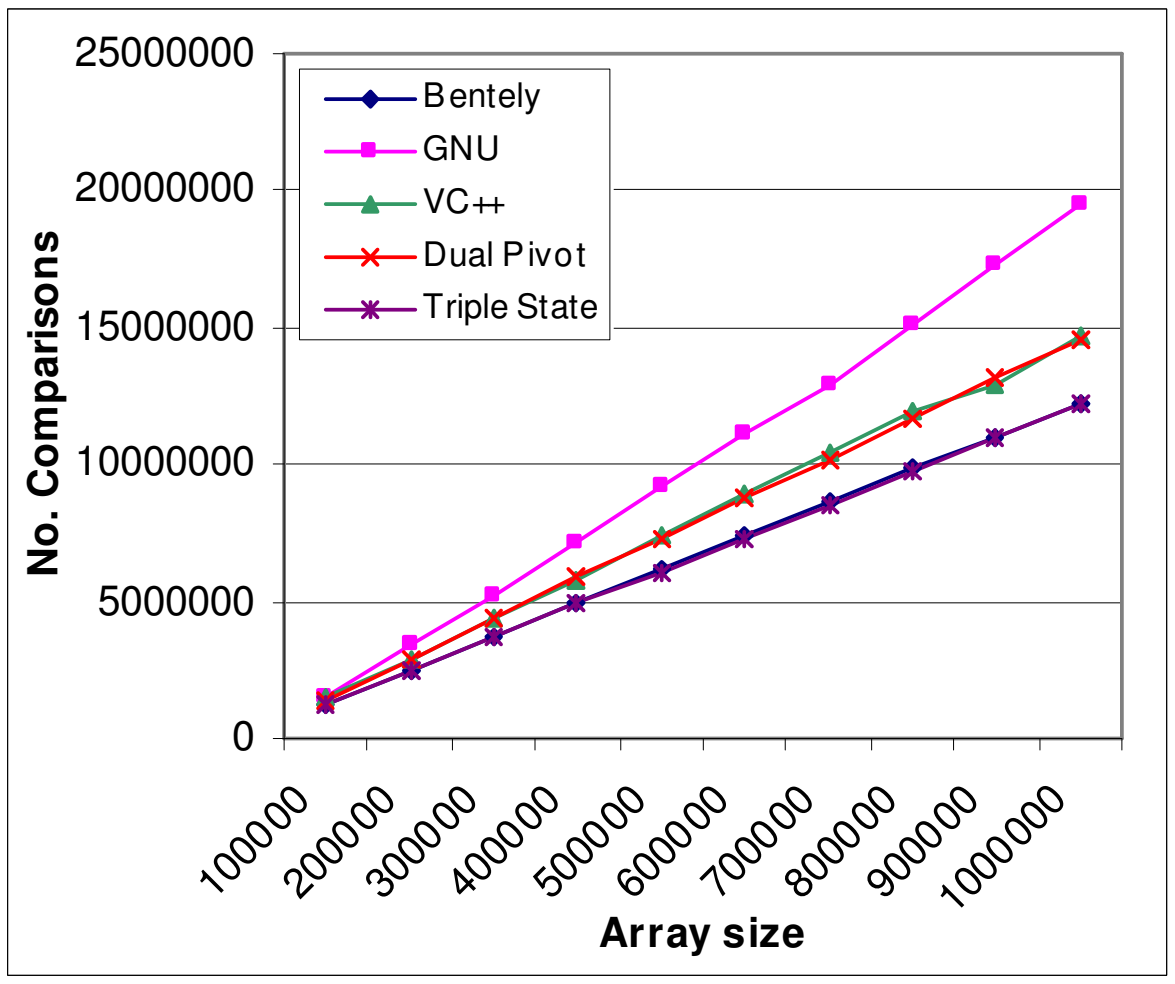

Fig. 38. Number of comparisons. Reversed input, random elements. Large arrays sizes.

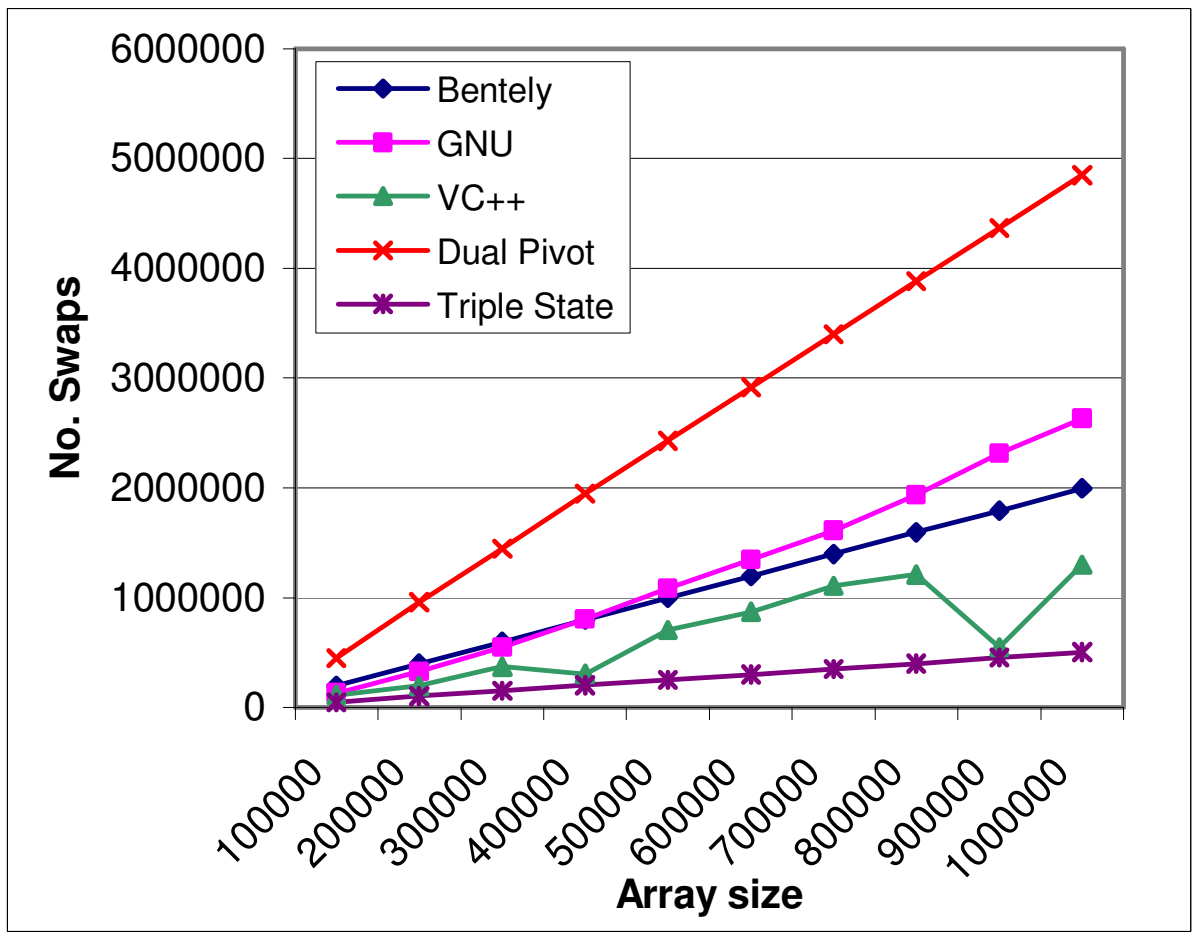

Fig. 39. Number of swaps. Reversed input, random elements. Large arrays sizes.

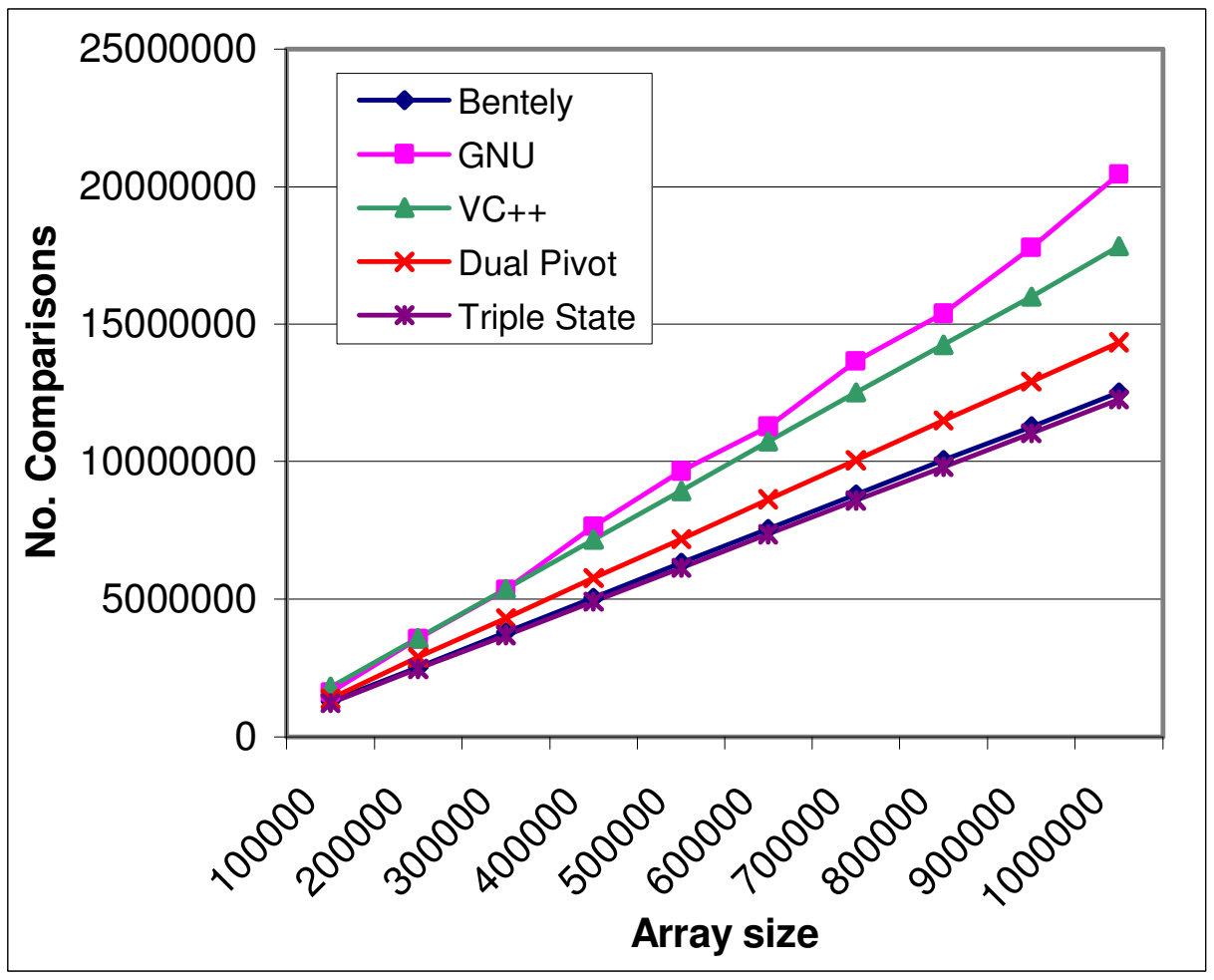

Fig. 40. Number of comparisons. Back half reversed input, random elements. Large arrays sizes.

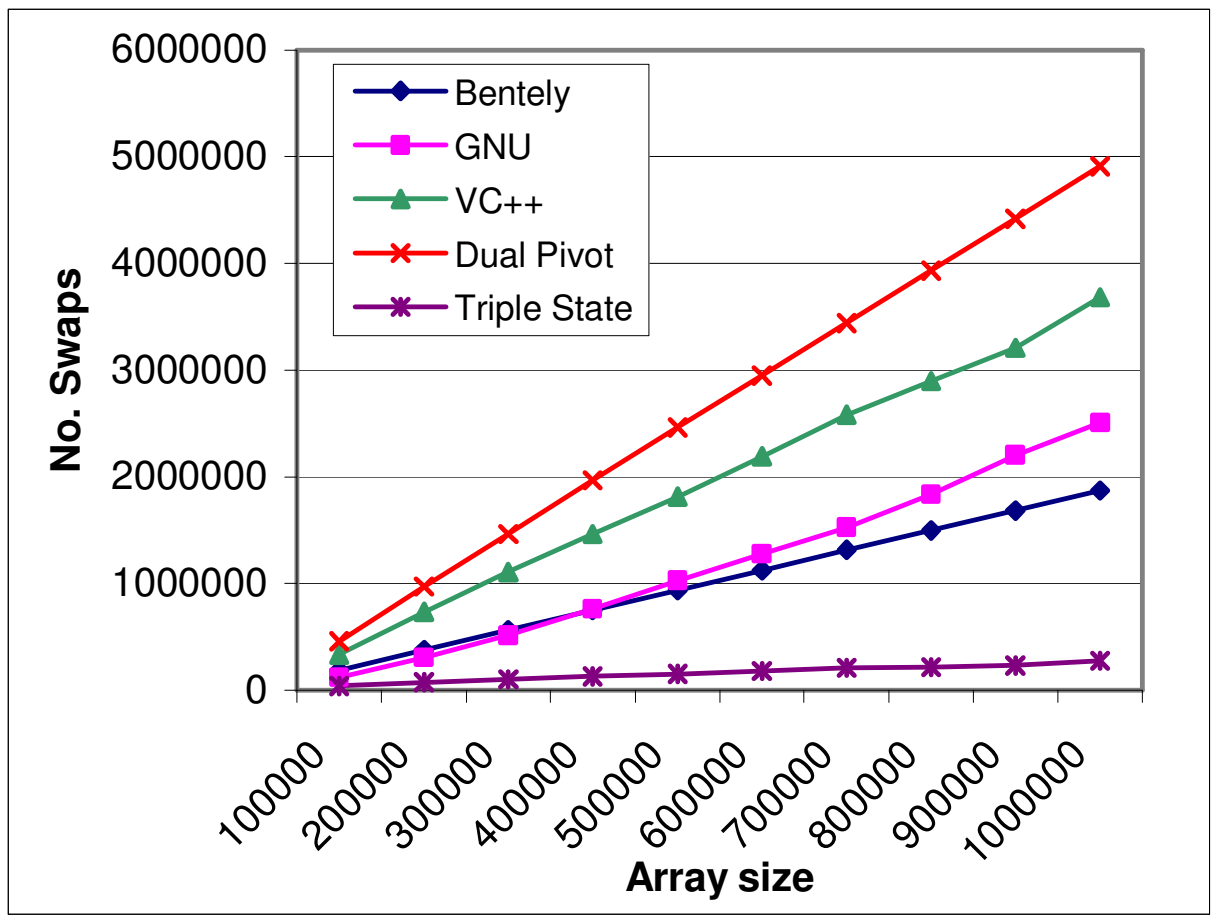

Fig. 41. Number of swaps. Back half reversed input, random elements. Large arrays sizes.

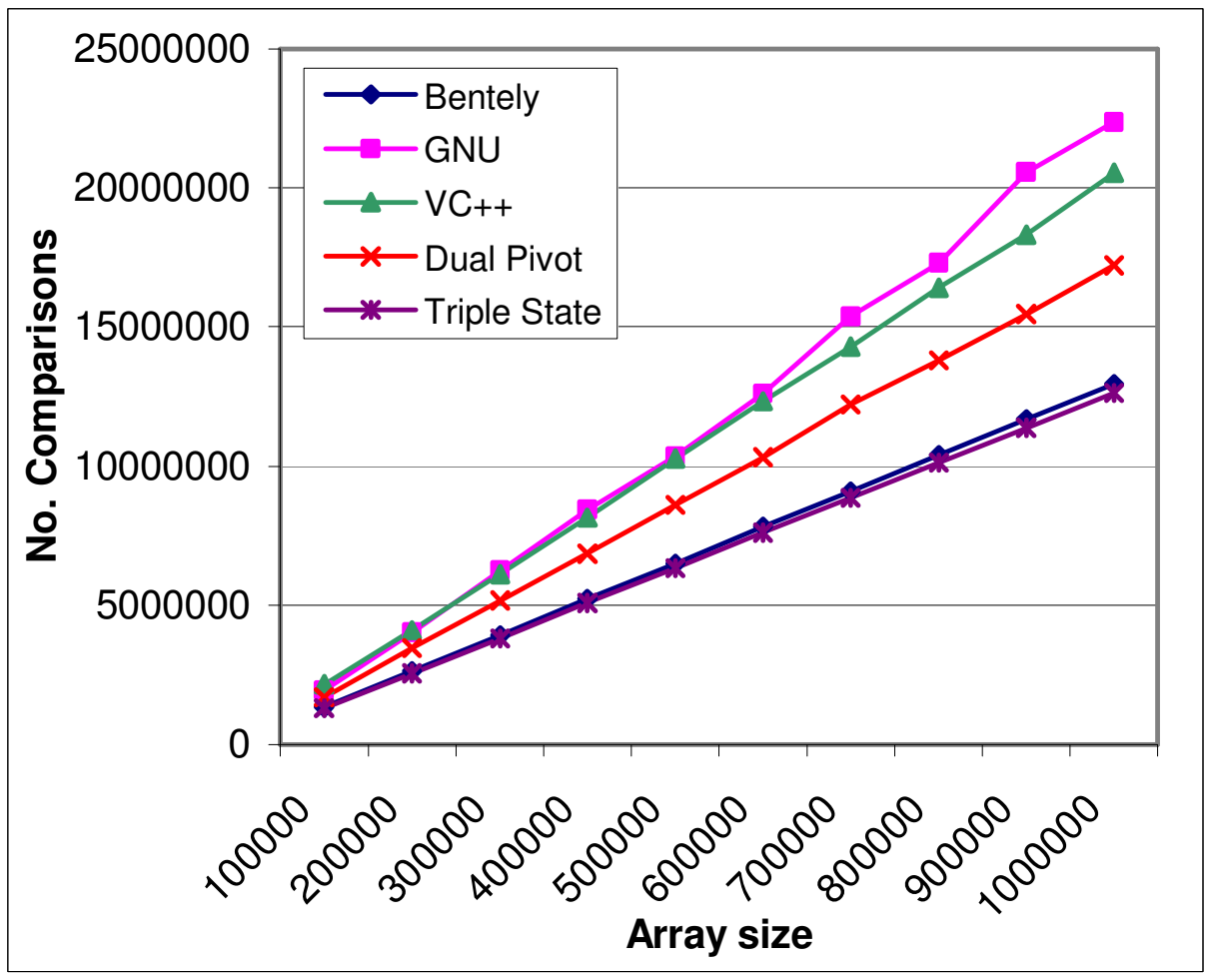

Fig. 42. Number of comparisons. Fort input, random elements. Large arrays sizes.

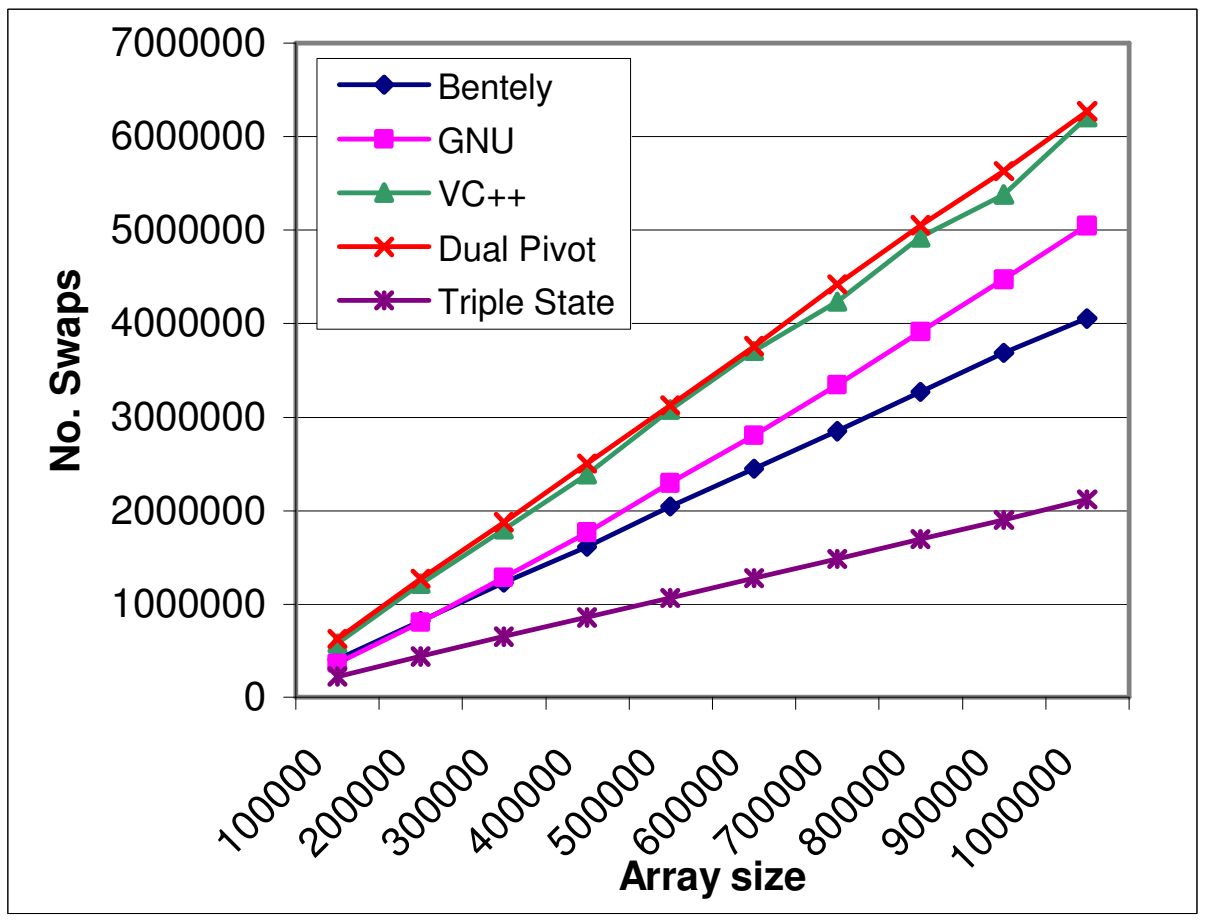

Fig. 43. Number of swaps. Fort input, random elements. Large arrays sizes.

## APPENDIX C

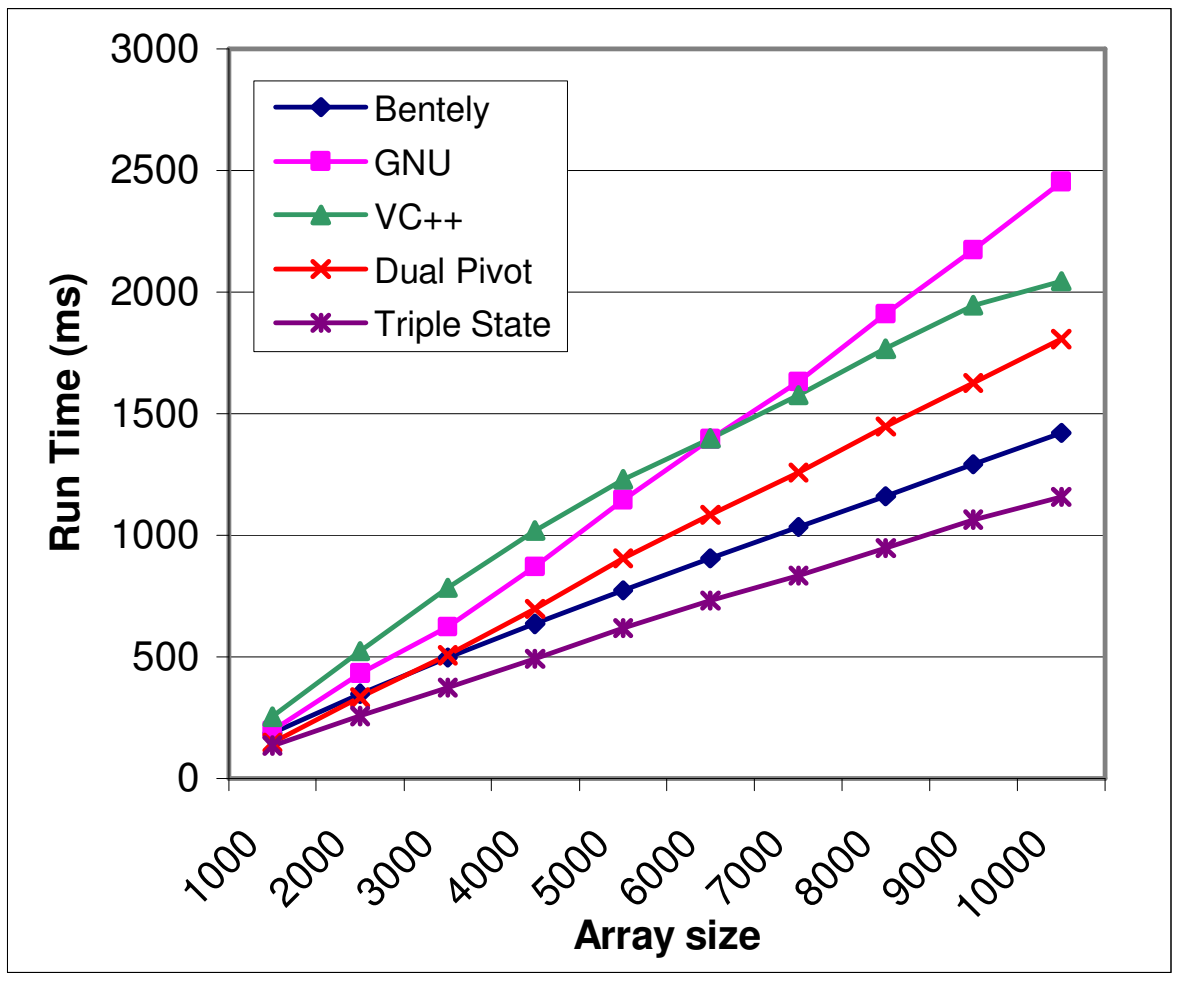

Fig. 44. Run time. Back half reversed. Random elements.

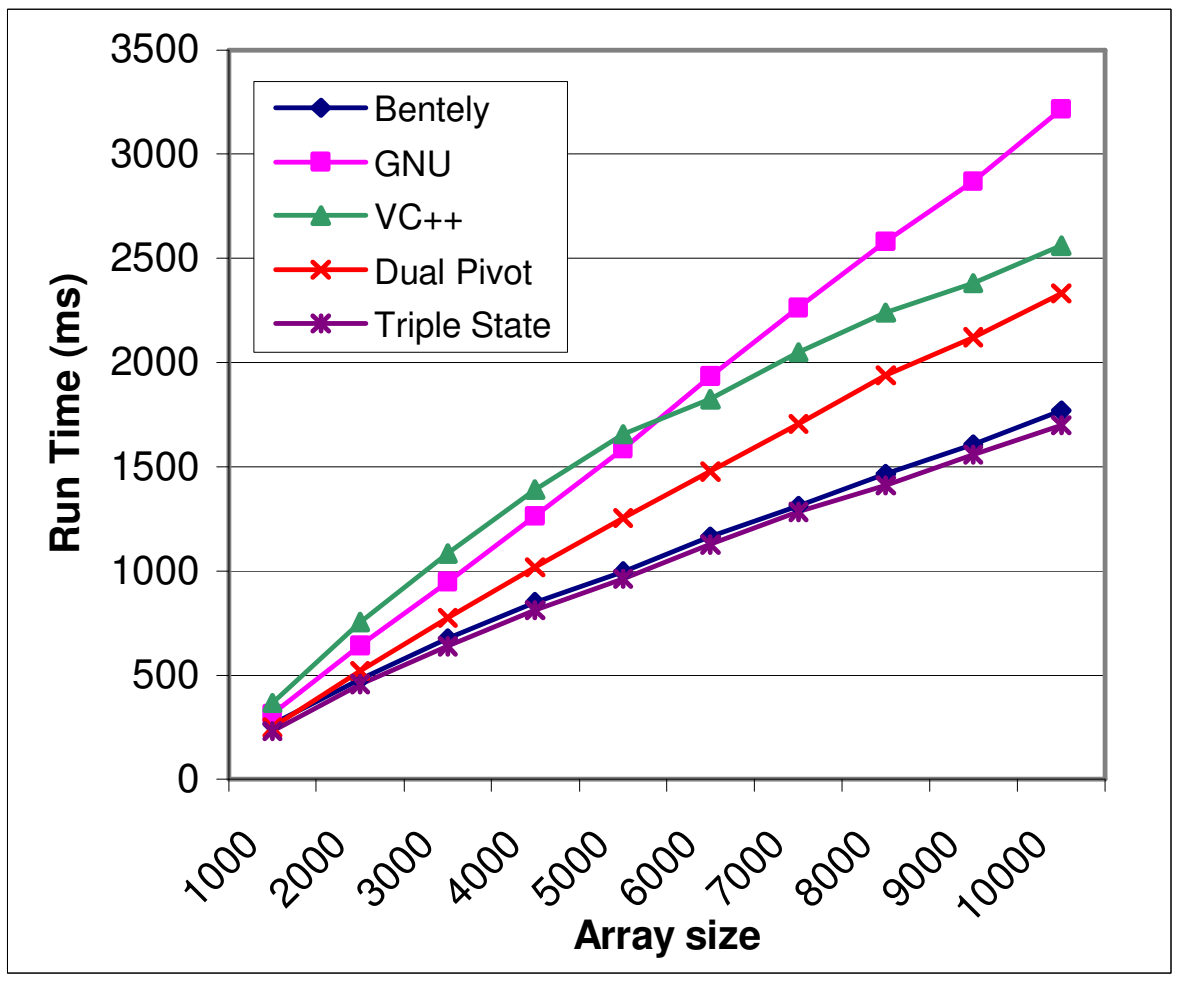

Fig. 45. Run time. Fort input. Random elements.

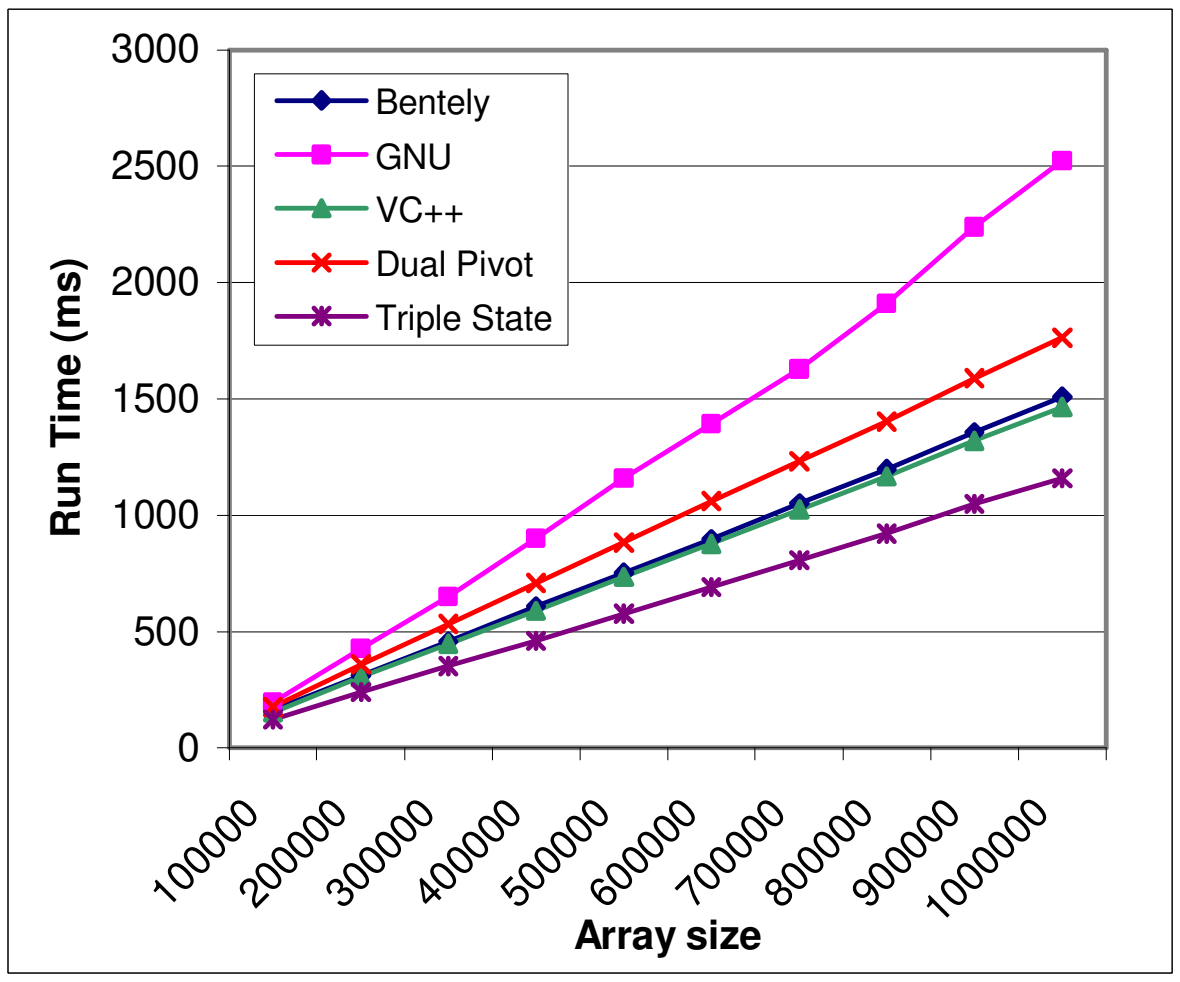

Fig. 46. Run Time. Sorted input. Random elements. Large arrays sizes.

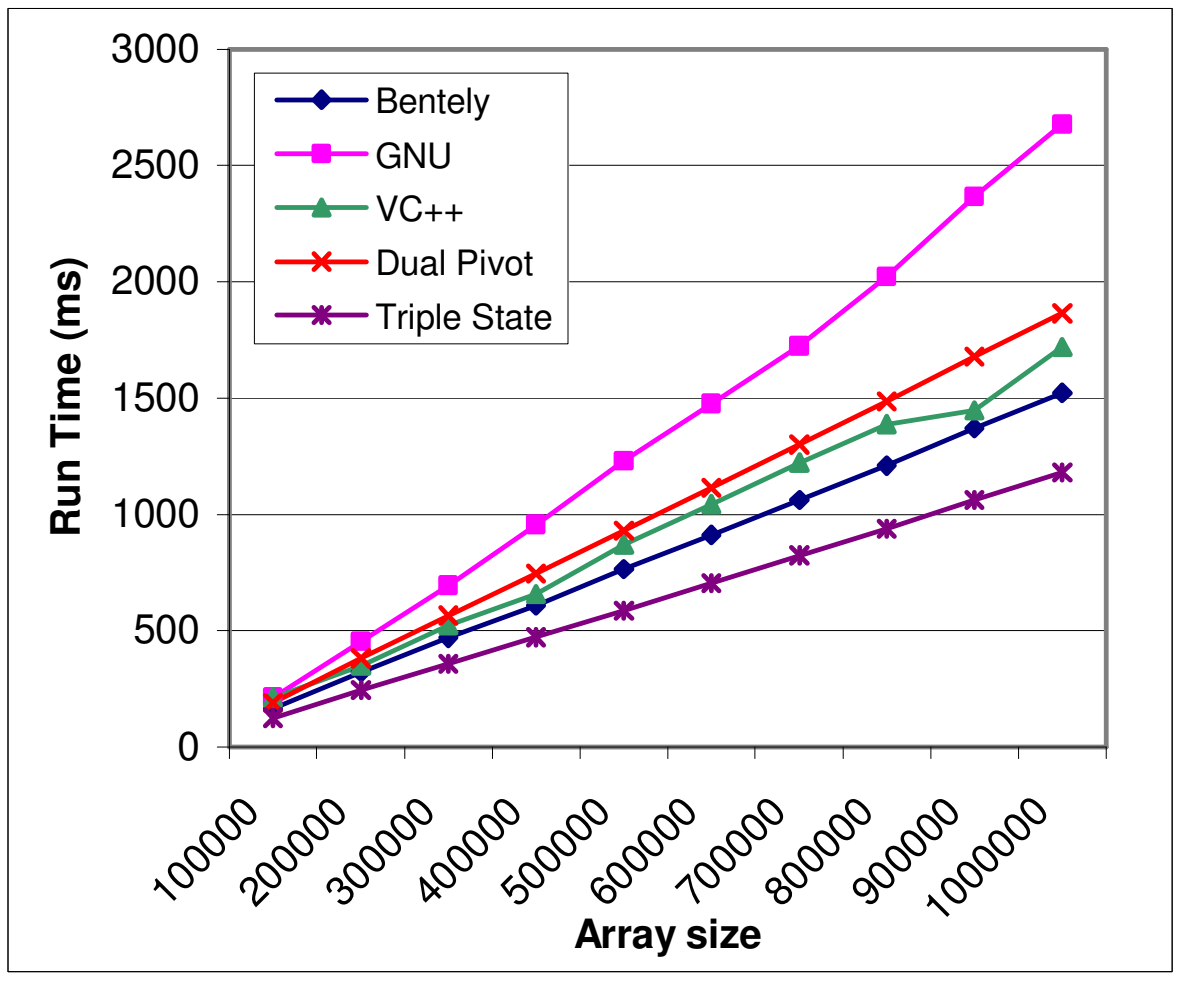

Fig. 47. Run Time. Reversed input. Random elements. Large arrays sizes.

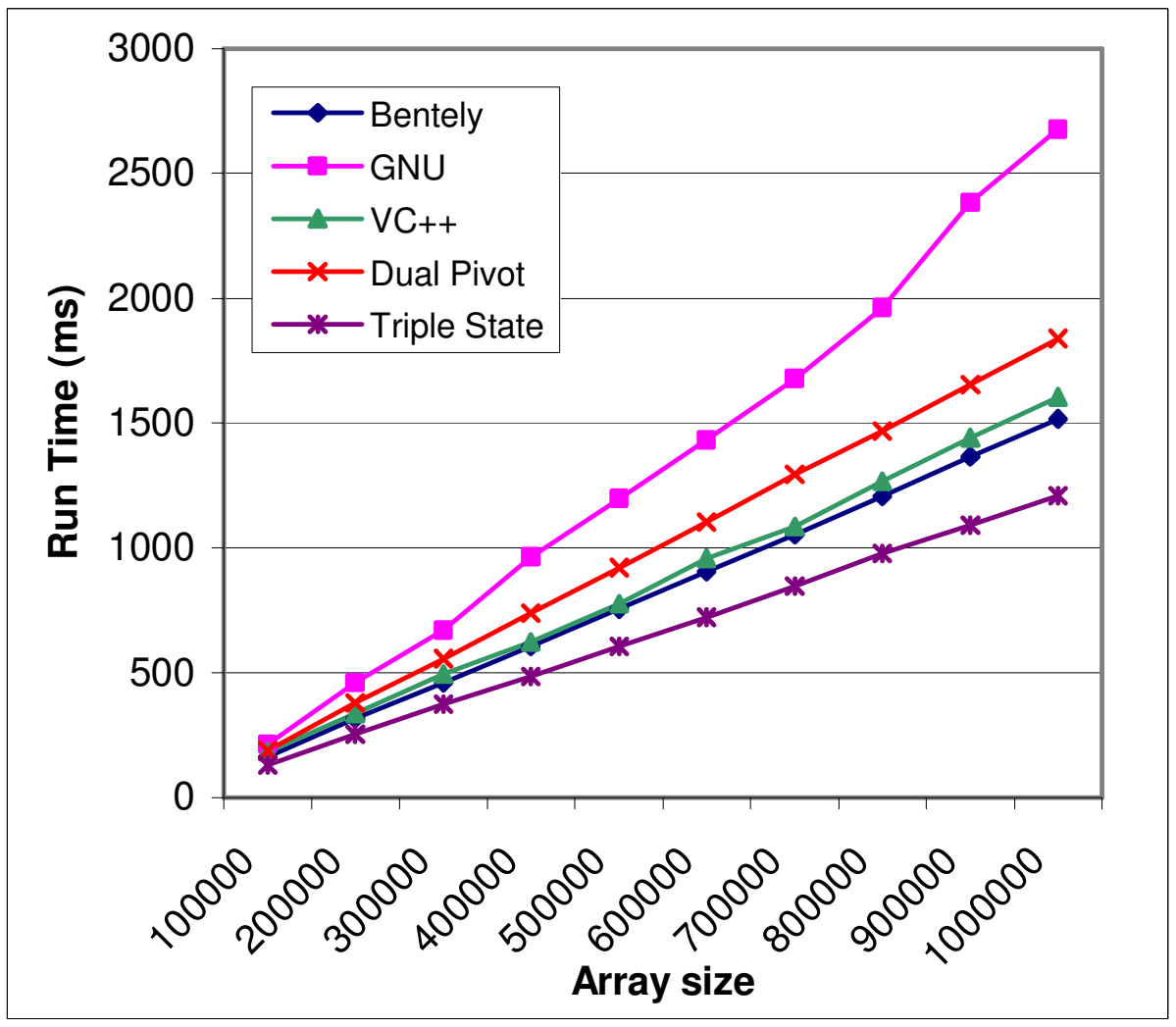

Fig. 48. Run Time. Front half reversed input. Random elements. Large arrays sizes.

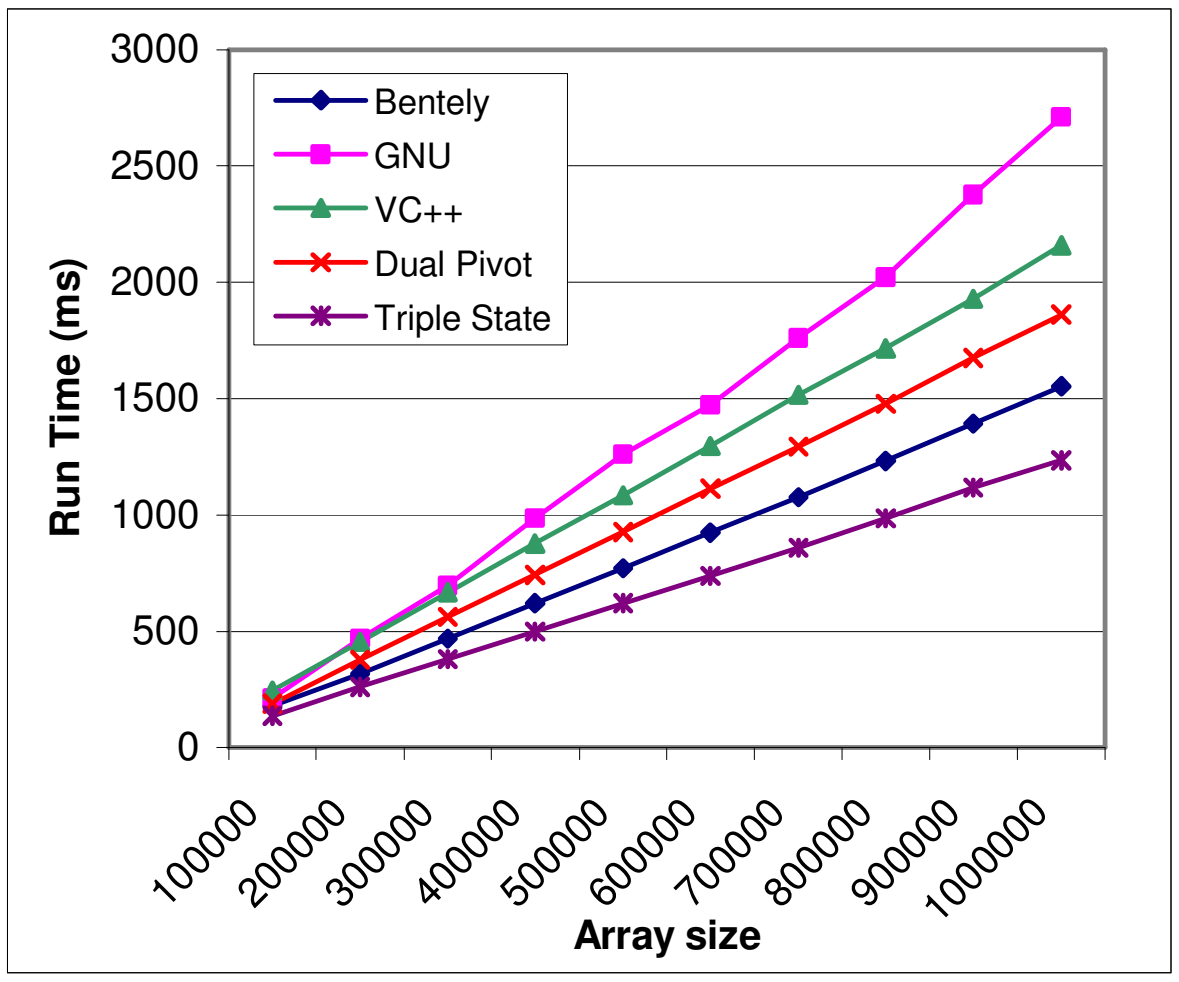

Fig. 49. Run Time. Back half reversed input. Random elements. Large arrays sizes.